\documentclass[bibyear]{aa}
\usepackage[varg]{txfonts}
\usepackage{graphicx}
\usepackage{longtable}
\usepackage{epstopdf}
\usepackage{lscape}
\newcommand{\e}{et al.\ }
\newcommand{\ha}{H$\alpha$}
\newcommand{\rha}{$r-H\alpha$}

\begin{document}

\title{X-ray survey of the North-America and Pelican star-forming complex
(NGC7000/IC5070)
}

\date{Received date / Accepted date}

\author{Francesco Damiani\inst{1}, Ignazio Pillitteri\inst{1,2},
and Loredana Prisinzano\inst{1}
}
\institute{INAF - Osservatorio Astronomico di Palermo G.S.Vaiana,
Piazza del Parlamento 1, I-90134 Palermo, Italy \\
\email{damiani@astropa.inaf.it}
\and
Harvard-Smithsonian Center for Astrophysics, 60 Garden St., Cambridge MA
02138 USA.
}

\abstract{}{
We present the first extensive X-ray study of
the North-America and Pelican star-forming region (NGC~7000/IC~5070),
with the aim of finding and characterizing the young population of this
cloud.
}{
X-ray data from Chandra (four pointings) and XMM-Newton (seven
pointings) were reduced and source detection algorithm applied to each image.
We complement the X-ray data with optical and near-IR data from the
IPHAS, UKIDSS, and 2MASS catalogs, and with other published optical and
Spitzer IR data. More than 700 X-ray sources are detected, the majority
of which have an optical/NIR counterpart.
This allows to identify young stars in different stages of formation.
}{
Less than 30\% of X-ray sources are identified with a previously
known young star.
We argue that most X-ray sources with an optical/NIR counterpart,
except perhaps for a few tens
at near-zero reddening, are likely candidate members of the star-forming
region, on the basis of both their optical/NIR magnitudes and colors,
and of X-ray properties like spectrum hardness or flux variations.
They are characterized by a wide range of extinction, and sometimes
near-IR excesses, both of which prevent derivation of accurate stellar
parameters. The optical color-magnitude diagram suggests ages between
1-10~Myrs.
The X-ray members have a very complex spatial distribution with some
degree of subclustering,
qualitatively similar to that of previously known members. The detailed
distribution of X-ray sources relative to the objects with IR excesses
identified with Spitzer
is sometimes suggestive of sequential star formation, especially
near the ``Gulf of Mexico'' region, probably triggered by the O5 star which
illuminates the whole region. We confirm that around this latter star no
enhancement in the young star density is found, in agreement with
previous results.
Thanks to the precision and depth of the IPHAS and UKIDSS data used, we
also determine the local optical/IR reddening law, and compute an updated
reddening map of the entire region.
}{}


\titlerunning{X-ray observations of NGC7000/IC5070}
\authorrunning{Damiani et al.}
\maketitle

\section{Introduction}
\label{intro}

\begin{figure*}
\sidecaption
\includegraphics[width=12cm]{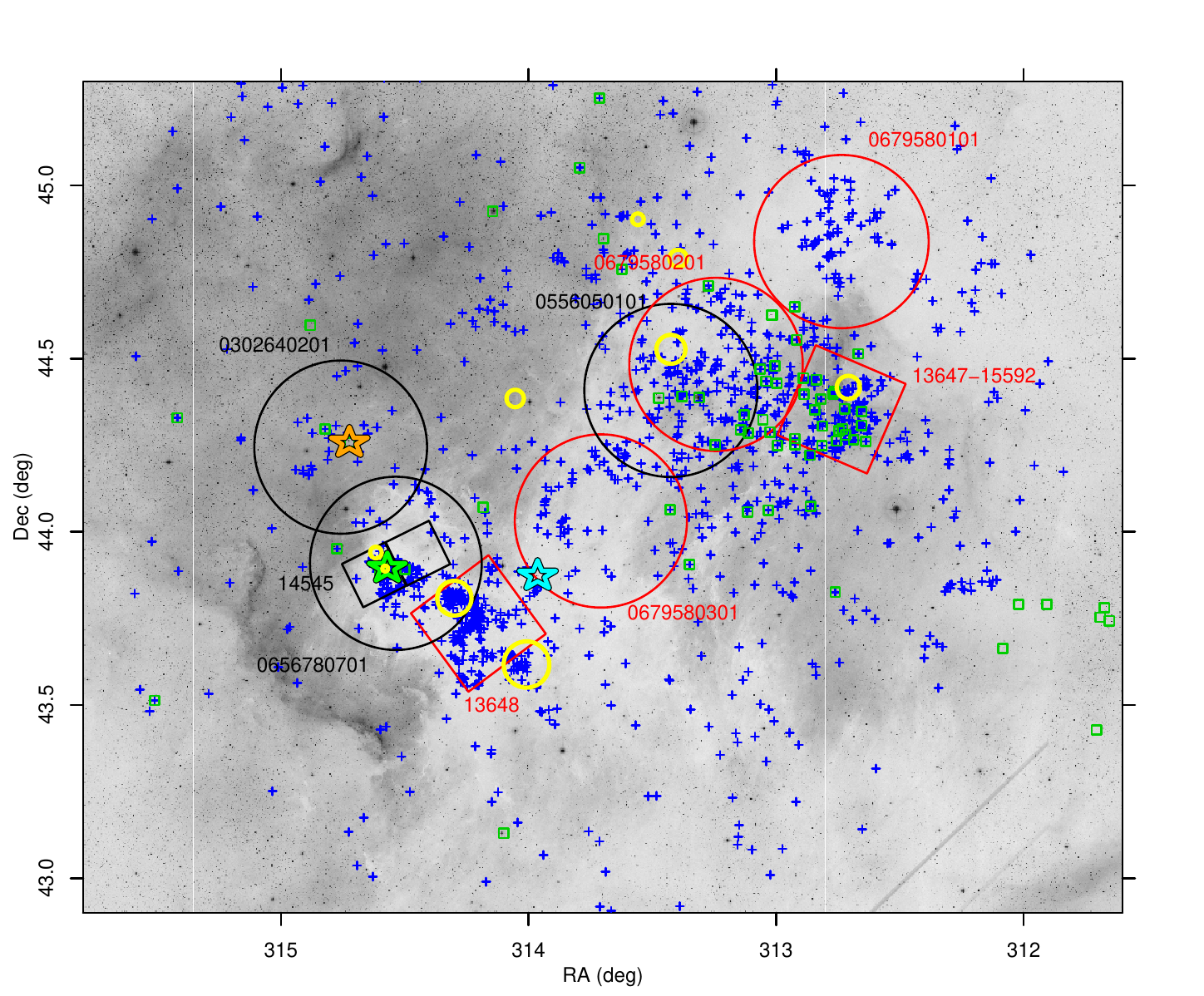}  
\caption{
Digital-Sky-Survey II red image of the North-America and
Pelican nebulae (inverted greyscale), with overlaid the regions covered
by XMM-Newton observations (big circles), and Chandra ACIS observations
(squares), each labeled with its ObsId number.
X-ray fields of view are drawn in red for new data, and black for archive data.
ACIS-I ObsIds 13467 and 15592 share the same pointing direction and roll
angle. Archive XMM ObsIds 0656780701 and 0656781201 also share the same pointing
direction, but with a slightly different roll angle; only 0656780701 is
labeled in the Figure.
Small green squares are the H58 emission-line stars; blue
crosses are YSOs from RGS11.
Yellow circles indicate positions and approximate sizes of the Cambr\'esy
\e (2002) clusters of 2MASS NIR sources.
The three bigger star symbols indicate respectively the O5 star 2MASS
J20555125+4352246 (cyan), the FU~Ori object candidate HBC722 (= V2493~Cyg =
2MASS J20581702+4353433, green), and the FU~Ori
star V1057~Cyg (= 2MASS J20585371+4415283, orange).
\label{dss-fov-yso}}
\end{figure*}

The star-formation region associated with the North-America (NGC~7000)
and Pelican (IC~5070) Nebulae in Cygnus have received increased
attention in the last decades. The two nebulae are catalogued as
separate objects because of their visual appearance, but are now
commonly regarded as parts of the same physical gas and dust cloud.
The properties of the region were reviewed by Reipurth and Schneider
(2008).
The ionizing source of the nebulae has remained a mystery for decades,
but was at last identified with a highly reddened O5 star (2MASS
J20555125+4352246, $A_V \sim 9.6$) by Comer\'on and Pasquali (2005), hidden
by the dark dust cloud LDN935, lying between NGC~7000 and IC~5070.
This star is remarkably isolated according to these authors: there are
no other massive stars close to it, or they would have been easily
detected even if highly obscured.

Since the work of Herbig (1958, henceforth H58) it was known that several
tens young pre-main-sequence (PMS) stars are found in this region.
Welin (1973) presented also an objective-spectra study of emission-line
stars in the region, and a few more young stars were studied spectroscopically
by Corbally, Strai\v{z}ys and Laugalys (2009). 
Narrow-band filter studies have also revealed many emission-line star
candidates (e.g.\ Witham \e 2008, Armond \e 2011), as well as many
Herbig-Haro flows (Bally and Reipurth 2003, Armond \e 2011, and also
Bally \e 2014). The strong
obscuration towards many parts of the cloud undoubtedly prevents many
other PMS stars to be found optically, but was no obstacle for the
sensitive survey for Young Stellar Objects (YSOs) made with the Spitzer
Space Observatory
(IRAC camera, Guieu \e 2009; IRAC$+$MIPS, Rebull \e 2011, henceforth
RGS11).
Cambr\'esy \e (2002) also studied this region using 2MASS data, tracing
spatial extinction variations, and finding nine clusters of near-IR
(NIR) objects.

The distance of the cloud was estimated to be in the range 500-600~pc by
Laugalys \e (2006, 2007). The line of sight towards Cygnus is tangent to
the corresponding spiral arm, and therefore objects at vastly different
distance from the Sun can appear visually close in the sky: the Cyg~OB2
association for example, only a few degrees away from NGC7000,
lies at a much larger distance $\sim 1.5$~kpc.
In the direction of the North-America/Pelican (NAP), Strai\v{z}ys, Corbally
and Laugalys (2008) find a larger-than-typical
$E(J-H)/E(H-K)$ ratio of 2.0.

The NAP region is also known to host two PMS stars in
the FU~Ori class: V1057~Cyg and V2493~Cyg. This latter was only known as
a Classical T~Tauri star (CTTS) named HBC722 before its outburst in
2010, and intensively studied thereafter (Semkov \e 2010, Covey \e 2011,
Aspin 2011, Miller \e 2011, Green \e 2011, K\'osp\'al \e 2011, 2013,
Dunham \e 2012, Semkov \e 2012, Liebhart \e 2014, and Lee \e 2015).
V1057~Cyg was observed in X-rays (but not detected) by Skinner \e
(2007), while a positive X-ray detection of V2493~Cyg was presented by
Liebhart \e (2014). No other published X-ray study exists of this
region, to our knowledge.

The strong and nonuniform obscuration towards the cloud, and its large
size ($\sim 2^{\circ}$) are formidable difficulties when attempting to
characterize the global young-star population of this star-forming
region. The Spitzer RGS11 study of YSOs is extremely important in this respect,
since it covered basically all the cloud, and was sensitive to even
highly obscured objects. Since the selection of YSOs in RGS11 is based on the
existence of non-photospheric excess IR emission from circumstellar dust
(disks), it is however biased against diskless PMS stars (Class~III
objects in the SED-based nomenclature, or Weak-line T~Tauri stars -
WTTS, as opposed to CTTS - in the spectroscopic nomenclature).
At typical ages of PMS stars,
both CTTS and WTTS are however bright X-ray sources (with X-ray
luminosities in the range $\log L_X \sim 29-31$ erg/s).
Therefore, we have obtained new X-ray observations at selected positions
in the NAP Nebula, using both XMM-Newton and Chandra
X-ray observatories, with the purpose of obtaining a more complete and
unbiased census of the PMS population of this star-forming region.
In addition, we analyze here other X-ray datasets from these
observatories, which were only partially examined in the published
literature.

Moreover, we take advantage of two important wide-area optical and NIR
surveys: IPHAS (Drew \e 2005) and UKIDSS (Lawrence \e 2007), covering
the entire cloud, with limiting magnitudes far deeper than most other
optical/NIR catalogues available in the same region.

This paper is structured as follows: Section~\ref{obs} describes the
X-ray data presented, and Section~\ref{xdata} their analysis.
Section~\ref{optir} describes the optical/NIR data also used in this work.
Section~\ref{results} presents all results obtained.
Section~\ref{discuss} is a discussion of the results, and
Section~\ref{concl} summarizes our conclusions.

\begin{figure}
\resizebox{\hsize}{!}{
\includegraphics[]{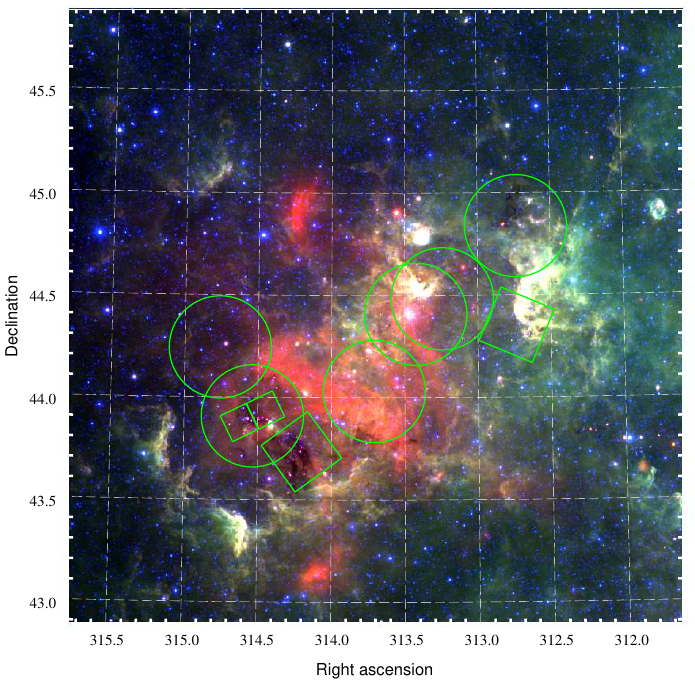}}
\caption{
True-color WISE image of the region. Red: 22$\mu$, green: 12$\mu$, and
blue: 4.6$\mu$ channels, respectively. The big circles and squares
indicates the X-ray FOVs as in Fig.~\ref{dss-fov-yso}.
North is up and East to the left.  At a distance of 560~pc, the 3-degree
sky region shown measures 29.3~pc on a side.
\label{wise}}
\end{figure}

\begin{figure}
\resizebox{\hsize}{!}{
\includegraphics[width=17.5cm]{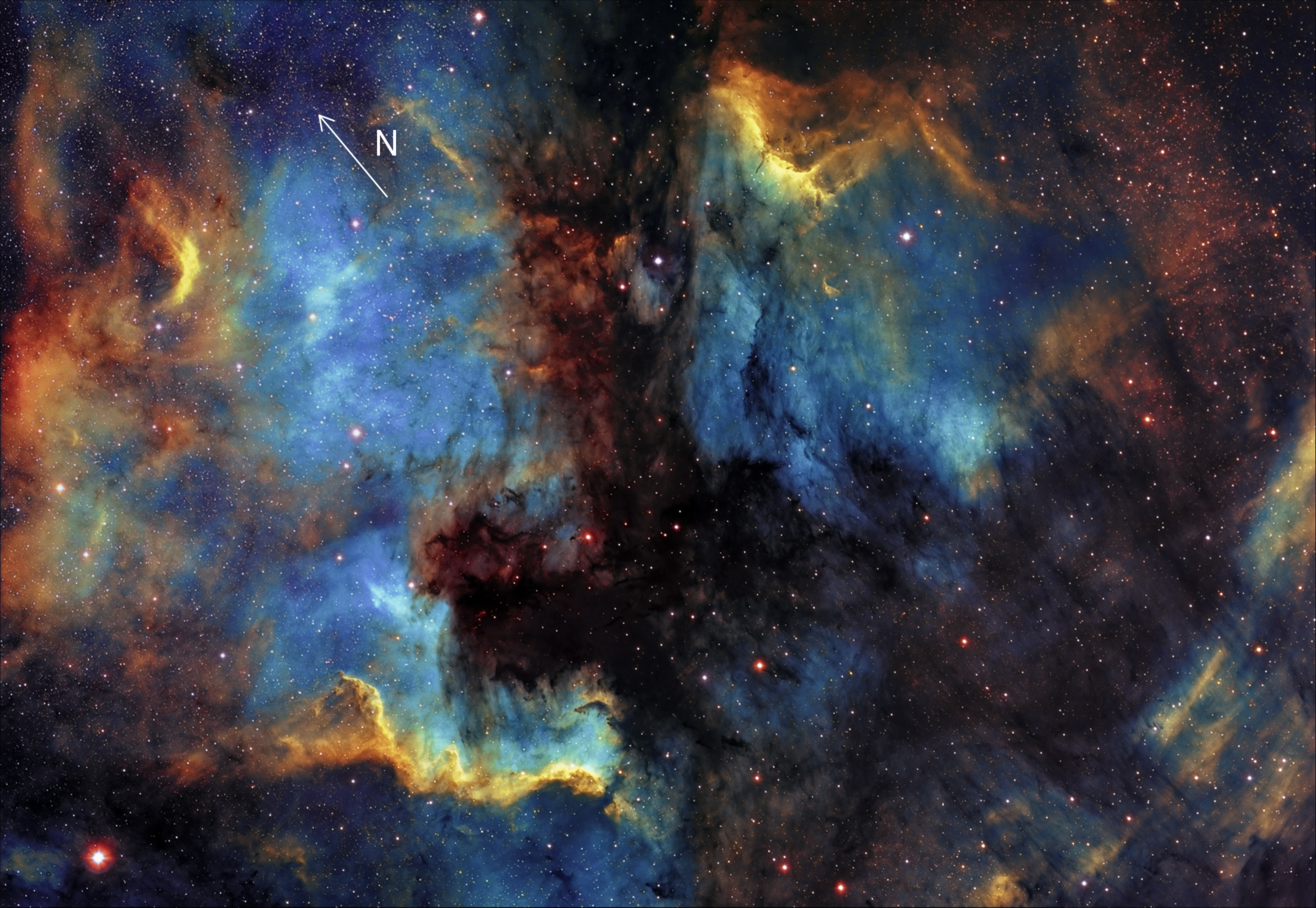}
}
\caption{
Optical emission-line image of part of the region (courtesy
Ond\v{r}ej Podluck\'y), in the lines of [O III] (blue), \ha\ (green) and
[S II] (red), showing increasing ionization towards a location behind
the obscuring dust.
The white arrow indicates North.
\label{ngc7000_ic5067_hst_final}}
\end{figure}

\section{X-ray observations}
\label{obs}

\begin{table*}[ht]
\centering
\caption{XMM-Newton observing log.
Dates are UT, exposure times $t_{exp}$ in seconds.}
\label{table.xmm}
\begin{tabular}{cccccccc}
  \hline
ObsId & RA & Dec & PI & PN start date & PN $t_{exp}$ & MOS start date & MOS $t_{exp}$ \\ 
  \hline
0679580101 & 312.7004 & 44.85204 & Damiani & 2011-11-24,00:13:45 & 25591 & 2011-11-23,23:51:28 & 30239 \\ 
  0679580201 & 313.2083 & 44.49669 & Damiani & 2011-11-24,10:32:48 & 13750 & 2011-11-24,09:34:44 & 19092 \\ 
  0679580301 & 313.6737 & 44.04564 & Damiani & 2011-11-24,18:48:44 & 23827 & 2011-11-24,18:26:25 & 28239 \\ 
  0302640201 & 314.7238 & 44.25789 & Skinner & 2005-11-26,21:12:19 & 21027 & 2005-11-26,20:50:00 & 23733 \\ 
  0556050101 & 313.4454 & 44.38394 & Motch & 2008-05-02,11:27:25 & 14498 & 2008-05-02,10:58:53 & 17392 \\ 
  0656780701 & 314.5710 & 43.89539 & Schartel & 2010-11-25,19:29:52 & 15968 & 2010-11-25,19:07:30 & 18237 \\ 
  0656781201 & 314.5710 & 43.89539 & Schartel & 2011-05-26,12:47:35 & 17560 & 2011-05-26,12:25:15 & 21327 \\ 
   \hline
\end{tabular}
\end{table*}

\begin{table*}[ht]
\centering
\caption{Chandra observing log.
Dates are UT, exposure times $t_{exp}$ in seconds.}
\label{table.acis}
\begin{tabular}{ccccccc}
  \hline
ObsId & RA & Dec & PI & Detector & Start date & $t_{exp}$ \\ 
  \hline
13648 & 314.2114 & 43.74551 & Damiani & ACIS-I & 2011-12-26,20:43:50 & 44588 \\ 
  13647 & 312.7507 & 44.35863 & Damiani & ACIS-I & 2012-11-20,03:37:14 & 22766 \\ 
  15592 & 312.7507 & 44.35863 & Damiani & ACIS-I & 2012-11-21,17:52:27 & 20792 \\ 
  14545 & 314.5673 & 43.89382 & Guedel & ACIS-S & 2013-07-17,06:16:36 & 29677 \\ 
   \hline
\end{tabular}
\end{table*}

\begin{figure*}
\includegraphics[angle=0,width=8.9cm]{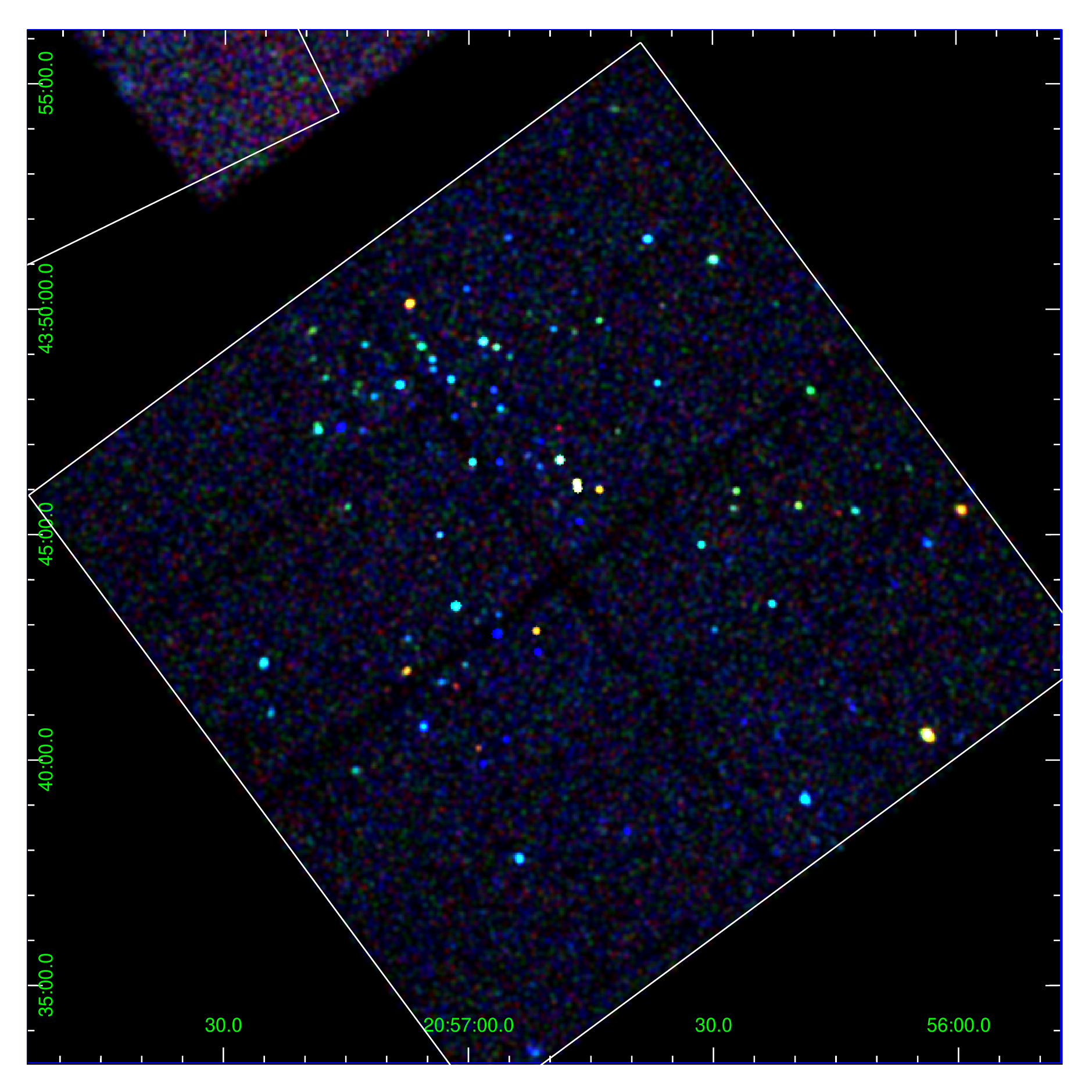}
\includegraphics[angle=0,width=8.9cm]{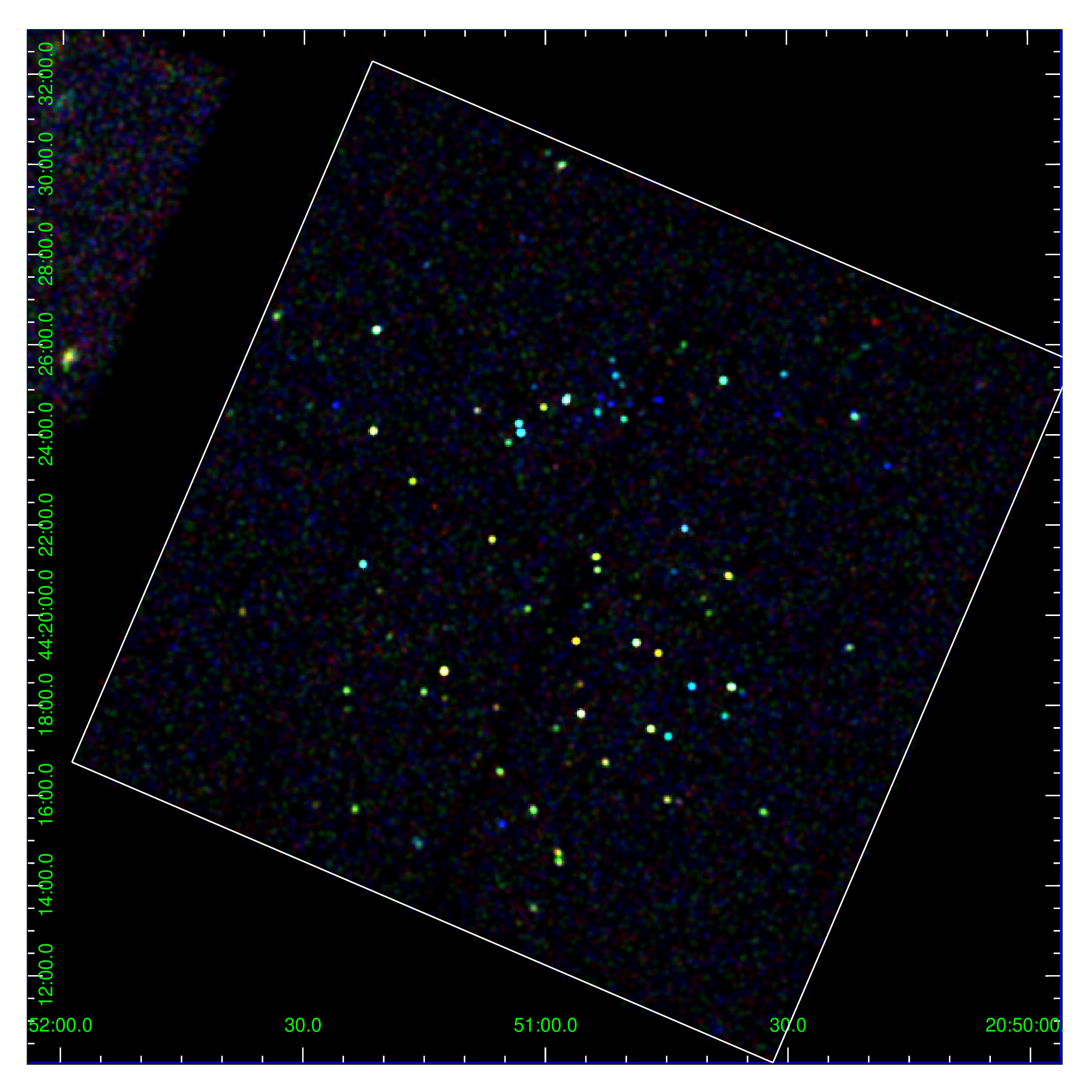} \\
\includegraphics[angle=0,width=8.9cm]{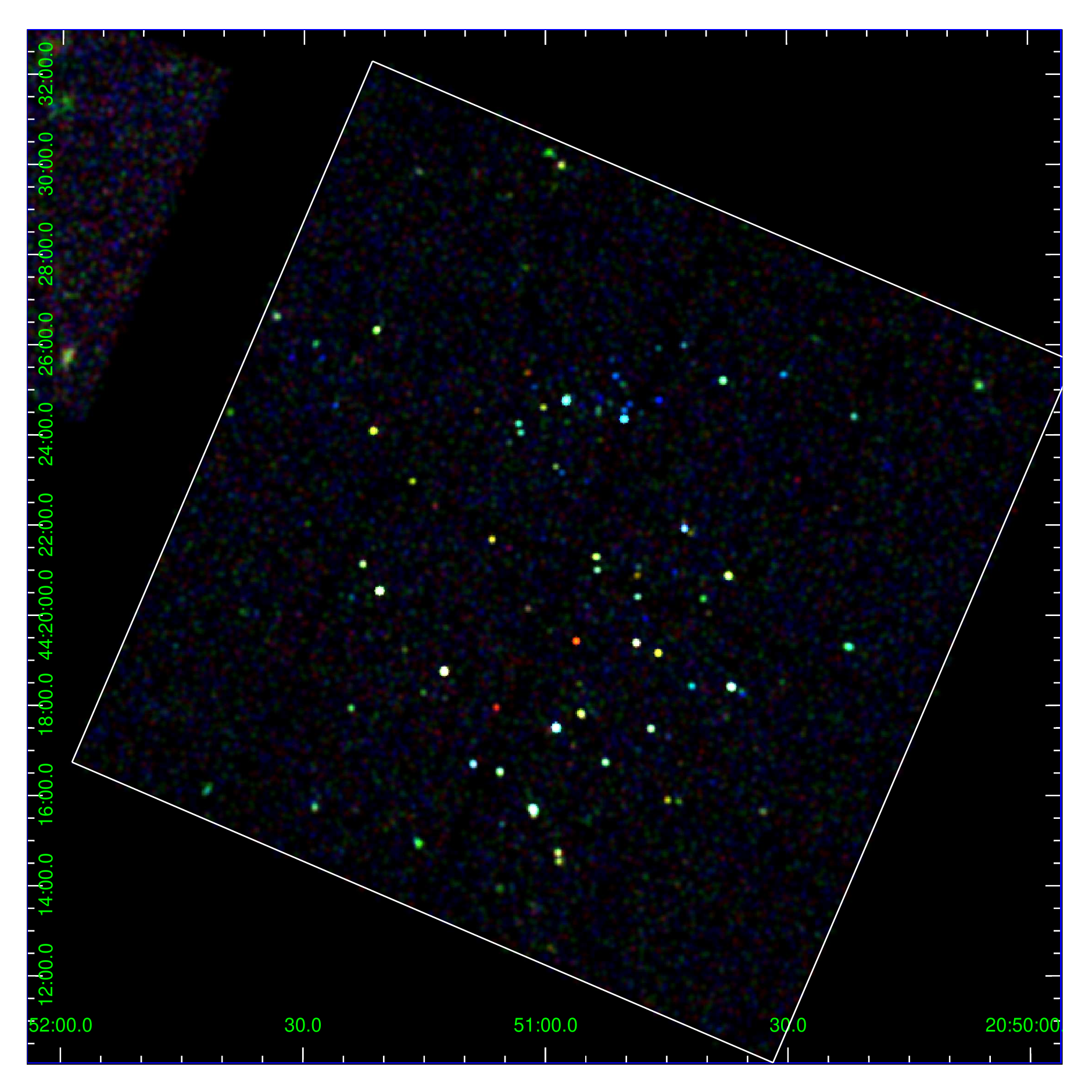}
\includegraphics[angle=0,width=8.9cm]{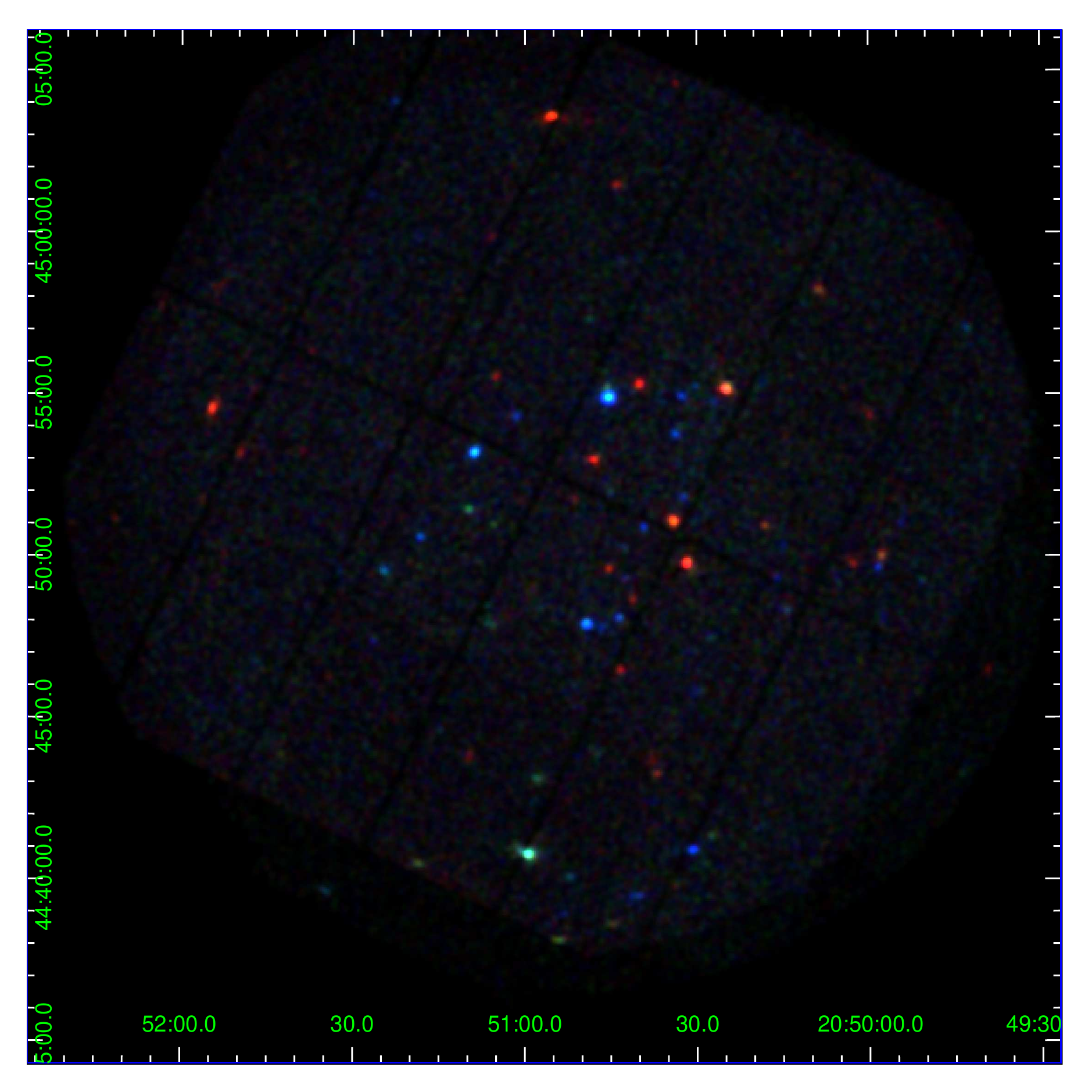}
\caption{
True-color X-ray images of individual observations,
slightly smoothed to emphasize point sources.
Red: 0.4-1.2 keV; green: 1.2-2.4 keV; blue: 2.4-7.9 keV.
Panel $a$, upper left:
ACIS-I ObsId 13648 (in the ``Gulf of mexico'').
The ACIS-I FOV size is 16.9 arcmin on a side.
$b$, upper right:
ACIS-I ObsId 13647 (``Pelican'').
$c$, lower left:
ACIS-I ObsId 15592, with same pointing as ObsId 13647 (panel $b$).
Comparison with this latter ObsId makes source variability very evident.
$d$, lower right:
XMM-Newton ObsId 0679580101 (``Pelican hat''). In all XMM-Newton images
shown here, MOS1,2 and PN data are combined together.
\label{xray-img-1}}
\end{figure*}

\begin{figure*}
\includegraphics[angle=0,width=8.9cm]{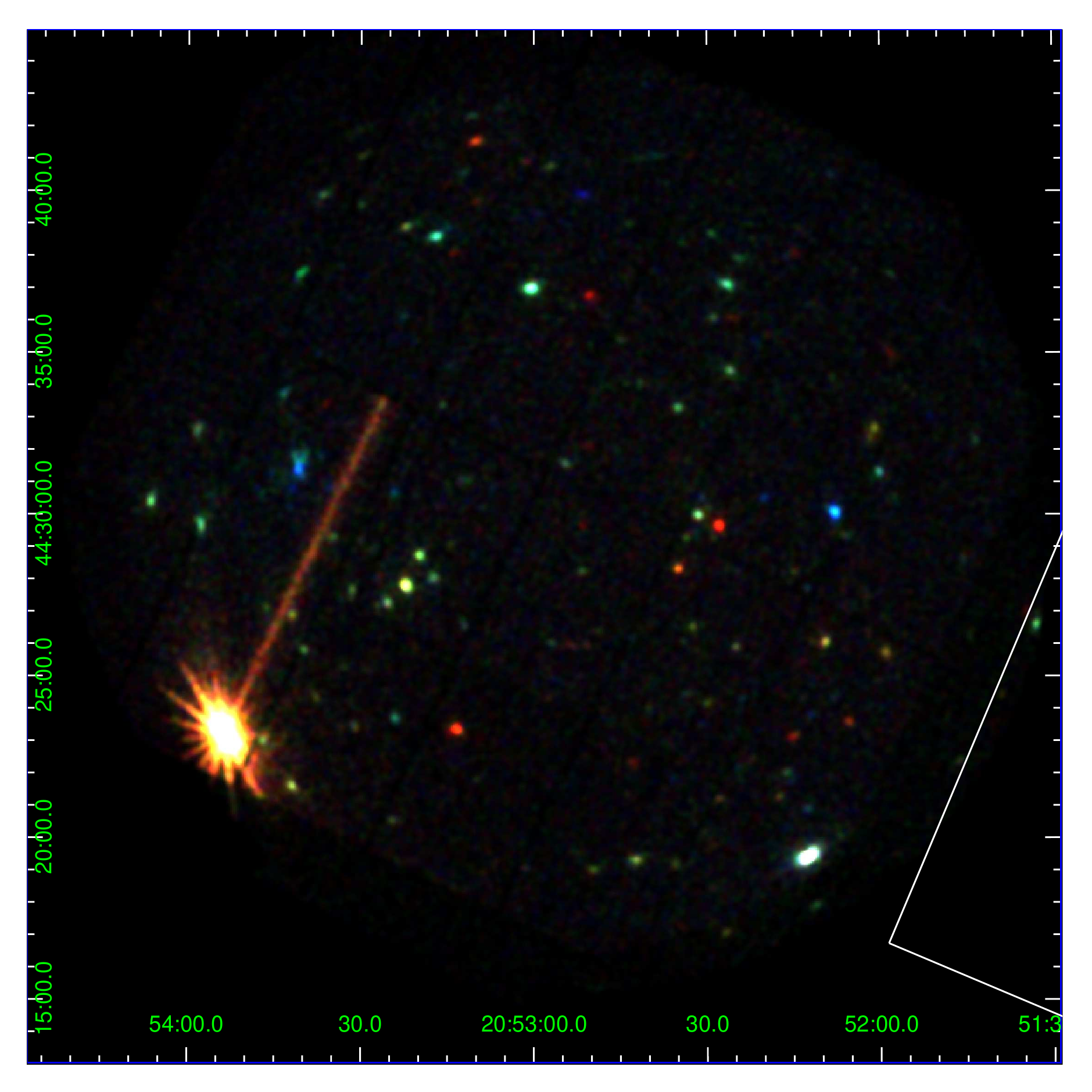}
\includegraphics[angle=0,width=8.9cm]{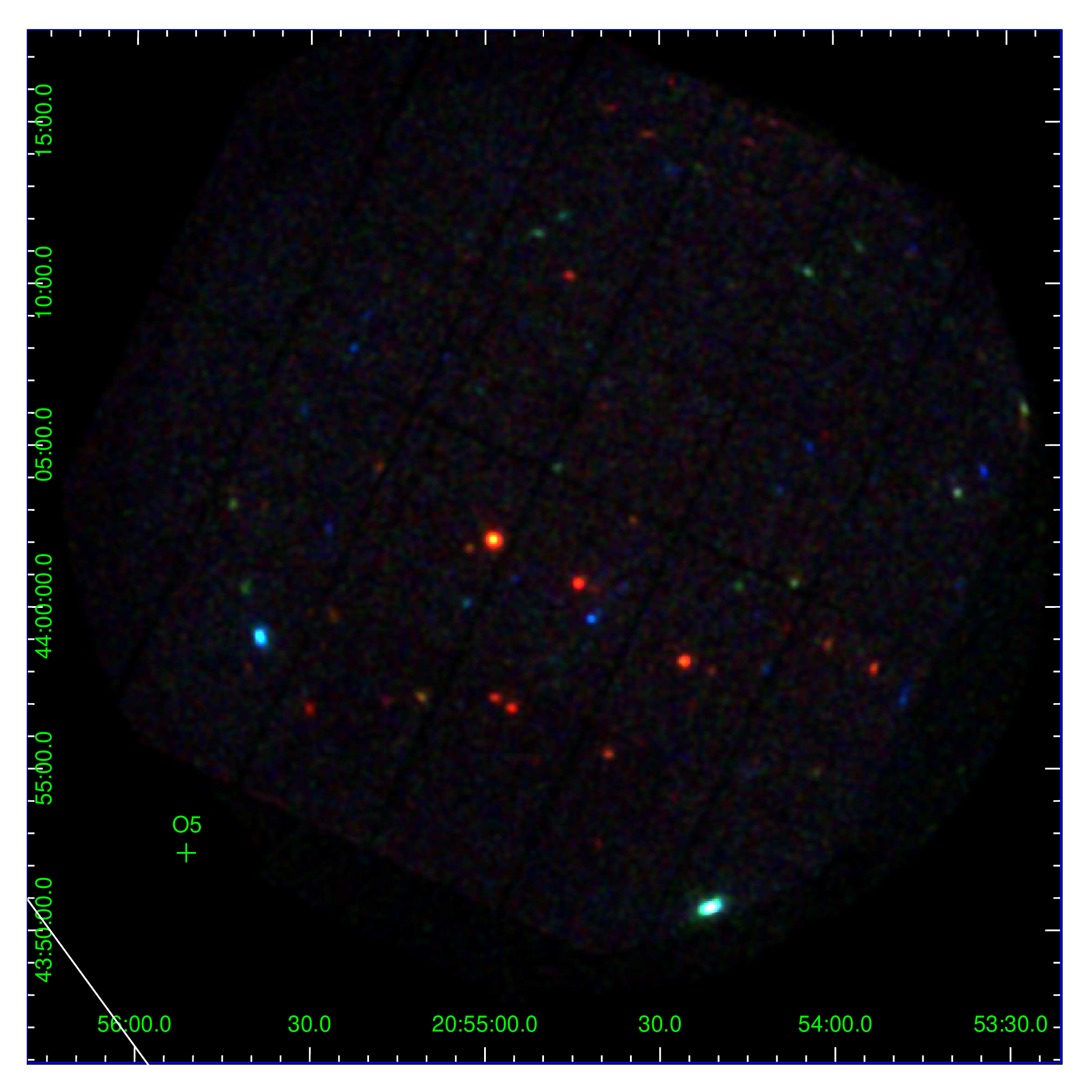} \\
\includegraphics[angle=0,width=8.9cm]{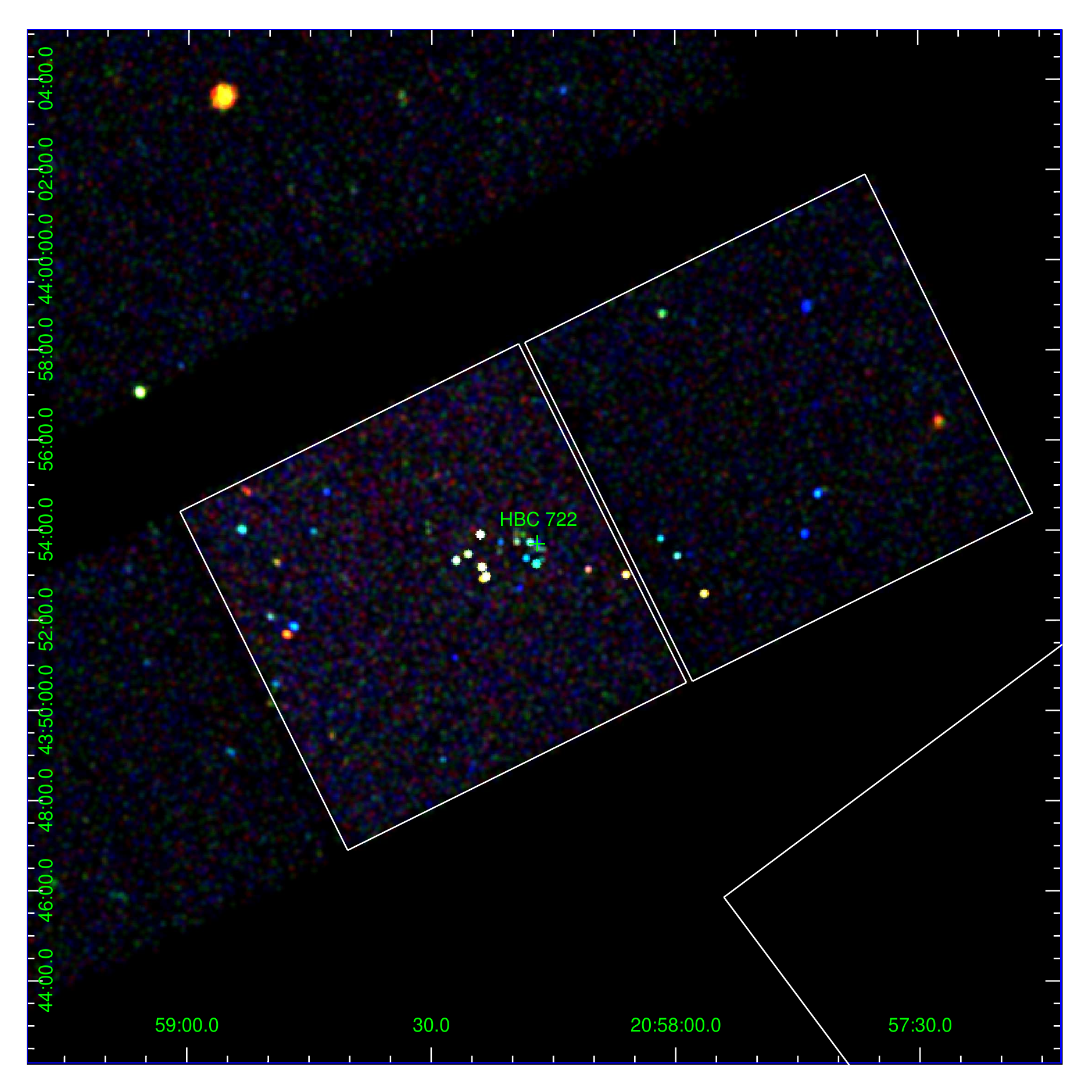}
\includegraphics[,width=8.9cm]{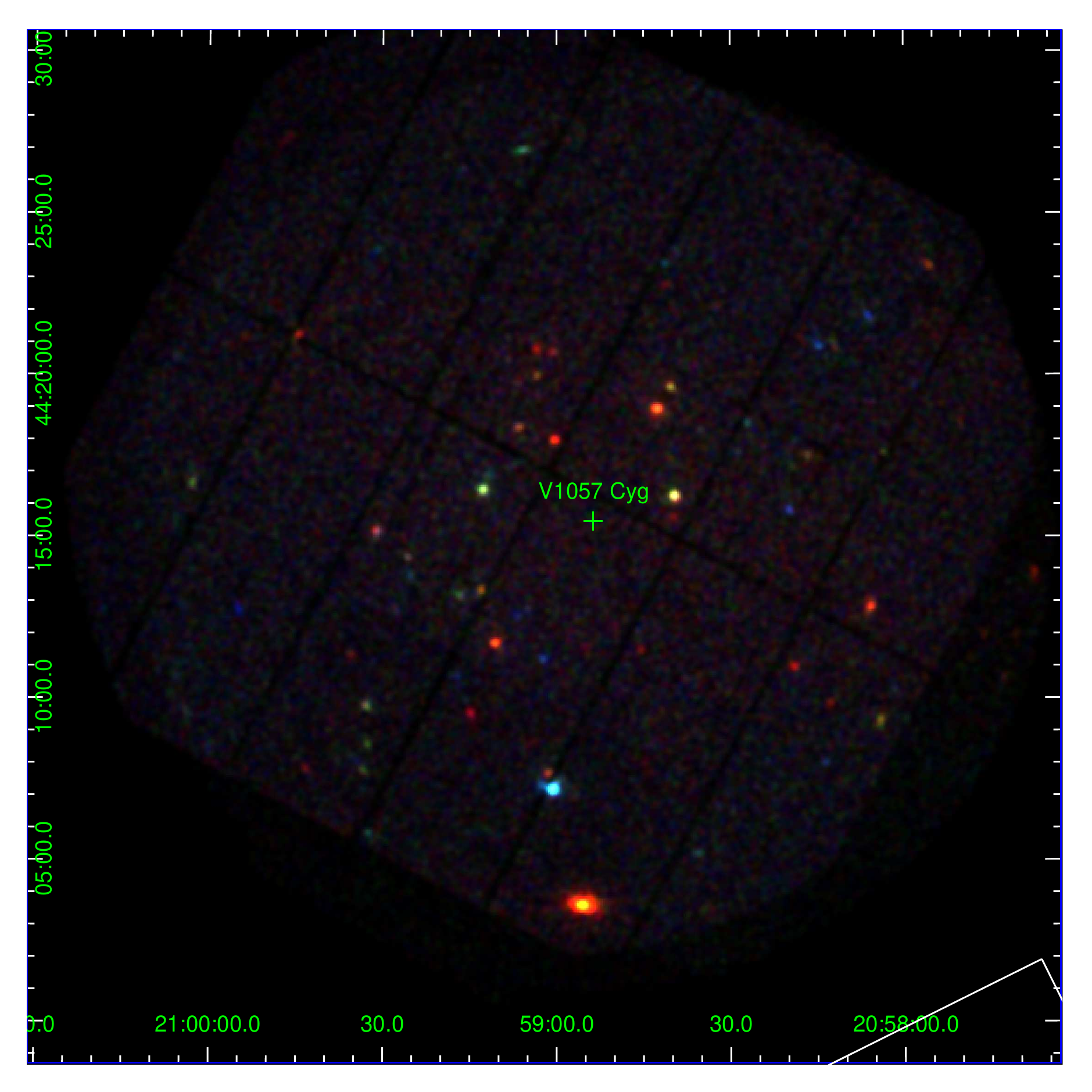}
\caption{
X-ray images as in Fig.~\ref{xray-img-1}.
Panel $a$, upper left:
XMM-Newton ObsId 0679580201.
The brightest source (2XMM~J205347.0$+$442301) (classified as a star by
Lin \e 2012) produces both diffraction spikes and a stripe of
out-of-time events, accumulated during CCD readout.
The white line outlines the eastern part of ACIS ObsId 13647 FOV.
$b$, upper right:
XMM-Newton ObsId 0679580301, towards the highly obscured region
LDN~935. The plus sign labeled ``O5'' is the O star
2MASS~J20555125$+$4352246 from Comer\'on and Pasquali (2005).
$c$, lower left:
Archive ACIS-S ObsId 14545.
The two white squares indicate ACIS-S chips 6 (front-illuminated, right)
and~7 (back-illuminated, left), which have different sensitivities and
are analyzed separately. This ObsId also has data from chips 2, 3, and
8, shown in the same Figure; however the sensitivity and/or spatial
resolution of these chips is inferior, and they are not considered
further here.
Near the center of the image, the green plus sign indicates the new
FU~Ori candidate HBC~722.  To the SW a part of ACIS 13648 FOV can be seen.
$d$, lower right:
Archive XMM-Newton ObsId 0302640201, just north of ACIS-S ObsId
14545 (seen in lower-right corner).
Near image center the FU~Ori star V1057~Cyg is labeled.
\label{xray-img-2}}
\end{figure*}

The log of the X-ray observations used in this work is reported in
Tables~\ref{table.xmm} and~\ref{table.acis}. We obtained three new
XMM-Newton EPIC observations and two new Chandra ACIS-I observations
(ACIS-I ObsIds 13647 and 15592 are in fact the same observation split in
two segments, while 13648 is a distinct pointing). Four archive
XMM-Newton observations are also analyzed (of which ObsIds 0656780701
and 0656781201 nearly co-pointed), as well as one ACIS-S observation:
this latter has a different field of view (FOV) than the ACIS-I detector,
and a slightly different sensitivity. All XMM-Newton observations, made
with the EPIC camera, are composed by a set of three simultaneous
observations with the PN, MOS1, and MOS2 detectors, approximately
co-pointed but having different FOV shapes; PN is the most sensitive of
the three, while the MOS1 (MOS2) effective area is comparable to that of ACIS-I.
All XMM-Newton observations were made using the Medium filter.
The Chandra ACIS images have an on-axis PSF much narrower ($\sim 0.5$
arcsec) than XMM-Newton EPIC images ($\sim 4$ arcsec): this implies not
only a better spatial resolving power, but also that less counts are
needed for a significant source detection ($\sim 4$ X-ray counts vs.\
$\sim 40$ for the PN+MOS combination)
because the background contribution in the EPIC images is
usually non-negligible, while it is in ACIS images. Far from optical
axis, the PSF degrades sensibly in ACIS images, while that of XMM-Newton
images is rather uniform across the FOV.

We tried to optimize our X-ray pointings, guided by the
large-scale distribution and local spatial density of YSOs from Guieu \e
(2009). Therefore, we have chosen ACIS-I where the YSO density was
highest, but preferred XMM-Newton with its larger FOV for the sparser
regions. Figure~\ref{dss-fov-yso} shows a Digital-Sky-Survey (DSS) red
image of the region, with indicated all the X-ray observations studied
here, and also the RGS11 YSOs, and H58 PMS stars; the illuminating O5
star, and the two FU~Ori objects are also indicated, as well as the NIR
clusters found by Cambr\'esy \e (2002). The sub-regions with the
highest YSO density are the so-called ``Gulf of Mexico'', covered by our
ACIS field 13648, and the ``Pelican'', covered by our ACIS fields 13647
and 15592. The less-obscured part of the cloud containing V1057~Cyg,
covered by XMM-Newton archive ObsId 0302640201, contains only few YSOs.

Besides the studies mentioned above, more evidences can be presented in
favour of the North-America and Pelican Nebulae being part of the same
cloud; this is especially important in this part of the sky because of
the mentioned tangent-arm geometry, and in this particular nebula to
ascertain that the foreground obscuring cloud is physically associated
with the bright nebulae and is not much closer along the line of sight.
For example, the WISE (Wright \e 2010) images in four NIR bands at 3.4,
4.6, 12, and 22$\mu$ demostrate the presence of dense, hot dust behind
the dark nebula LDN935 (Figure~\ref{wise}), with a geometry
approximately centered on the illuminating O5 star. The distance of the
dark cloud from us is therefore the same as the distance of the bright
nebulae oscured by it. The region with the highest YSO density, just S-E
of the O5 star and inside our ACIS 13648 field, is also seen in the WISE
image to be one of highest obscuration.

The same qualitative conclusion that the North-America and Pelican
Nebulae are the same physical cloud is reached by inspecting an optical
emission-line image, obtained using narrow-band filters centered at the
[O~III], \ha, and [S~II] optical lines, shown in
Figure~\ref{ngc7000_ic5067_hst_final}. The ionization (highest for
[O~III]) has a global pattern, centered close to the O5 star; there is
no other obvious ionization center, based on the optical-line image.

\begin{figure*}
\includegraphics[angle=0,width=8.9cm]{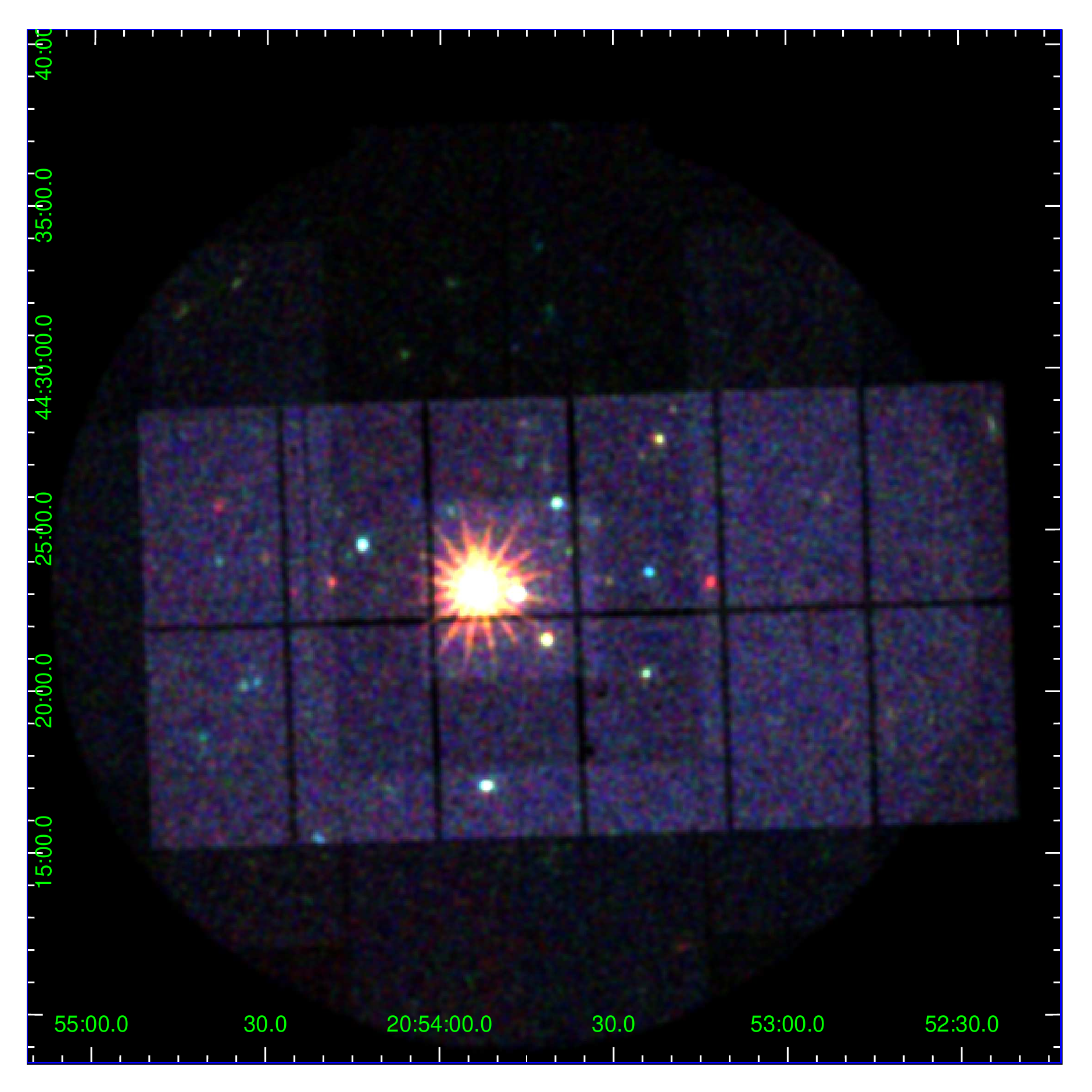}
\includegraphics[,width=8.9cm]{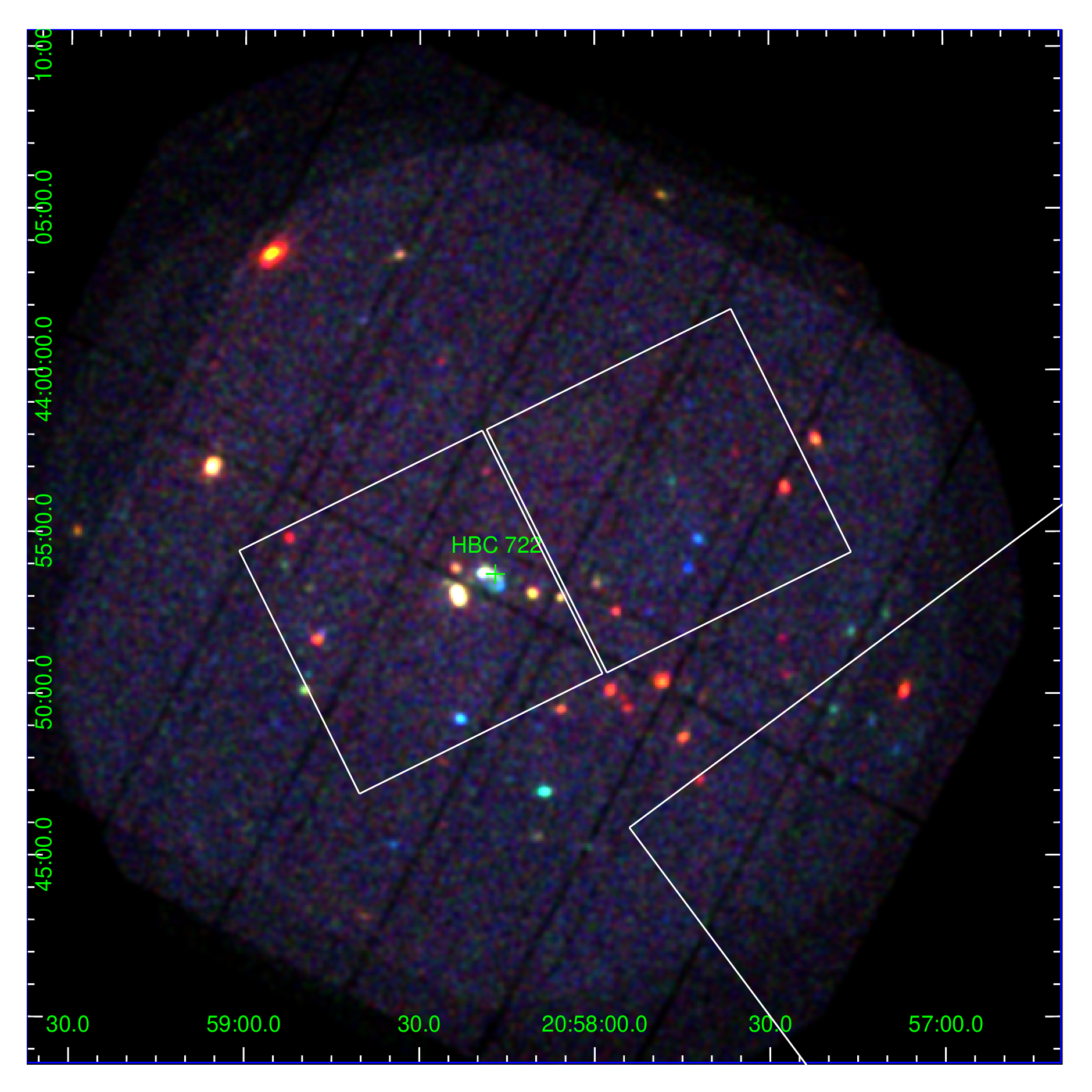}
\caption{
X-ray images as in Fig.~\ref{xray-img-1}.
Panel $a$, left:
Archive XMM-Newton ObsId 0556050101, overlapping
XMM ObsId 0679580201 of Fig.~\ref{xray-img-2}, panel~$a$.
The bright source is the same as in that Figure. The distribution of
background emission seen here is caused by the small-window mode used
here for the MOS cameras (not for the PN camera).
$b$, right:
Archive XMM-Newton ObsId 0656780701$+$0656781201.
The slight pointing difference between the two ObsIds causes the
inter-chip gap pattern seen in the Figure. The image overlaps completely
with the ACIS 14545 FOV, and partially with ACIS 13648 FOV in the lower
right corner. The tight group of X-ray sources surrounding HBC~722
(Fig.~\ref{xray-img-2}, panel~$c$) is unresolved here.
\label{xray-img-3}}
\end{figure*}

All X-ray images studied here are shown in Figures~\ref{xray-img-1},
\ref{xray-img-2}, and~\ref{xray-img-3}. In these true-color images, the
red (green, blue) intensity is proportional to detected X-ray counts in
the soft (medium, hard) X-ray band. Blue sources are typically highly
absorbed objects, with a strong low-energy cutoff in their X-ray
spectrum. The Chandra FOVs are indicated, as well as the O5 star and the
FU~Ori objects.

\section{X-ray data analysis}
\label{xdata}

\begin{figure*}
\sidecaption
\includegraphics[width=12cm]{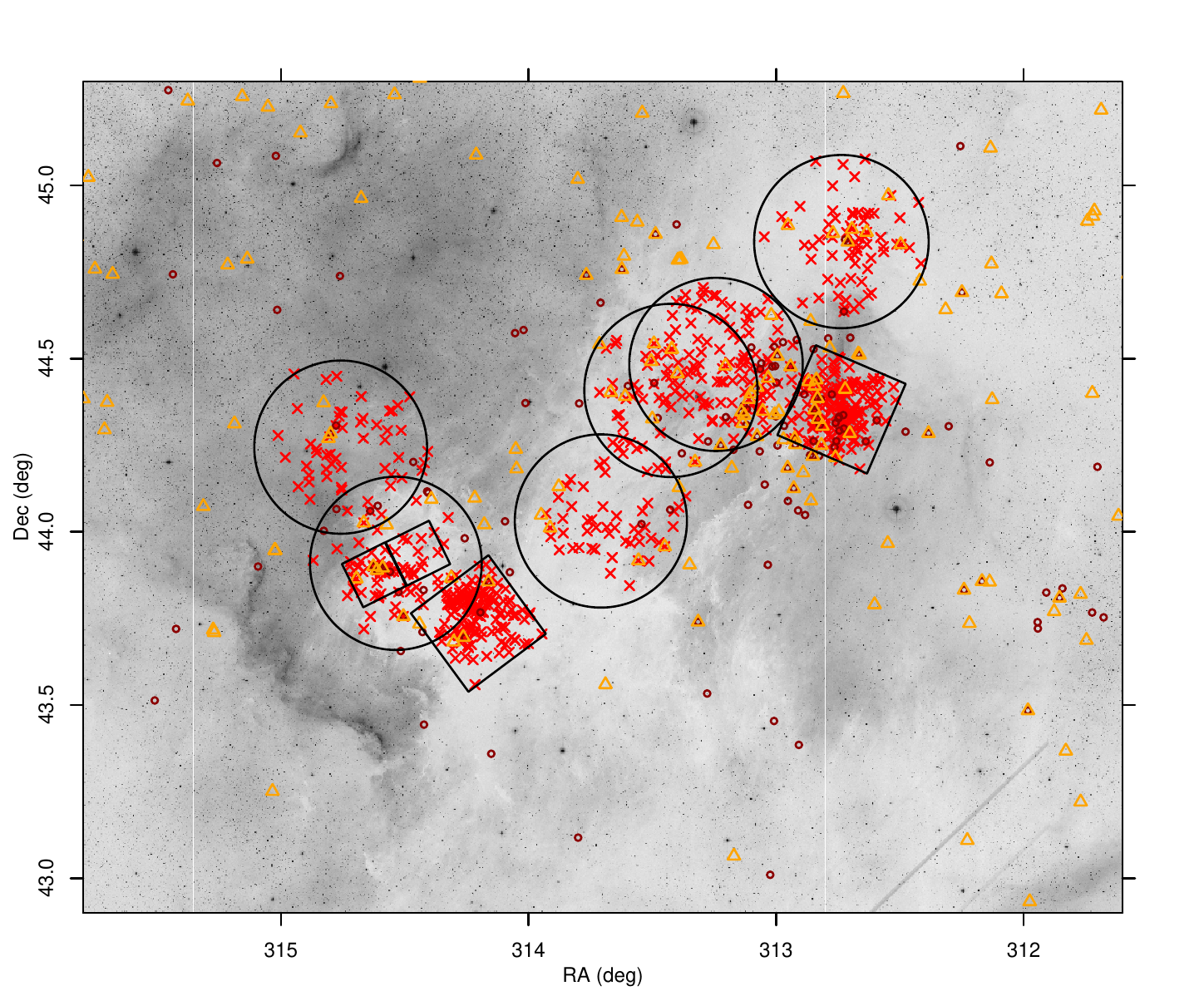}  
\caption{
Same field and background image as in Fig.~\ref{dss-fov-yso}, with
small symbols now indicating: X-ray detected sources (red crosses), near-IR
excess objects (orange triangles), and \ha-excess objects (dark-red
circles). Big circles/squares are X-ray FOVs as in
Fig.~\ref{dss-fov-yso}.
\label{dss-ctts-xdet}}
\end{figure*}

The X-ray datasets were processed using standard software packages (CIAO
v4.8 for Chandra and SAS v11 for XMM-Newton data), and exposure maps at a
representative energy of 1~keV were computed using the same packages.
Source detection was made using PWDetect (v1.3.2) for Chandra data, and
its XMM-Newton analogue PWXDetect (Damiani \e 1997a,b). The
front- and back-illuminated chips of ACIS-S ObsId 14545 were analyzed
separately. ACIS-I ObsIds 13647 and 15592, sharing the same pointing
direction and roll angle, were only analyzed jointly. This was also the
case for archive XMM ObsIds 0656780701 and 0656781201.
Spurious detections associated with diffraction spikes or
out-of-time events (Fig.~\ref{xray-img-2}, panel $a$;
Fig.~\ref{xray-img-3}, panel $a$) were manually eliminated.
We then merged all individual X-ray detection lists according to
computed position errors, separately for XMM-Newton and Chandra, giving
priority to highest-significance detections. Thus, each detected X-ray
source receives its IAU-recommended instrument-specific designation.
The match and merging of
XMM-Newton and Chandra detection lists was made in a subsequent step,
taking in mind the much better spatial resolution of this latter, and
therefore giving priority to the Chandra source positions and properties
in ambiguous cases (see e.g.\ the tight group of sources near HBC~722 in
Fig.~\ref{xray-img-2}, panel $c$, which is unresolved by XMM-Newton in
Fig.~\ref{xray-img-3}, panel $b$). The final X-ray source list comprises
721 objects, of which 378 ACIS detections (of which 34 with an XMM-Newton
counterpart), and 343 XMM-Newton-only detections.
The chosen detection threshold, corresponding to approximately one
spurious detection per field, ensures that no more than $\sim 10$ of the
721 detections are spurious.
The positions of all X-ray detections are shown in
Figure~\ref{dss-ctts-xdet}.

\section{Auxiliary data: optical and near-IR}
\label{optir}

We have complemented the X-ray data with data taken from the IPHAS DR2
(Barentsen \e 2014), 2MASS PSC, and UKIDSS DR10plus GPS (Lawrence \e 2007;
table ``reliableGpsPointSource'').
All of these surveys cover the entire spatial region considered here.
We extracted sources from a region of radius $1.5^{\circ}$ centered on
$(RA,Dec) = (313.7,44.4)$. This yielded
1117062 IPHAS sources, however after applying the recommended screening
(keeping only sources with flags {\em a10=1} and {\em a10point=1},
i.e.\ point-like sources with high signal-to-noise ratio in all bands)
their number reduces to 161635 sources.

The screened 2MASS PSC source list (flag {\em ph\_qual} different from {\em
E,F,U,X}, flag {\em cc\_flg} different from {\em p,d,s})
contains 339774 sources.
The UKIDSS source list instead contains 980770
sources. This is much larger than the number of 2MASS sources because of
the greater depth of the UKIDSS survey; however, the UKIDSS data do not
supersede the 2MASS data, since they saturate at the bright end ($J \leq 12$),
where 2MASS
still contains reliable data. We therefore use both sets of data, after
an examination of their agreement on the sources common to both
(see Appendix~\ref{ukidss-2mass}).
For these latter, the photometric errors are much
lower in the UKIDSS than in the 2MASS magnitudes, so we adopt the UKIDSS
ones whenever available. At large reddening, often found in this region,
the UKIDSS photometry differs from 2MASS, so a slight recalibration was
applied to agree with the 2MASS system (Appendix~\ref{ukidss-2mass}).

We matched each of these source catalogs with our X-ray source list, up to
a maximum offset of $4\sigma$, where $\sigma$ is the combined positional
error from our X-ray detection code, from the 2MASS PSC error {\em err\_maj},
or set fixed to 0.07$^{\prime\prime}$ for the IPHAS and UKIDSS sources.
There are 568 2MASS, 235 IPHAS, and 300 UKIDSS counterparts to our X-ray
sources.
{
The number of total counterparts using the combined NIR catalog
is instead 548, lower than the 2MASS counterparts alone because of
incomplete sets of magnitudes for several stars.
}
115 X-ray sources remained unidentified.
The number of spurious matches between X-ray and optical/NIR sources
is difficult to evaluate, since assumptions usually made to compute it
(uniform spatial distributions of the two matched sets) are not fulfilled
here. As we have seen above, X-ray sources are found near regions of strong
obscuration, where the optical/NIR source density is much smaller than
average, and changes on very small spatial scales. Moreover, the PSF (and
source position error) of the XMM-Newton data is much larger than that of
the Chandra ACIS data, so that even if the optical/NIR spatial density were
uniform there is a higher chance of spurious matches for XMM-Newton X-ray
sources than for the Chandra ACIS sources.

Finally, we positionally matched our X-ray source list with the YSOs
from RGS11, obtaining 184 matches out of all 802 YSOs falling in our X-ray
FOVs (by class: 33/189 for flat-spectrum YSOs, 23/195 for Class~I,
123/370 for Class~II, 4/23 for Class~III, and 1/25 for unclassified YSOs).
Of the H58 stars, 45 were falling in the X-ray FOVs and 31 were X-ray detected.
It is worth remarking that the ACIS fields yielded larger detection
percentages than the XMM-Newton fields, whose higher background resulted in
a lower sensitivity.
The median threshold X-ray luminosity for ACIS detection was found to be
$\log L_X \sim 29.9$ erg/s, while the corresponding value for XMM-Newton
is $\log L_X \sim 30.4$ erg/s, at the average absorption values inferred for
these YSOs (see below).
Considering all ACIS FOVs, 115 YSOs were detected out
of 373 observed (by class: 25/94 flat-spectrum, 17/118 Class~I, 72/132
Class~II, and 1/8 Class~III). The corresponding numbers for the XMM-Newton
FOVs are 83 detections out of 551 observed YSOs (by class: 11/136
flat-spectrum, 10/115 Class~I, 58/274 Class~II, and 3/17 Class~III).

Our X-ray source properties and identifications are reported in
Tables~\ref{table1} and~\ref{table2}.

\begin{longtab}
\begin{landscape}
\begin{longtable}{rcccccccccccc}
\caption{ \label{table1}
Detected X-ray sources in the NAP. Full table in electronic format
only.} \\
  \hline \hline
X-ray & CXO Id & XMM Id & RA & Dec & Pos.\ err.\  & Count rate &
Rate error & HR1 & HR2 & HR1 & HR2 & $\log L_X$ \\
no.\  & (CXOU) & (XMMU) & (J2000) & (J2000) & (arcsec) & (cts/ksec) &
(cts/ksec) & ACIS & ACIS & XMM & XMM & (erg/sec) \\
  \hline
\endfirsthead
\caption{continued.}\\
\hline\hline
X-ray & CXO Id & XMM Id & RA & Dec & Pos.\ err.\  & Count rate &
Rate error & HR1 & HR2 & HR1 & HR2 & $\log L_X$ \\
no.\  & (CXOU) & (XMMU) & (J2000) & (J2000) & (arcsec) & (cts/ksec) &
(cts/ksec) & ACIS & ACIS & XMM & XMM & (erg/sec) \\
  \hline
\endhead
\hline
\endfoot
  1 & J205004.0+442249 &  & 312.51677 & 44.38037 & 1.87 & 0.428 & 0.226 & 0.33 & 0.60 &     &     &     \\ 
    2 & J205006.2+442505 &  & 312.52597 & 44.41824 & 2.34 & 2.067 & 0.322 & 0.31 & -0.06 &     &     & 30.09 \\ 
    3 & J205013.5+441910 &  & 312.55631 & 44.31957 & 2.56 & 0.402 & 0.161 & 0.00 & 0.88 &     &     &     \\ 
    4 & J205017.6+442318 &  & 312.57358 & 44.38850 & 1.48 & 0.759 & 0.239 & 1.00 & 0.81 &     &     &     \\ 
    5 & J205019.0+442630 &  & 312.57958 & 44.44189 & 2.45 & 0.550 & 0.208 & -0.90 & 0.50 &     &     &     \\ 
    6 & J205020.3+442557 &  & 312.58466 & 44.43271 & 1.73 & 0.488 & 0.185 & 0.45 & -0.07 &     &     & 29.99 \\ 
    7 & J205021.7+442425 &  & 312.59051 & 44.40696 & 1.08 & 1.937 & 0.264 & 0.48 & -0.07 &     &     & 30.76 \\ 
    8 & J205021.8+442101 &  & 312.59113 & 44.35048 & 1.12 & 0.192 & 0.103 & -0.20 & 0.20 &     &     & 29.02 \\
\end{longtable}
\end{landscape}
\end{longtab}

\begin{longtab}
\begin{landscape}
\begin{longtable}{rccccccccc}
\caption{ \label{table2}
Optical/NIR photometry for X-ray sources.} \\
  \hline \hline
X-ray & $i$ & $r-i$ & $r-H\alpha$ & $J$ & $H$ & $K$ & IR & \ha\ & RGS11 \\
no.\ & & & & & & & excess & excess & number \\
  \hline
\endfirsthead
\caption{continued.}\\
\hline\hline
X-ray & $i$ & $r-i$ & $r-H\alpha$ & $J$ & $H$ & $K$ & IR & \ha\ & RGS11 \\
no.\ & & & & & & & excess & excess & number \\
\hline
\endhead
\hline
\endfoot
  1 &  &  &  &  &  &  &  &  &  \\ 
    2 &  &  &  & 13.52 & 12.70 & 12.35 &  &  &  \\ 
    3 &  &  &  &  &  &  &  &  &  \\ 
    4 &  &  &  &  &  &  &  &  &  \\ 
    5 &  &  &  &  &  &  &  &  &  \\ 
    6 &  &  &  & 14.21 & 12.81 & 12.10 &  &  &  \\ 
    7 &  &  &  & 13.22 & 11.53 & 10.51 &  &  &  98 \\ 
    8 &  &  &  & 15.82 & 15.15 & 14.60 &  &  &  \\
\end{longtable}
\end{landscape}
\end{longtab}

\section{Results}
\label{results}

\subsection{Distance}
\label{dist}

\begin{figure}
\resizebox{\hsize}{!}{
\includegraphics[]{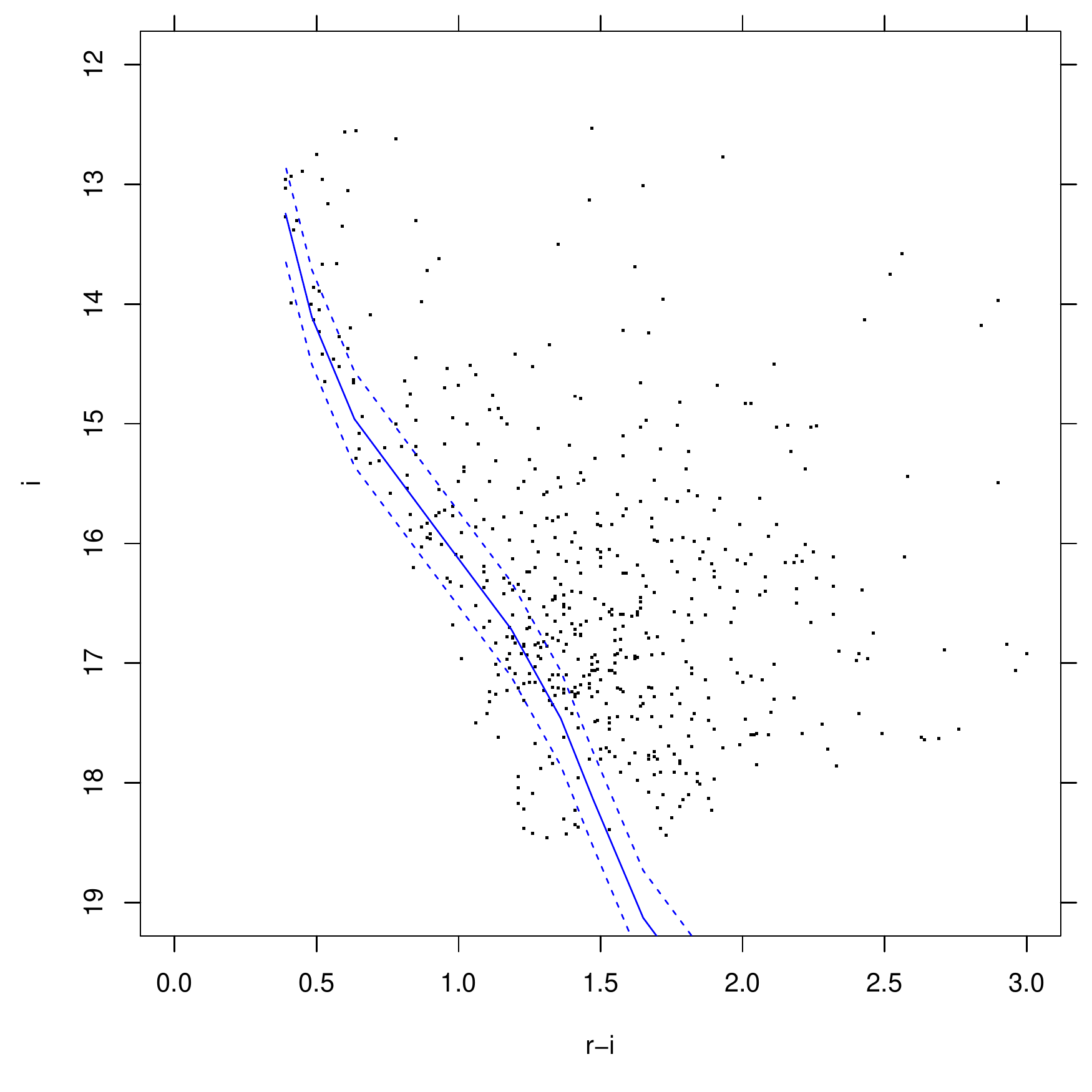}}
\caption{
An $(i,r-i)$ CMD of stars towards two heavy-obscuration regions in
LDN~935, with a 10-Gyr isochrone from BHAC at zero
reddening and distance of 560~pc (blue { solid} curve).
{ Blue dashed curves correspond to distances smaller and larger by
20\%, respectively.}
Only datapoints with errors $<0.05$ in both $i$ and $r-i$ are shown.
\label{distance}}
\end{figure}

The deep IPHAS photometry enables us to perform a new determination of
distance, relying upon the assumption that the obscuring dark cloud LDN935
lies immediately before the star-forming region, as argued above. The
method (already used by Prisinzano \e 2005 to determine the distance of the
young cluster NGC~6530) fits the lower envelope of datapoints in the
color-magnitude diagram (CMD) to the shape of the main sequence of field
stars, which are the lowest-luminosity field stars at each given color.
It is therefore independent of any assumption on the cluster age or
specific isochrone set. To apply the test we have selected two
heavily obscured regions in LDN~935 (12-arcmin radius circles around
$(RA,Dec)=(314.2,43.6)$ and $(312.9,44.9)$, respectively), whose CMD using
IPHAS data is shown in Figure~\ref{distance}. Also shown is a 10-Gyr
isochrone from Baraffe \e (2015; henceforth BHAC) at zero
reddening and distance of 560~pc, which provides a good fit to the
lower envelope of datapoints
{ (solid), at least for $r-i<1$.
The dashed curves ($\pm 20$\% in distance) moreover suggest that
the distance is unlikely to be larger than $\sim 650$~pc.}
Therefore, the new data are in good
agreement with the accepted distance of the complex
(560-600~pc, Laugalys \e 2006, 2007).
Since the BHAC tracks are not currently available from the IPHAS filter set
we have recalibrated them for this filter set starting from those for the
Johnson/Cousin filters, as explained in Appendix~\ref{iphas-vphas}.

\subsection{\ha-emission stars}
\label{halpha}

\begin{figure}
\resizebox{\hsize}{!}{
\includegraphics[angle=0]{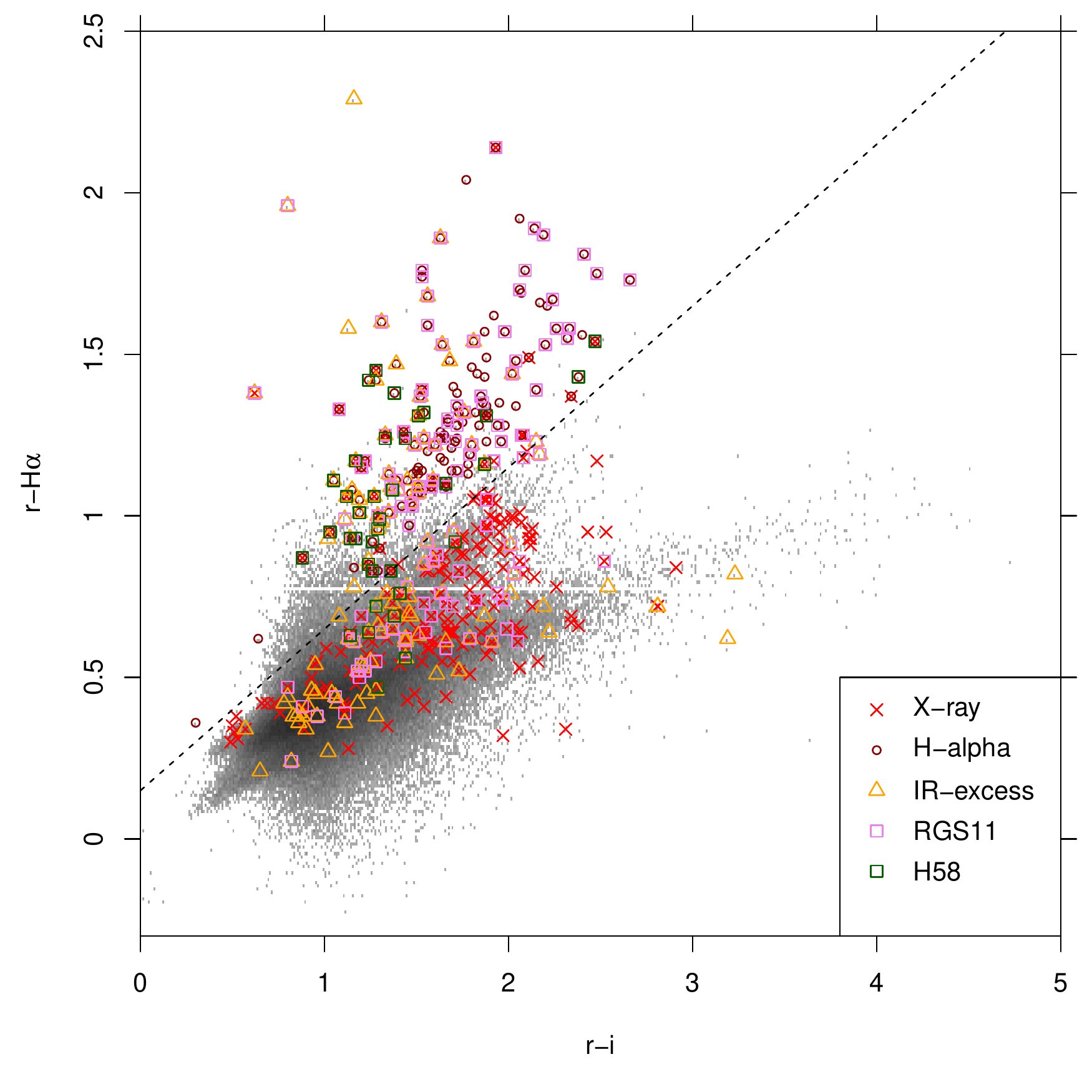}}
\caption{
IPHAS $(r-H\alpha,r-i)$ diagram.  The grayscale background is
a 2-D histogram of all IPHAS objects in the sky region of
Fig.~\ref{dss-fov-yso}, with bin sizes of 0.01 in both axes.
The dashed line is the CTTS-star limit.
Symbols as in Fig.~\ref{dss-ctts-xdet}, with the addition of YSOs from
RGS11 (violet squares), and emission-line stars from H58
(dark-green squares). Hundreds of faint stars above the dashed
line are not selected here as \ha-emission stars because of their
position in the CMD.
\label{ri-rha}}
\end{figure}

The IPHAS survey also provides a \rha\ index, useful to select
\ha-emission objects over wide sky areas.
A \rha\ vs.\ $r-i$ diagram for NAP is shown in Figure~\ref{ri-rha}.
Datapoints in the upper part of the diagram are emission-line stars.
In particular, we show with a dashed line the color-dependent limit used
by Kalari \e (2015) to select CTTS stars.
Almost all of the H58 PMS stars, and most of the RGS11 YSOs fall above that
limit. We have therefore selected all stars significantly above that line
($4 \sigma$ above errors) as candidate CTTS. However, this choice yielded
hundreds objects falling too low in the optical CMD ($i>18$, $r-i \sim
1$, much below the ZAMS at the NAP distance),
and mostly in the N-E quadrant of the surveyed region, i.e.\
spatially far from
the known YSOs, that are thus unlikely to be young stars in the NAP. These are
probably unrelated emission-line stars at much larger distances, and are
therefore not included in our emission-star list, which comprises 153
stars (of which 30 detected in X-rays).
The spatial distribution of accepted emission-line stars selected
from the \rha\ index is shown in Figure~\ref{dss-ctts-xdet}.

\subsection{Optical and near-IR color-color diagrams}
\label{ccd}

\begin{figure}
\resizebox{\hsize}{!}{
\includegraphics[angle=0]{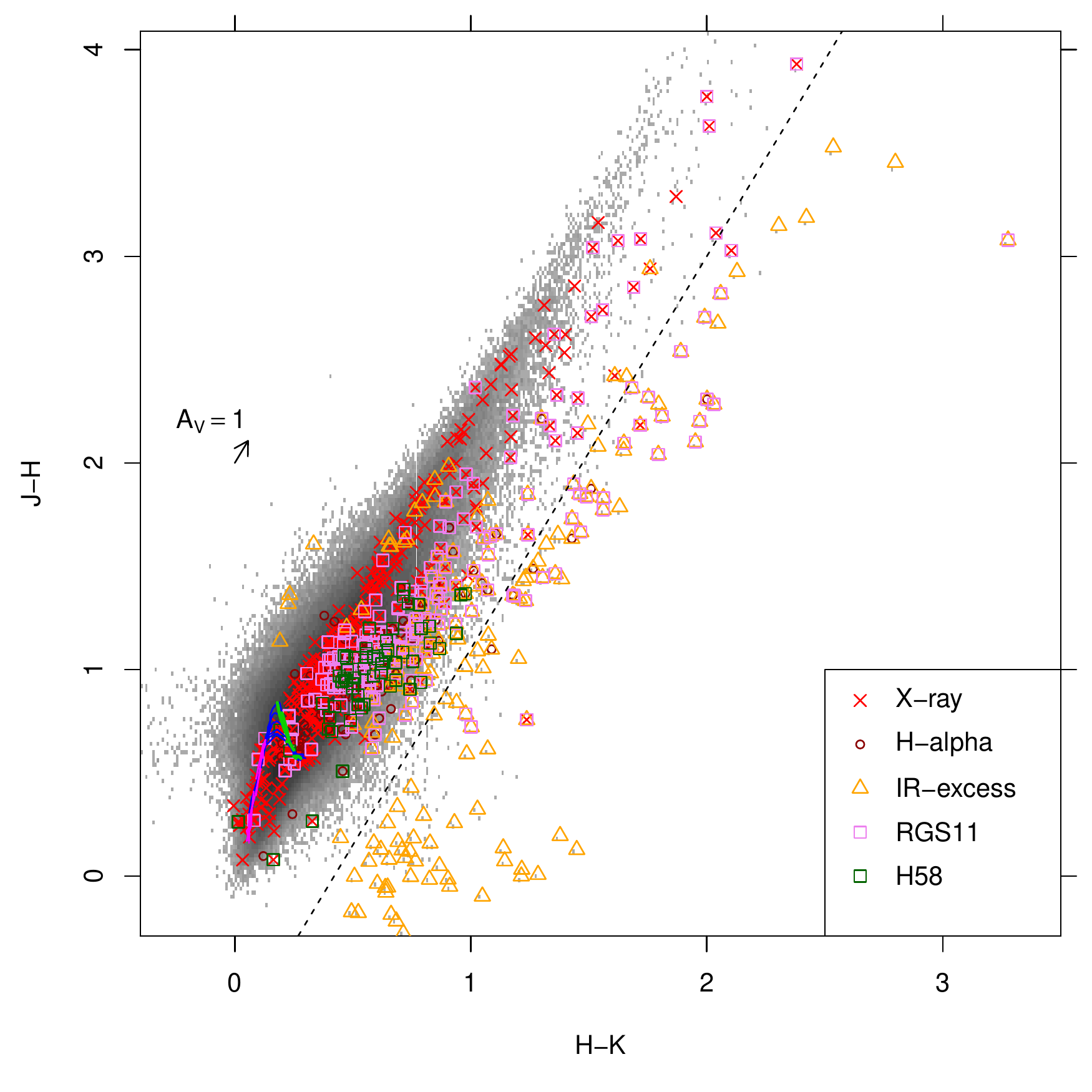}}
\caption{
Near-IR color-color diagram $(J-H,H-K)$, using both 2MASS and UKIDSS data.
Symbols as in Fig.~\ref{ri-rha}.
The arrow indicates the reddening vector corresponding to $A_V=1$. The dashed
line is the adopted limit to select NIR-excess objects from this
diagram. The blue lines in the lower-left part are unreddened BHAC isochrones
for ages of 1, 3, 10, 30, 100~Myr, and 10~Gyr, and masses between $0.08-1.4
M_{\odot}$. BHAC evolutionary tracks for masses of 0.1, 0.3,
and~0.5~$M_{\odot}$ are shown in green.
{
In this and following color-color diagrams,
only data with errors less than 0.1 mag on each color are shown.
}
\label{jh-hk}}
\end{figure}

Different color-color diagrams can be used to estimate the amount of
reddening and the possible presence of non-photospheric color excesses,
likely originated in circumstellar dusty disks.
The $(J-H,H-K)$ diagram for the entire NAP region is shown in
Figure~\ref{jh-hk}.
In agreement with Strai\v{z}ys, Corbally and Laugalys (2008),
we find that the Rieke and Lebofsky (1985)
reddening-vector slope does not adequately describe the slope of datapoints
in this diagram; therefore, we devote Appendix~\ref{redden-law} to the
careful determination of the reddening vector slopes for all independent color
pairs in the NAP optical/NIR catalog.
This redetermined reddening vector is shown in Fig.~\ref{jh-hk} and in
all following color-color diagrams.
To the right of the main locus of reddened normal photospheres, stars with
K-band excess emission are found. We have set a fiducial limit to define
such stars as the dashed line in the Figure. Stars significantly
($>4\sigma$) to the right of this line are selected as NIR-excess objects,
or candidate CTTSs (orange triangles in the Figure). The complete list of
NIR-excess objects includes stars selected from other optical/NIR
color-color diagrams (see below), hence the presence of several of them
also to the left of the fiducial limit in Fig.~\ref{jh-hk}.
It should be remarked that this diagram, although widely used, is not a
very efficient tool to select CTTSs: none of the CTTSs found by H58
(green squares) falls to the right of the limiting line in this diagram.
Also very few of the RGS11 YSOs are found in the CTTS region of the diagram.

\begin{figure}
\resizebox{\hsize}{!}{
\includegraphics[]{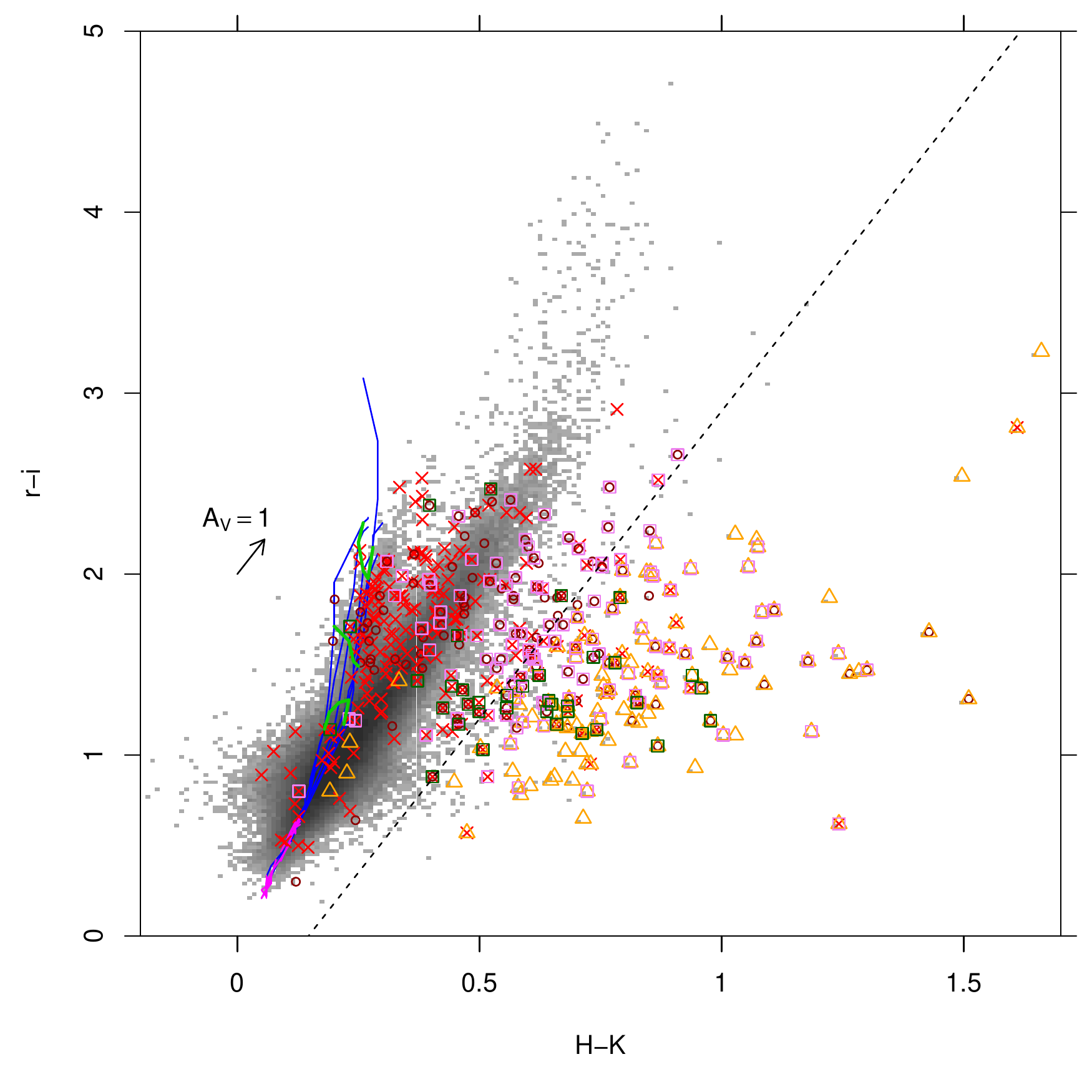}}
\caption{
Optical-IR color-color diagram $(r-i,H-K)$.
Symbols as in Figs.~\ref{ri-rha} and~\ref{jh-hk}.
The dashed line is the adopted limit for selection of NIR-excess objects
from this particular diagram.
The magenta thick line in the lower-left part indicates the location of
unreddened stars in the mass range $1-1.4 M_{\odot}$. Much larger
fractions of the H58 stars and of the RGS11 YSOs show a NIR excess using
this diagram, compared to that of Fig.~\ref{jh-hk}.
\label{ri-hk}}
\end{figure}

We have studied the properties of several mixed optical/NIR diagrams in the
young cluster NGC~6530 (Damiani \e 2006), which enable the selection of
many more NIR-excess stars than the $(J-H,H-K)$ diagram.
Therefore, we use similar mixed optical/NIR diagrams also for the NAP
stars. Figure~\ref{ri-hk} is a $(r-i,H-K)$ diagram, where we again have
drawn a fiducial line (with the slope of the reddening vector, see
Appendix~\ref{redden-law}), sufficiently distant from the bulk of stars,
separating normal reddened photospheres (to its left) from NIR-excess
stars, whose NIR colors are too red compared to the respective $r-i$ colors
for any reddening value. The usefulness of this diagram is immediately seen
by considering that here most of the H58 CTTSs and of the optically
detected RGS11 YSOs are found to the right of our fiducial limit, together
with other tens NIR-excess stars (orange triangles) not present in
previously-known catalogs. The bulk of normal field stars are instead found
along a reddened-photosphere locus, well reproduced by the available BHAC
models (blue lines) and reddening vector. The magenta segment of the BHAC
isochrones represents the locus of (unreddened) stars with $M>1 M_{\odot}$;
the upper envelope of datapoints is instead populated by the lowest-mass
dwarf stars (isochrones are shown down to a minimum mass $M=0.08 M_{\odot}$).
Most of the X-ray sources in the NAP region do not show NIR excesses,
neither from Fig.~\ref{ri-hk} nor from Fig.~\ref{jh-hk}.
Some of them follow well the zero-reddening isochrones, down to the
substellar-mass limit, while in general Figs.~\ref{jh-hk} and~\ref{ri-hk}
suggest that they are characterized by a wide range of reddening.
The same conclusion was reached in Section~\ref{obs} above from
the visual inspection of
the true-color X-ray images of Figs.~\ref{xray-img-1}-\ref{xray-img-3}.

\begin{figure}
\resizebox{\hsize}{!}{
\includegraphics[]{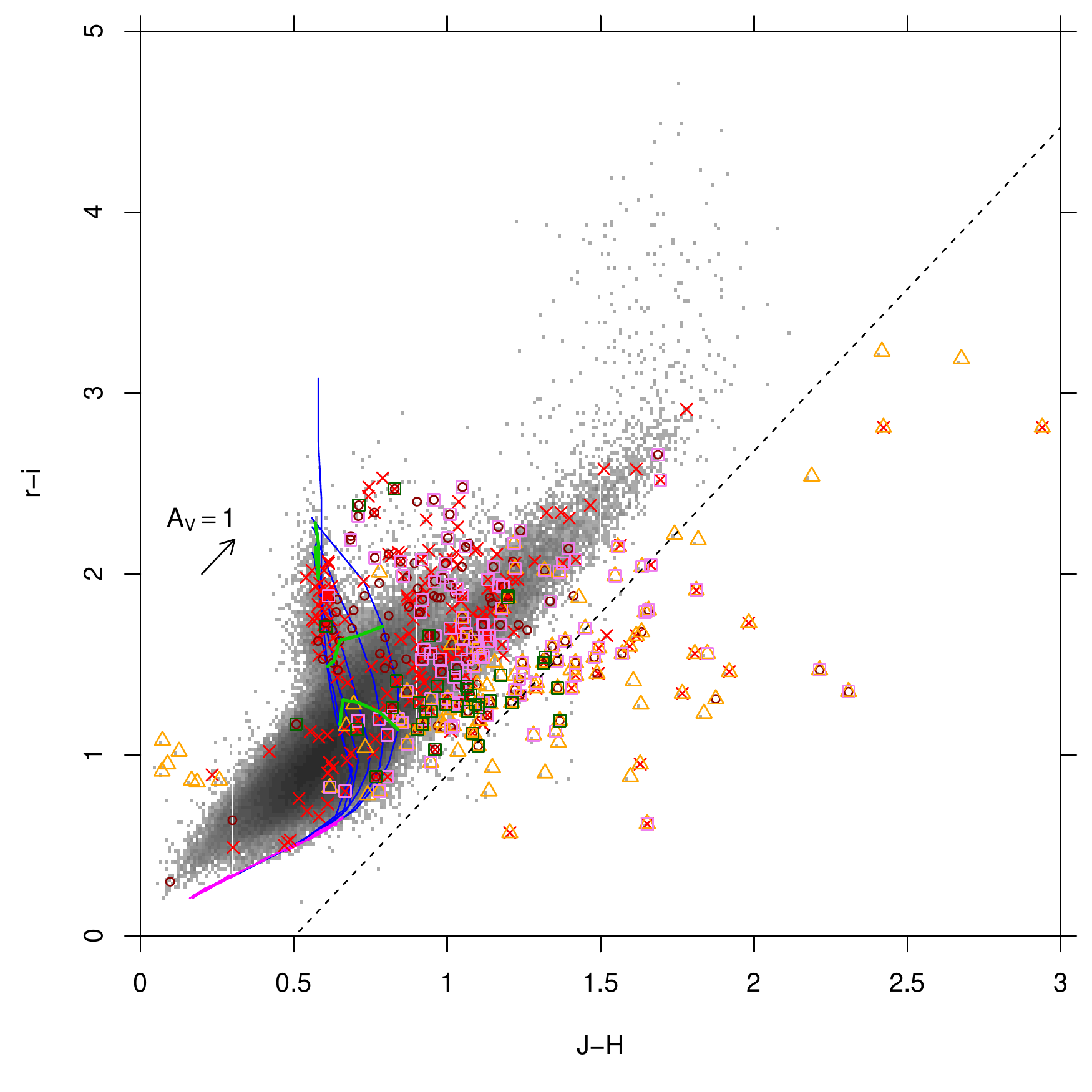}}
\caption{
Optical-IR color-color diagram $(r-i,J-H)$.
Symbols as in Figs.~\ref{ri-rha} and~\ref{jh-hk}.
The dashed line is the adopted limit for selection of NIR-excess objects
from this particular diagram.
\label{ri-jh}}
\end{figure}

\begin{figure}
\resizebox{\hsize}{!}{
\includegraphics[angle=0]{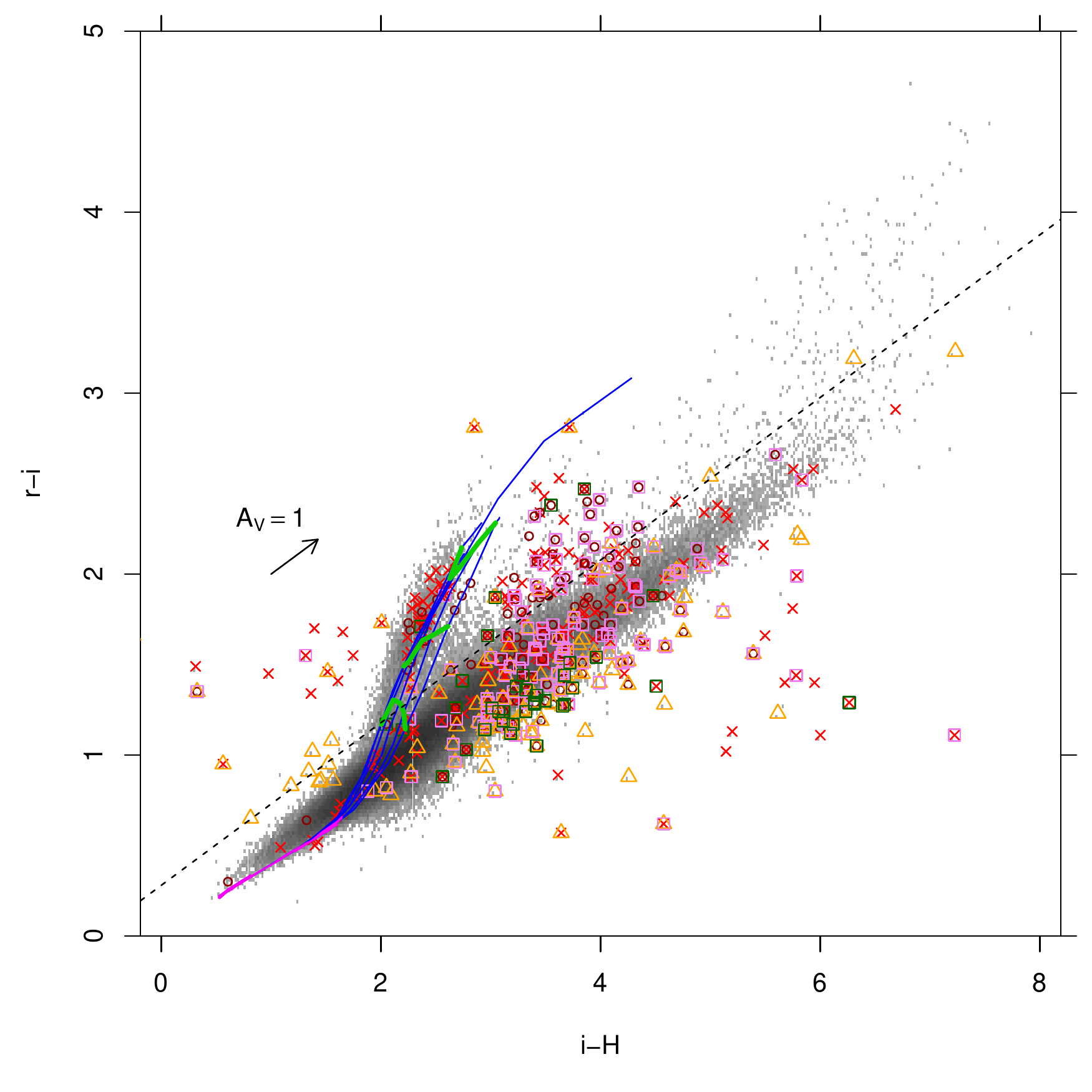}}
\caption{
Optical-IR color-color diagram $(r-i,i-H)$.
Symbols as in Figs.~\ref{ri-rha} and~\ref{jh-hk}.
Stars less massive than $0.5 M_{\odot}$ (M-type) fall above the dashed
line.
\label{ri-ih}}
\end{figure}

Another useful diagram is $r-i$ vs.\ $J-H$ (Fig.~\ref{ri-jh}) although in this
case the number of NIR-excess stars selected is smaller than from
Fig.~\ref{ri-hk}. However, this color combination enables us to identify
the lowest-mass stars more securely than previous diagrams, as it breaks
the near-degeneracy between intrinsic stellar color and reddening, for
masses below $\sim 0.3 M_{\odot}$ (green curves in Fig.~\ref{ri-jh}
are BHAC evolutionary tracks for the lowest masses), as long as a star has
photospheric colors.
An even more detailed, non-degenerate determination of mass and reddening for
individual stars is made possible by the rarely-used, but very interesting,
$(r-i,i-H)$ diagram shown in Figure~\ref{ri-ih}. The high-precision IPHAS
and UKIDSS photometry undoubtedly are a major ingredient for this
representation being so useful. Dwarf M stars, with $M \leq 0.5 M_{\odot}$,
stand out clearly above the main locus of more massive stars, and are
therefore effectively selected using this diagram.
Most of them are unreddened ($2<i-H<3$), of which 34 detected in X-rays.
This diagram indicates that the majority of X-ray sources having an
optical counterpart have instead larger reddening, in
the range $A_V \sim 2-4$. Only 3-4 of the H58 CTTSs have inferred masses $M
\leq 0.5 M_{\odot}$, while all others are more massive (if their $i-H$
color is photospheric, which may not be strictly true, see Fig.~\ref{ri-hk}).
The few X-ray sources to the left of the unreddened isochrones are not
easily justified by any current model: their discrepant colors might arise
from spurious matches between the IPHAS and 2MASS/UKIDSS catalogs, or from
variability in the optical or NIR bands.

{ We have estimated the number of foreground X-ray sources as in
Pillitteri \e (2013), by integrating the field-star X-ray luminosity
function (Favata and Micela 2003) up to the NAP distance, and down to the
limiting X-ray flux values appropriate to our X-ray observations (see
Sect.~\ref{sensitivity} and Fig.~\ref{sens-hist}). We estimate that about 67
foreground sources are detected cumulatively in all XMM pointings,
and 18 in all ACIS pointings, for a total of $\sim$85 foreground sources not
related to the NAP. Being older than the NAP members, these
are expected to be found among the softest detected X-ray sources
(G\"udel \e 1997).  The unreddened M stars found above, and many other
X-ray sources found along the unreddened sequence in the diagram of
Figure~\ref{ri-ih}, are therefore the best candidates as foreground
sources. }

The number of NIR-excess stars is 208 (of which 29 detected in X-rays,
and 38 common to the \ha-excess list),
and their spatial distribution can be seen in Fig.~\ref{dss-ctts-xdet},
with the maximum density found in the ``Pelican'' region.
{ The 273 X-ray undetected NIR-excess and \ha-excess stars are listed in
Table~\ref{table-ir}.}

\begin{table*}[ht]
\centering
\caption{Optical/NIR photometry for X-ray undetected stars with IR/H$\alpha$ excess. Full table in electronic format only.} 
\label{table-ir}
\begin{tabular}{rllccccccccc}
  \hline
Seq & RA & Dec & $i$ & $r-i$ & $r-H\alpha$ & $J$ & $H$ & $K$ & IR & \ha\ & RGS11 \\
no.\ & & & & & & & & & excess & excess & number \\
  \hline
722 & 311.40294 & 45.08670 &  &  &  & 14.27 & 11.12 & 8.82 & Y &  &  \\ 
  723 & 311.52945 & 44.78329 &  &  &  & 18.36 & 14.83 & 12.29 & Y &  &  \\ 
  724 & 311.53239 & 44.77889 & 17.24 & 1.39 & 1.47 & 14.09 & 12.99 & 11.90 & Y & Y &  \\ 
  725 & 311.55911 & 44.87168 &  &  &  & 16.97 & 14.68 & 12.89 & Y &  &  \\ 
  726 & 311.56617 & 43.70630 &  &  &  & 14.34 & 13.48 & 12.59 & Y &  &  \\ 
  727 & 311.57703 & 44.74357 & 18.29 & 1.23 & 0.45 & 14.51 & 12.68 & 11.83 & Y &  &  \\ 
  728 & 311.61675 & 44.04525 &  &  &  & 15.47 & 13.82 & 12.45 & Y &  &  \\ 
  729 & 311.67674 & 43.75285 & 17.36 & 1.98 & 1.28 & 15.02 & 14.15 & 13.66 &  & Y &  \\ 
  730 & 311.68683 & 45.21762 &  &  &  & 14.36 & 11.17 & 8.75 & Y &  &  \\ 
  731 & 311.70190 & 44.18731 & 16.78 & 2.17 & 1.66 & 13.53 & 12.46 & 11.95 &  & Y &  \\ 
  732 & 311.71429 & 44.92656 & 17.22 & 1.64 & 0.76 & 14.46 & 13.38 & 12.55 & Y &  &  \\ 
  733 & 311.71948 & 44.91088 & 15.76 & 1.25 & 0.55 & 13.46 & 12.46 & 11.66 & Y &  &  \\ 
  734 & 311.72259 & 44.39995 & 18.03 & 1.46 & 0.75 & 16.34 & 14.26 & 12.97 & Y &  &  \\ 
  735 & 311.72326 & 43.76692 & 17.89 & 2.21 & 1.65 & 15.23 & 14.54 & 14.07 &  & Y &  \\ 
  736 & 311.74237 & 44.89695 & 16.83 & 1.11 & 0.36 & 14.54 & 13.45 & 12.43 & Y &  &  \\ 
  737 & 311.74626 & 43.68682 & 16.51 & 3.19 & 0.62 & 12.88 & 10.21 & 8.16 & Y &  &  \\ 
  738 & 311.76965 & 43.22022 &  &  &  & 16.45 & 15.01 & 13.63 & Y &  &  \\ 
  739 & 311.77003 & 43.81984 & 17.55 & 1.16 & 2.29 & 15.18 & 14.26 & 13.56 & Y &  &  \\ 
  740 & 311.82935 & 43.36731 &  &  &  & 17.05 & 15.70 & 14.50 & Y &  &  \\ 
  741 & 311.84319 & 43.83651 & 15.38 & 1.16 & 0.84 & 13.70 & 12.73 & 12.41 &  & Y &  \\ 
   \hline
\end{tabular}
\end{table*}

\subsection{Color-magnitude diagrams}
\label{cmd}

\begin{figure}
\resizebox{\hsize}{!}{
\includegraphics[angle=0]{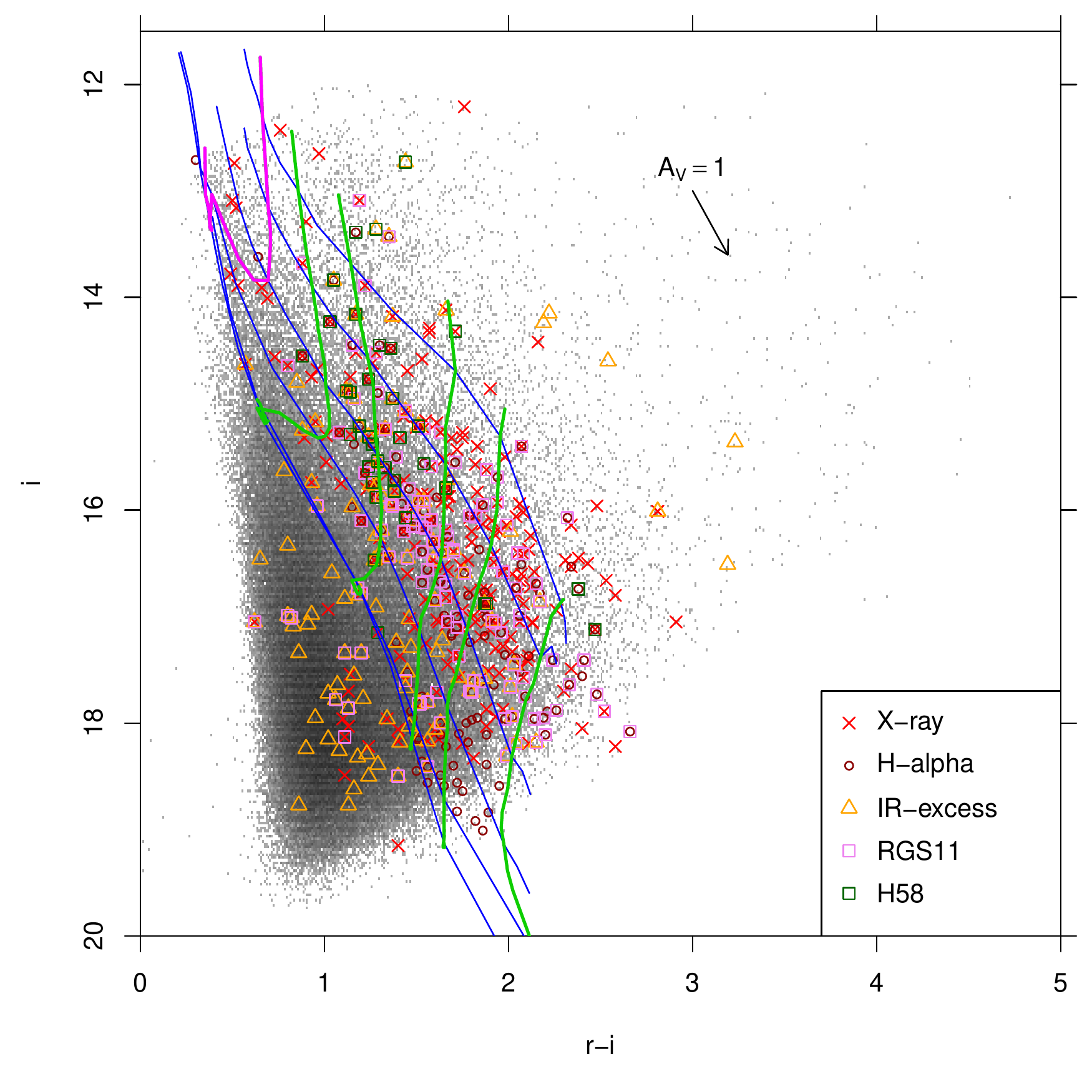}}
\caption{
Optical color-magnitude diagram $(i,r-i)$.
Symbols as in Figs.~\ref{ri-rha} and~\ref{jh-hk}.
Unreddened BHAC evolutionary tracks are here shown for masses of
0.1, 0.2, 0.3, 0.5, 0.7 $M_{\odot}$ (green), and~1 $M_{\odot}$ (magenta).
{
In this and following color-magnitude diagrams,
only data with errors less than 0.1 mag on each axis are shown.
}
\label{i-ri}}
\end{figure}

Figure~\ref{i-ri} is an optical CMD for the whole NAP region. Comparison
with Fig.~\ref{distance} shows that thousands background stars are observed
behind the cloud. Since no reddening correction was applied here, the
apparent masses of e.g.\ the H58 stars, which as seen above are reddened by
several magnitudes in the $V$ band, are much lower than their actual
values, in most cases larger than $0.5 M_{\odot}$ as discussed above.
The few tens X-ray sources and NIR-excess stars below the 10-Gyr isochrone
are not necessarily lying at larger distances: in fact, they share the same
position in this CMD as many of the RGS11 YSOs. Below-ZAMS NIR-excess objects
were already found in other young clusters like NGC~6530 (Damiani \e 2006)
or NGC~6611 (Guarcello \e 2010), and may be attributed to
photospheres obscured by edge-on disks, seen only in scattered light.
{
Since datapoints shown in Figure~\ref{i-ri} have all errors less
than 0.1 mag, attributing their position in the CMD to errors seems
instead unlikely.
}

\begin{figure}
\resizebox{\hsize}{!}{
\includegraphics[]{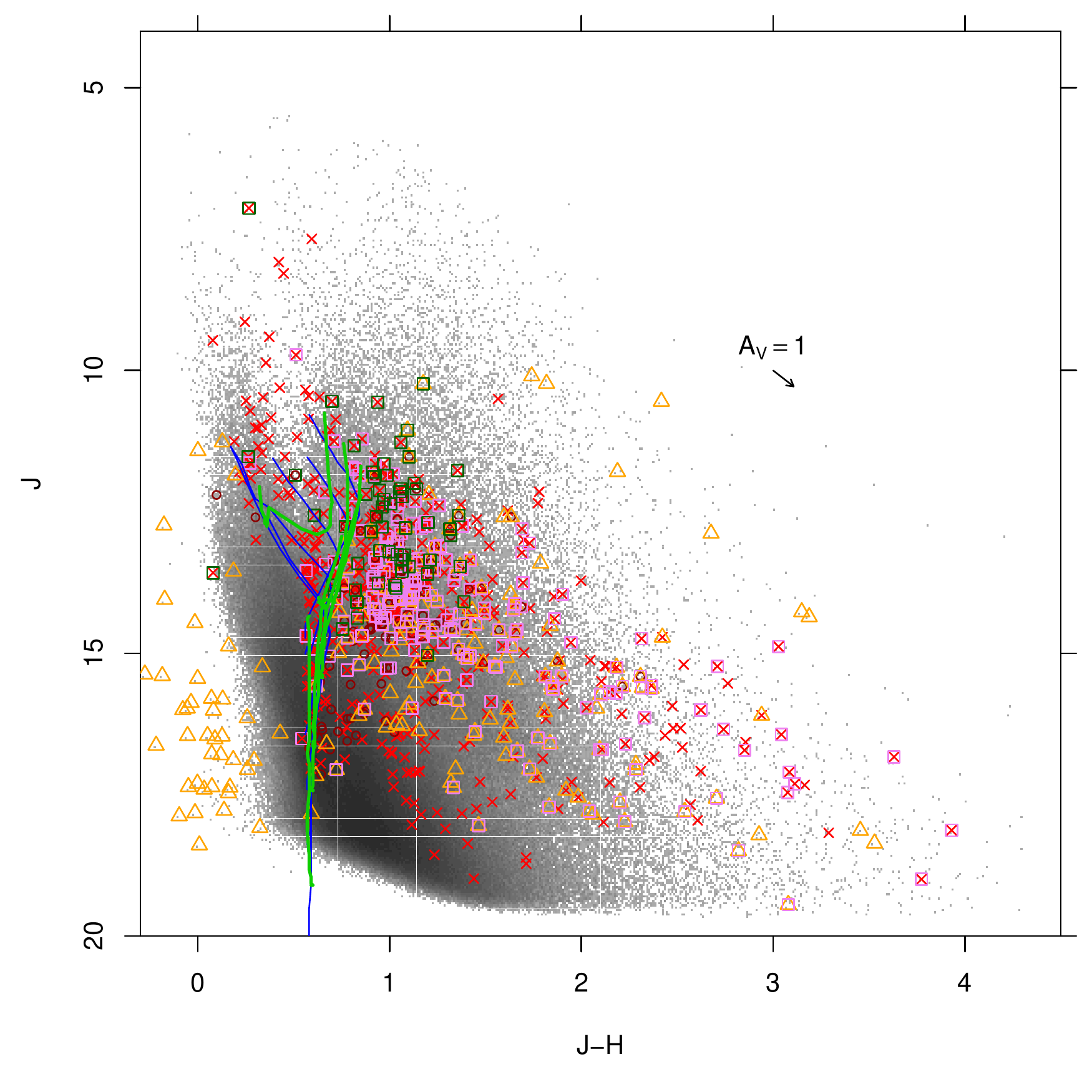}}
\caption{
NIR color-magnitude diagram $(J,J-H)$.
Symbols as in Figs.~\ref{ri-rha} and~\ref{jh-hk}.
\label{j-jh}}
\end{figure}

Considering X-ray sources, there is no clear sequence or cluster locus in
the CMD, probably because of the large scatter in extinction from star to
star. Estimating individual extinction values is subject to large uncertainties,
as we will discuss in the following. Two-thirds of X-ray sources lack an
IPHAS counterpart, probably because they are too highly reddened.
Figure~\ref{j-jh} shows therefore the NIR CMD $(J,J-H)$.
The frequency and intensity of non-photosperic dust emission in these NIR
bands is probably lower than in the $K$ band, from the comparison of
Figs.~\ref{ri-hk} and~\ref{ri-jh}.
The distribution of datapoints, elongated parallel to the reddening vector,
suggests that indeed reddening, rather than $H$-band excess, is responsible
for most of the spread towards red colors, by up to $A_V \sim 35$ for the
most reddened X-ray sources and YSOs with a NIR counterpart.
A considerable number of X-ray sources has inferred $A_V \geq 10$, if
indeed their $J-H$ color is photospheric.

\begin{figure}
\resizebox{\hsize}{!}{
\includegraphics[]{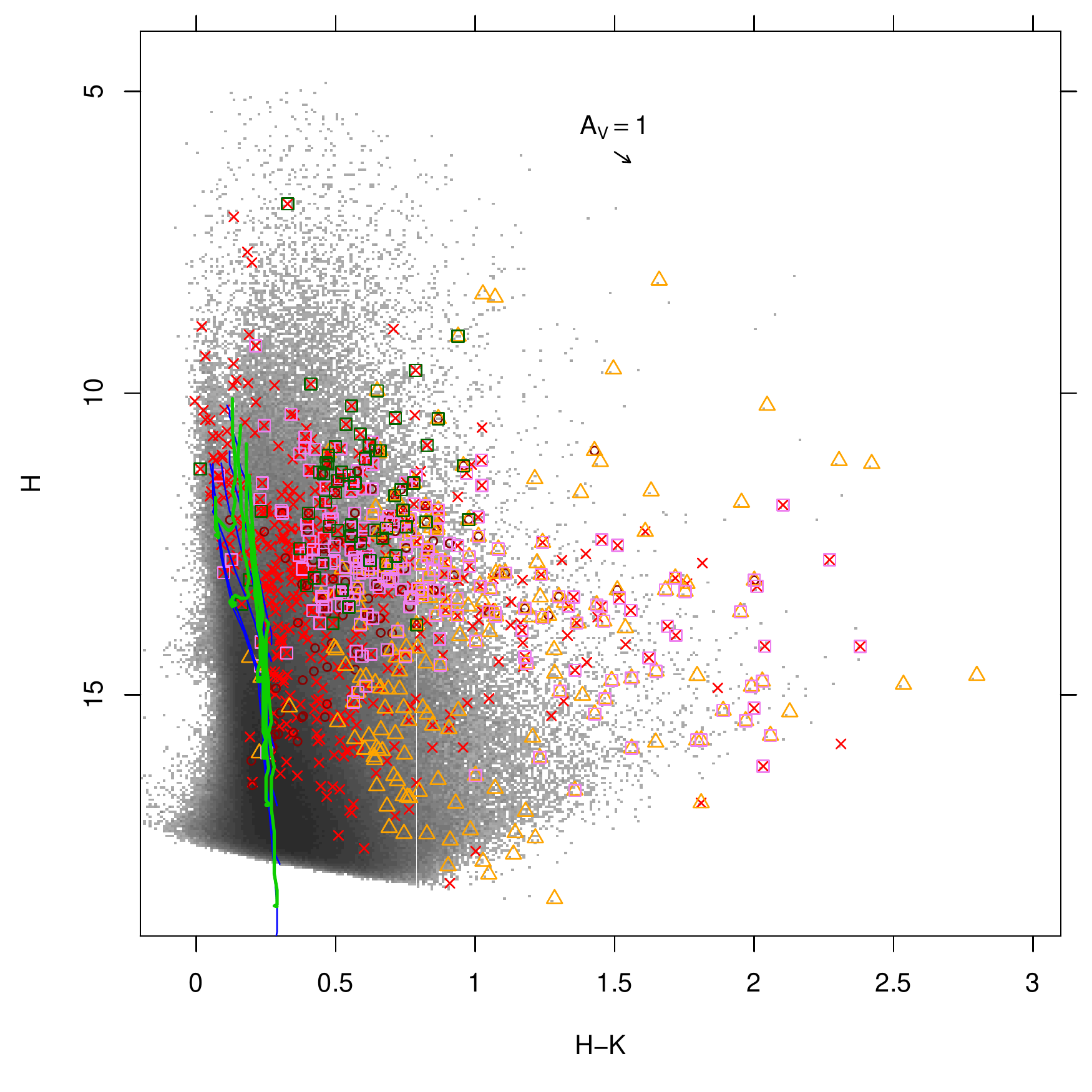}}
\caption{
NIR color-magnitude diagram $(H,H-K)$.
Symbols as in Figs.~\ref{ri-rha} and~\ref{jh-hk}.
\label{h-hk}}
\end{figure}

Finally, Figure~\ref{h-hk} shows the $(H,H-K)$ diagram.
The spread towards red colors is here probably due to a combination of
reddening and $K$-band excess emission. The diagram shows that several tens
new NIR-excess sources (not identified with a YSO from RGS11) are much
fainter objects (with $H>15$) than these latter.
The majority of X-ray sources has $H-K \leq 0.5$, no NIR excesses, and $H$
magnitudes fainter than the H58 stars.

{
In the presence of strong differential reddening, as we discuss below,
incompleteness affects the optical CMD much more than the NIR CMD: from
Figure~\ref{i-ri} we see that while for low reddening the IPHAS
photometry reaches down to the sub-stellar limit at the NAP distance and
ages $\leq 3$~Myr, the mass completeness limit rises rapidly for
increasing extinction (e.g.\ $\sim 1 M_{\odot}$ for $A_V \sim 5$).
Our combined NIR catalog reaches at least 3-4 magnitudes deeper than the
substellar limit at the same age, and is moreover much less sensitive to
extinction (Figures~\ref{j-jh} and~\ref{h-hk}), which accounts for the much
larger number of NIR counterparts
to X-ray sources (548) compared to optical counterparts (235).
}

\subsection{Extinction and reddening}
\label{redden}

\begin{figure}
\resizebox{\hsize}{!}{
\includegraphics[]{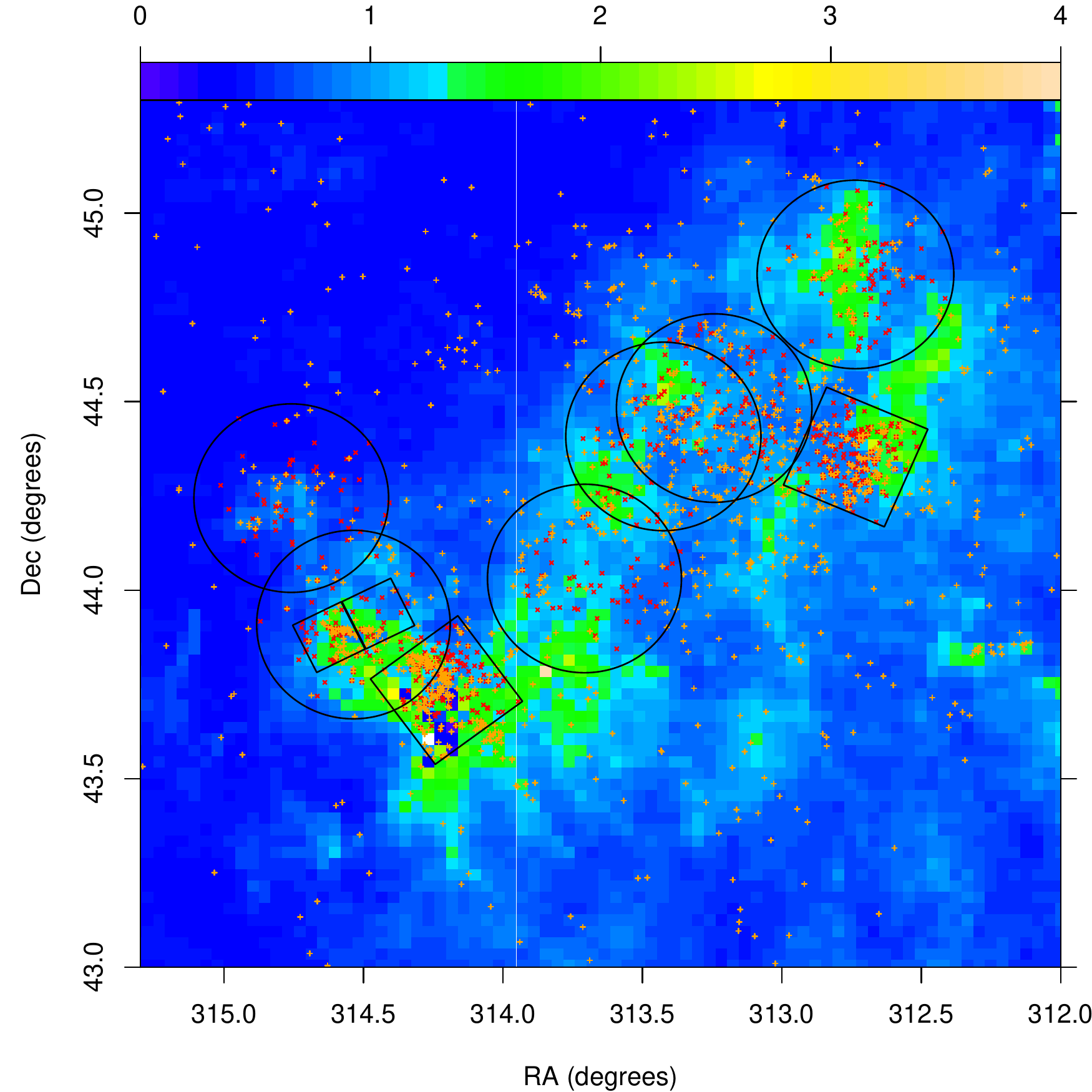}}
\caption{
Spatial map of average $J-H$ (scale at figure top) as a measure of extinction.
Indicated are the X-ray FOVs as in Fig.~\ref{dss-fov-yso}.
The datapoints indicate X-ray sources (red) and RGS11 YSOs (orange).
\label{spatial-jh-mean}}
\end{figure}

The wide range of extinction indicated by the above diagrams may have a
local component, arising in the individual circumstellar environments, but
also shows definite large-scale patterns, as shown by Cambr\'esy \e
(2002) using 2MASS data. Here we take advantage of the more precise UKIDSS
NIR data to build a more detailed extinction map of the NAP, based on
the average $J-H$ color, as shown in Figure~\ref{spatial-jh-mean}.
Optical extinction is approximately 10 times the average $J-H$ (minus
0.5), so that the most obscured regions correspond to $A_V \sim 35$, and
the ``green'' regions in the map correspond to $A_V \sim 15$.

\begin{figure}
\resizebox{\hsize}{!}{
\includegraphics[]{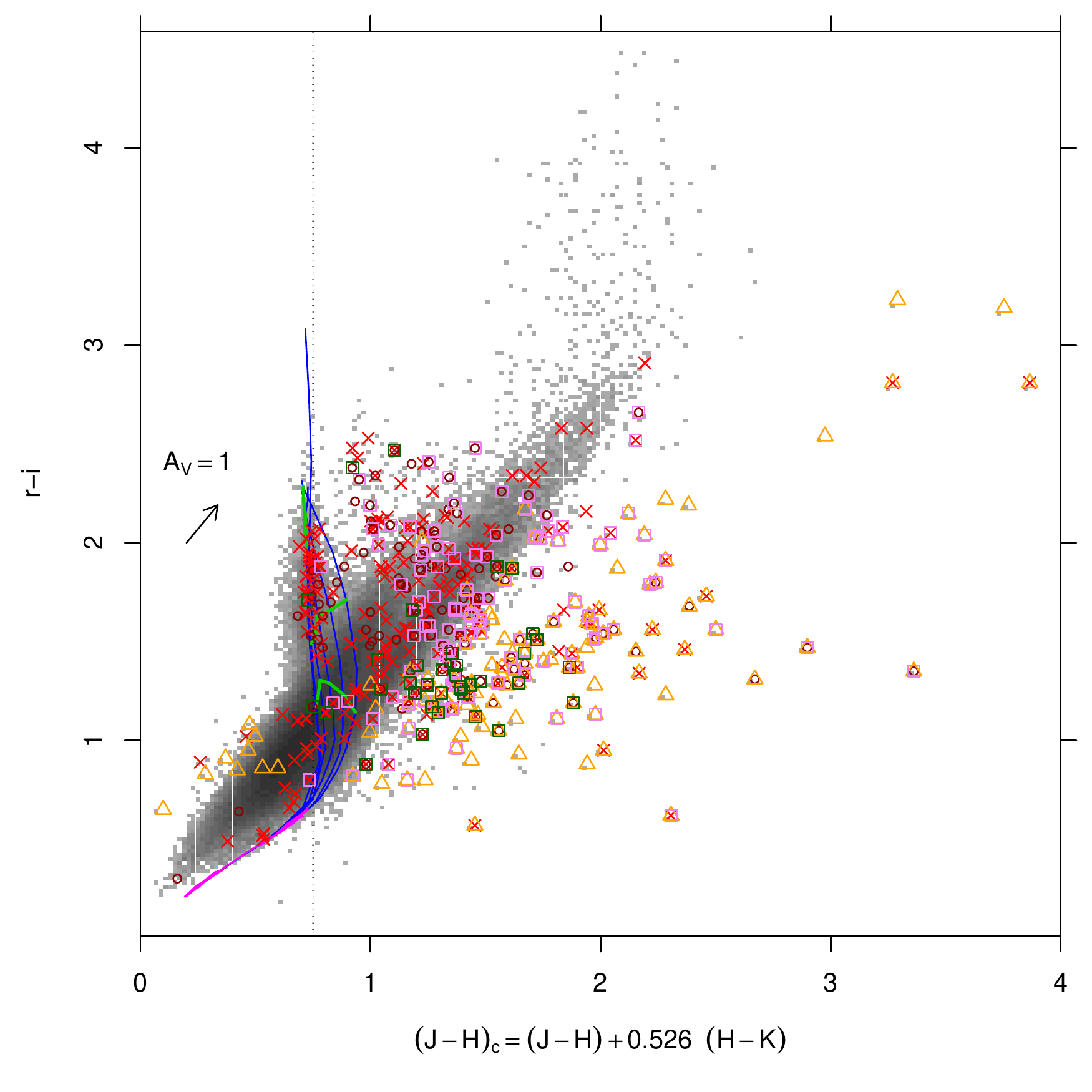}}
\caption{
Mixed color-color diagram of $r-i$ vs.\ corrected color $(J-H)_c$ (see text).
Symbols as in Figs.~\ref{ri-rha} and~\ref{jh-hk}.
The vertical dotted line indicates the value $(J-H)_c = 0.75$.
\label{ri-jh-hk}}
\end{figure}

The $J-H$ color may be considered as the best indicator of reddening for
our NAP candidate PMS members, either X-ray detected or from \ha- or NIR
excess: it is available for most candidate members (548/721 X-ray
sources, 817/994 X-ray/\ha/NIR-excess sources), is less affected by
non-photospheric emission than the $K$ band magnitudes, and normal stars
fall within a small $J-H$ color range (Fig.~\ref{ri-jh}), especially for
$M < 1 M_{\odot}$. The mass dependence of $J-H$ color may moreover be
partially compensated by using a ``corrected'' color, of the form:
\begin{equation}
(J-H)_c \equiv (J-H) +0.526 \; (H-K)
\end{equation}
for stars not classified as NIR-excess sources; the intrinsic unreddened
value of $(J-H)_c$ is $\sim$0.75 for all masses in the range
$0.08 - 1 M_{\odot}$,
and ages $\geq 20$~Myr, with slight deviations for younger low-mass stars
(Figure~\ref{ri-jh-hk}).
Individual visual extinction $A_V$, for stars with $(J-H)_c>0.75$,
can be estimated therefore as
\begin{equation}
A_V \sim ((J-H)_c -0.75) \times (A_V/(E(J-H) +0.526 \; E(H-K)))
\end{equation}
where the last factor above is the (constant) reddening vector slope,
equal to 7.318 (using the Rieke and Lebofsky 1985 reddening law).
For stars more massive than $1 M_{\odot}$ this
underestimates $A_V$, since their intrinsic $(J-H)_c$ color is lower
(bluer) than 0.75.
Extinction $A_V$ for stars with a NIR excess is instead estimated from
the $J-H$ color alone (this is again an underestimate for $M>1 M_{\odot}$).

\begin{figure}
\resizebox{\hsize}{!}{
\includegraphics[]{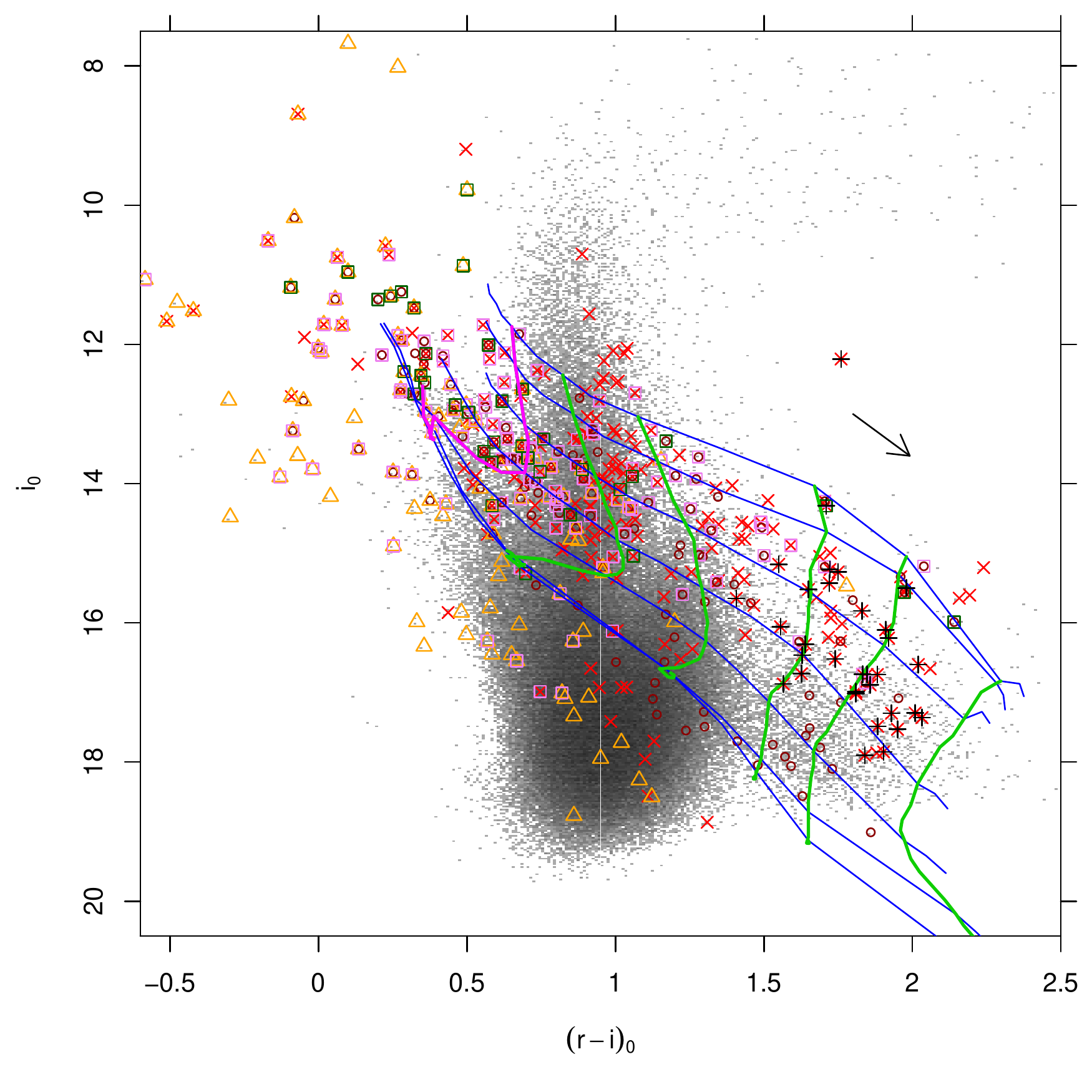}}
\caption{
De-reddened optical CMD. Symbols, isochrones and evolutionary tracks as
in Fig.~\ref{i-ri}. The black plus signs are the unreddened M stars
from Fig.~\ref{ri-ih} (with $2<(i-H)<3$).
\label{i-ri-0}}
\end{figure}

By using these individual $A_V$ values (converted to the IPHAS bands
$A_i$ and $E(r-i)$ as explained in Appendix~\ref{redden-law}) we have
built the dereddened optical CMD, shown in Figure~\ref{i-ri-0}.
The clustering of X-ray sources at $(r-i)_0 \sim 1$ is probably an
artifact resulting from our underestimated reddening for stars above $1
M_{\odot}$. The bulk of X-ray sources, if lying at the NAP distance, are
found in a 2-mag wide strip corresponding to the PMS band, between ages
1-10~Myr, also populated by the majority of H58 stars (on average
brighter than the rest of the X-ray-source counterparts) and
optically-visible YSOs. Tens of NIR-excess stars are found apparently
below the main sequence at the Nebula distance, as remarked above.
The unreddened X-ray-detected M stars selected from Fig.~\ref{ri-ih}
have especially reliable positions in this diagram, since they required
no dereddening: they do not form a proper sequence in this CMD, but are
mixed to other low-mass NAP X-ray sources. They might still be
foreground, active M dwarfs, 3 to 10 times closer than the NAP; this
diagram does not permit to confirm (nor exclude) their NAP membership.

\subsection{Spatial distributions}
\label{spatial}

\begin{figure*}
\includegraphics[width=8.9cm]{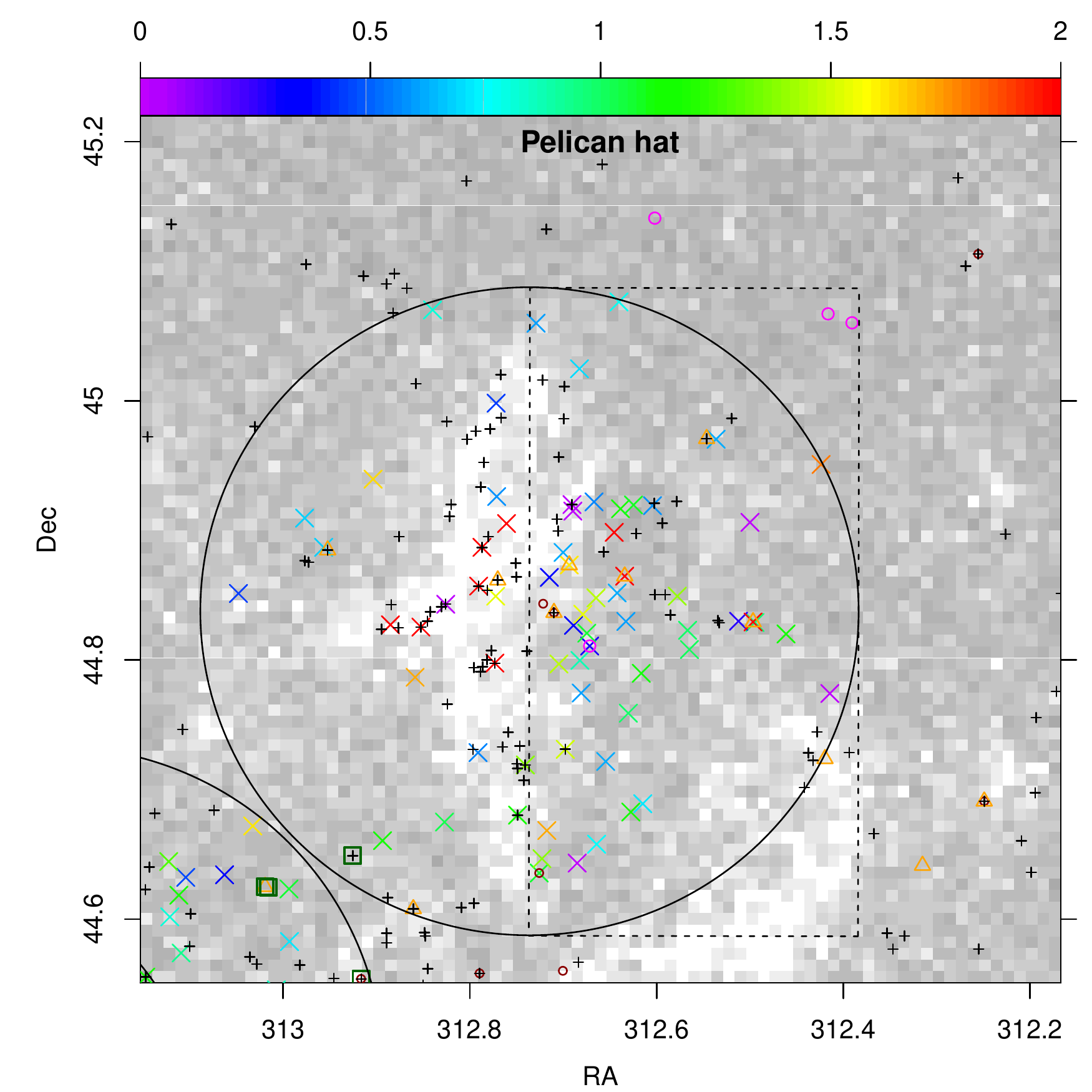}
\includegraphics[width=8.9cm]{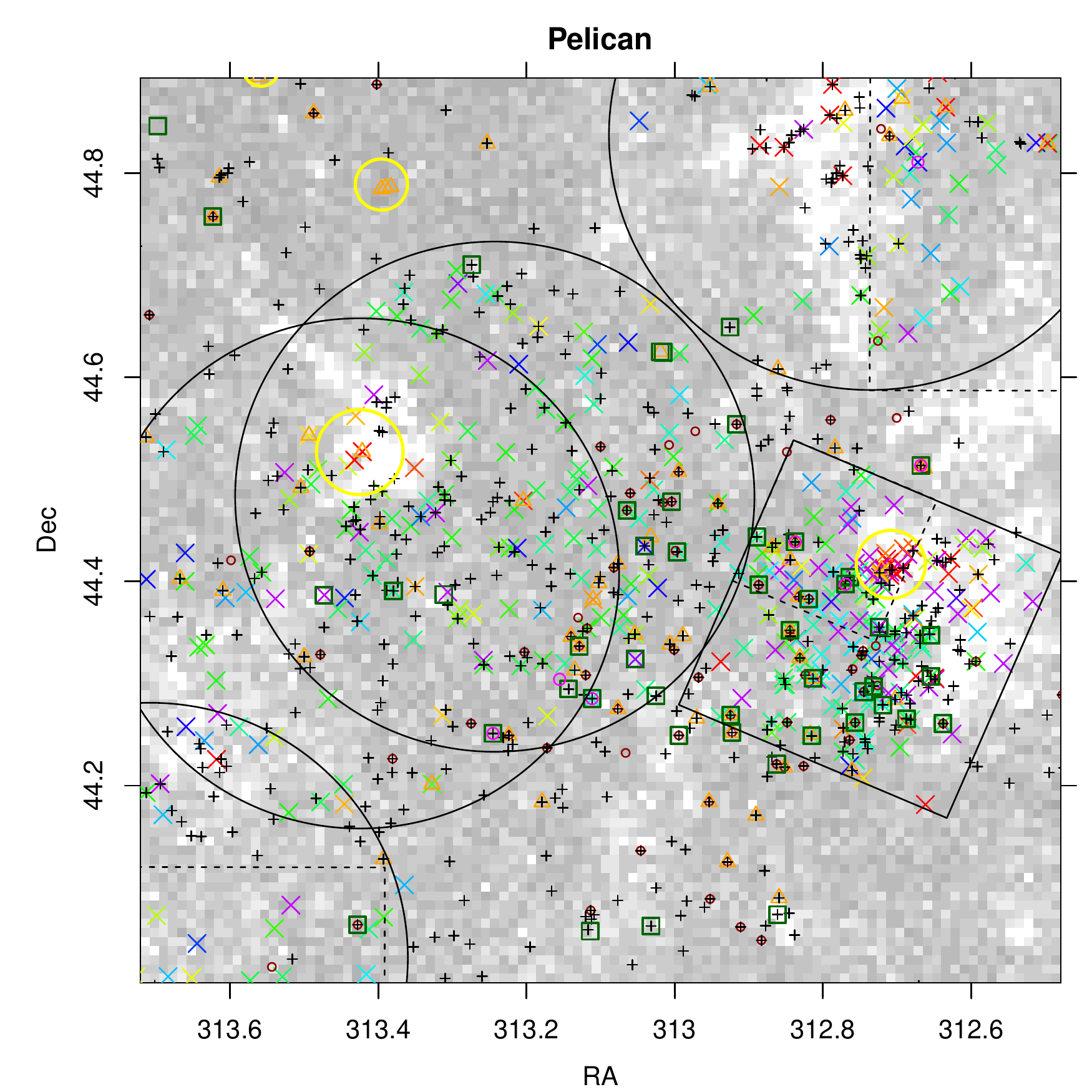} \\
\includegraphics[width=8.9cm]{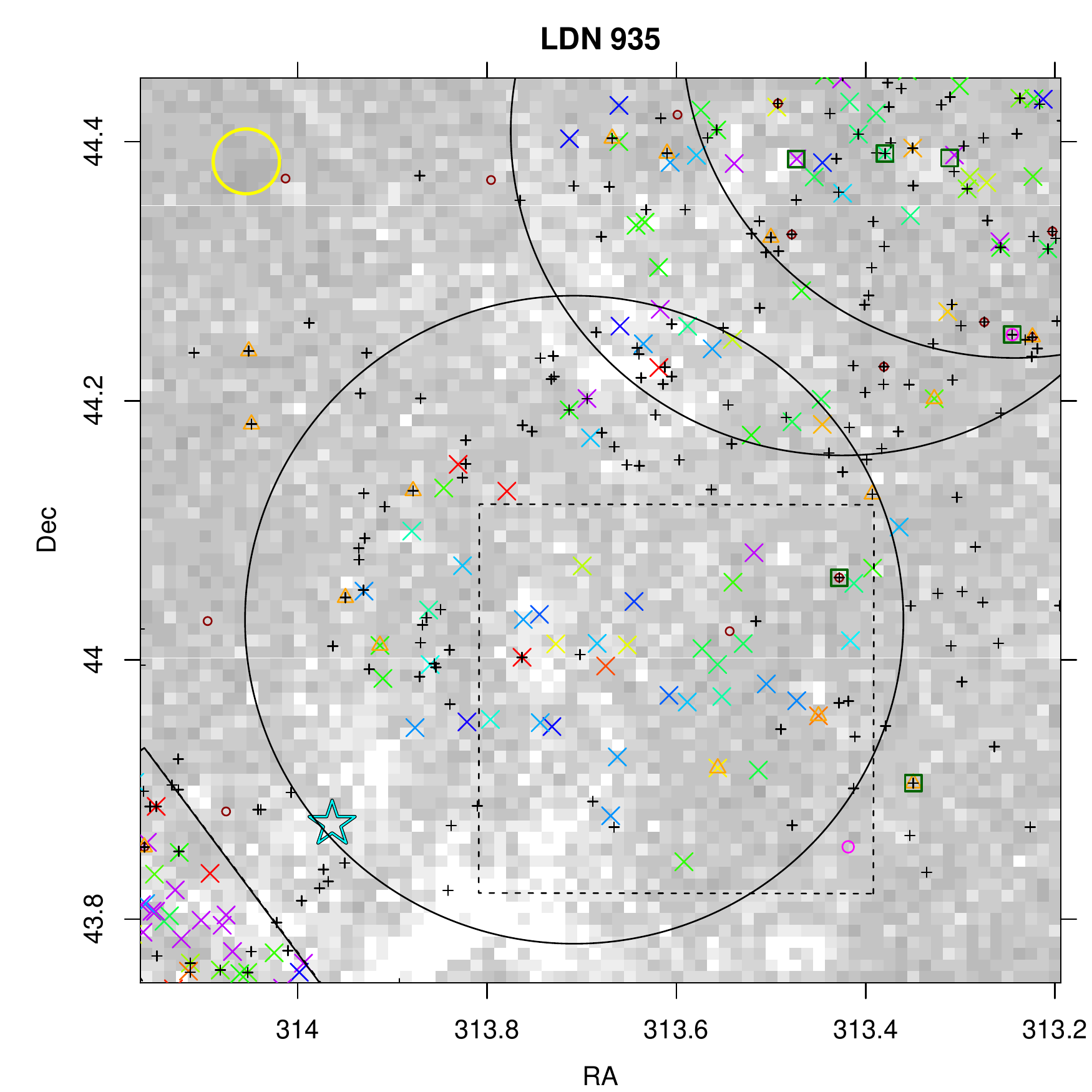}
\includegraphics[width=8.9cm]{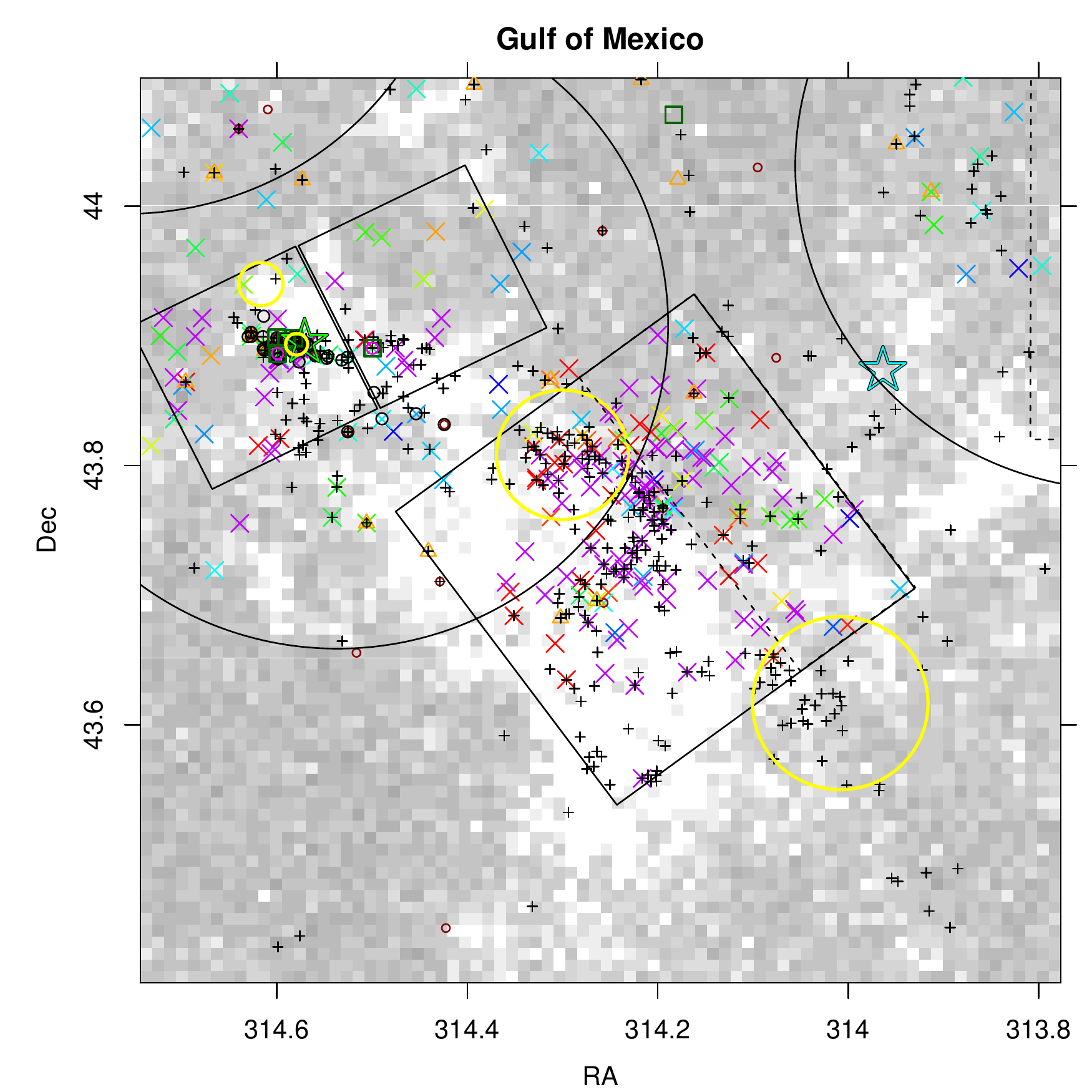}
\caption{
Comparison between spatial distributions of X-ray sources, YSOs and
other young stars in each X-ray FOV. The background grayscale image is a
2-D histogram of the combined optical-NIR object density, showing
clearly high-obscuration regions. Solid black circles/squares are
XMM/ACIS FOVs, respectively.
Symbols as in Figs.~\ref{ri-rha} and~\ref{jh-hk},
except that here X-ray sources (crosses) are color-coded according to their
$J-H$ color (with the color scale show in the top left panel).
X-ray sources without a NIR counterpart are shown in red.
Black plus signs indicate positions of all RGS11 YSOs, regardless of
their identification with X-ray or other catalogued objects.
Small empty circles are young stars from Armond \e (2011; black) and
Welin (1973; violet).
Yellow circles are 2MASS star clusters from Cambr\'esy \e (2002).
The black dashed rectangular regions are defined here as places with a
local overdensity of X-ray sources, not matched by an overdensity of
YSOs.
The three bigger star symbols indicate respectively the O5 star 2MASS
J20555125+4352246 (cyan), and FUOrs HBC722 (green) and V1057~Cyg (orange).
\label{spatial-a}}
\end{figure*}

As found by e.g.\ Cambr\'esy \e (2002) and RGS11, the spatial distribution
of young stars in the NAP is highly non-uniform. This non-uniformity
also holds for the X-ray source distribution (Figs.~\ref{xray-img-1}
to~\ref{xray-img-3}). Therefore, we examine here in more detail the
properties of each spatial sub-region.
The individual panels of Figures~\ref{spatial-a} and~\ref{spatial-b}
permit a comparison between the spatial distributions of all sets of
data considered here. The background greyscale image represents the
local density of all optical/NIR objects (in white the lowest
densities), which is inversely correlated with the mean reddening
(Cambr\'esy \e 2002)
shown in Fig.~\ref{spatial-jh-mean}, but is a somewhat more versatile
indicator (e.g.\ for sources with too noisy or missing $J$ magnitudes).
As in Figs.~\ref{dss-fov-yso} and~\ref{dss-ctts-xdet}, big solid
circles/squares are X-ray FOVs, while RGS11 YSOs are indicated with black
plus signs. X-ray sources are here instead color-coded, according to
their observed $J-H$ color, with the color scale shown in the upper
left panel of Fig.~\ref{spatial-a}. X-ray sources without a NIR
counterpart, or with missing $J-H$, are drawn with a red color, assuming
that high extinction is the reason for the missing measurement.

\begin{figure*}
\includegraphics[width=8.9cm]{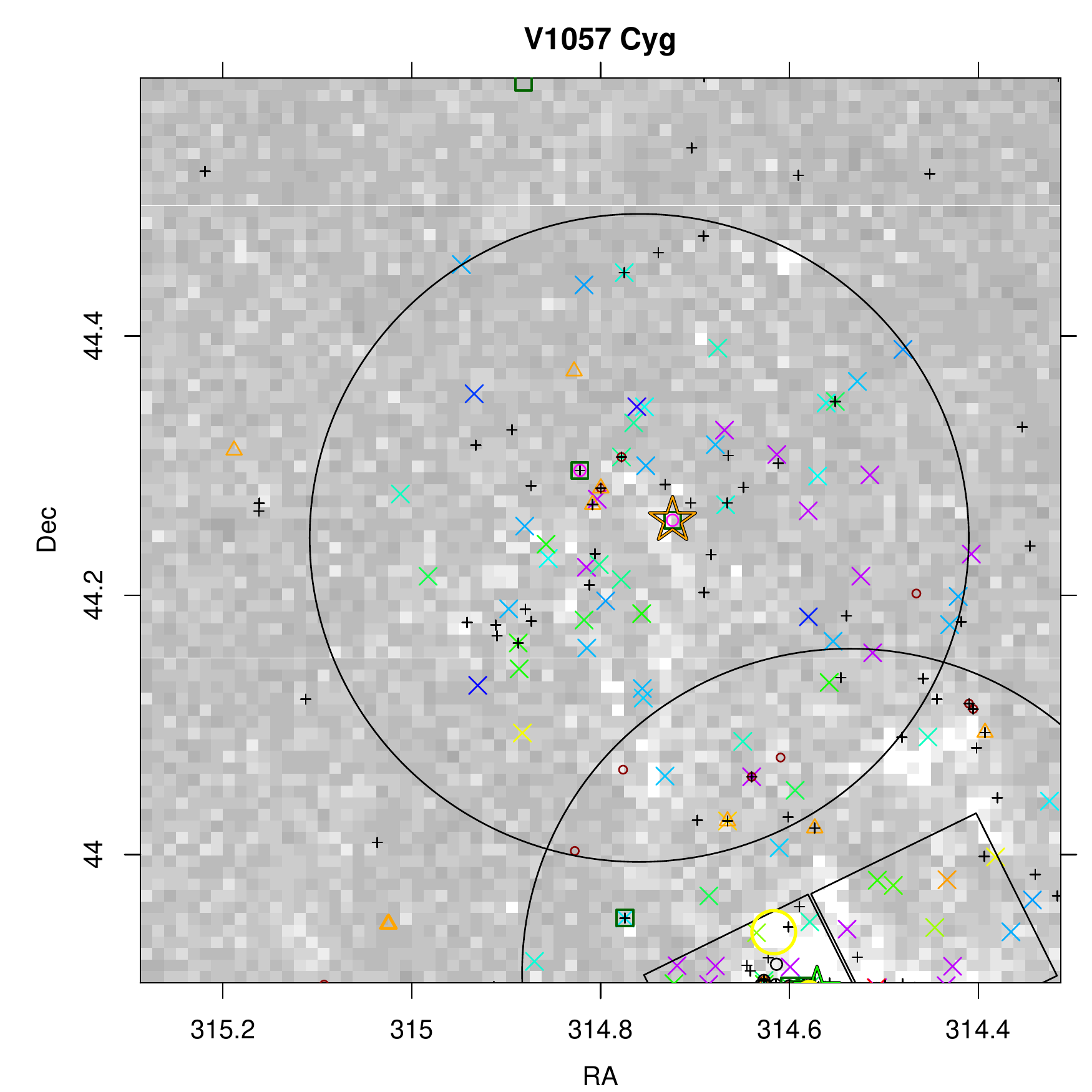}
\includegraphics[width=8.9cm]{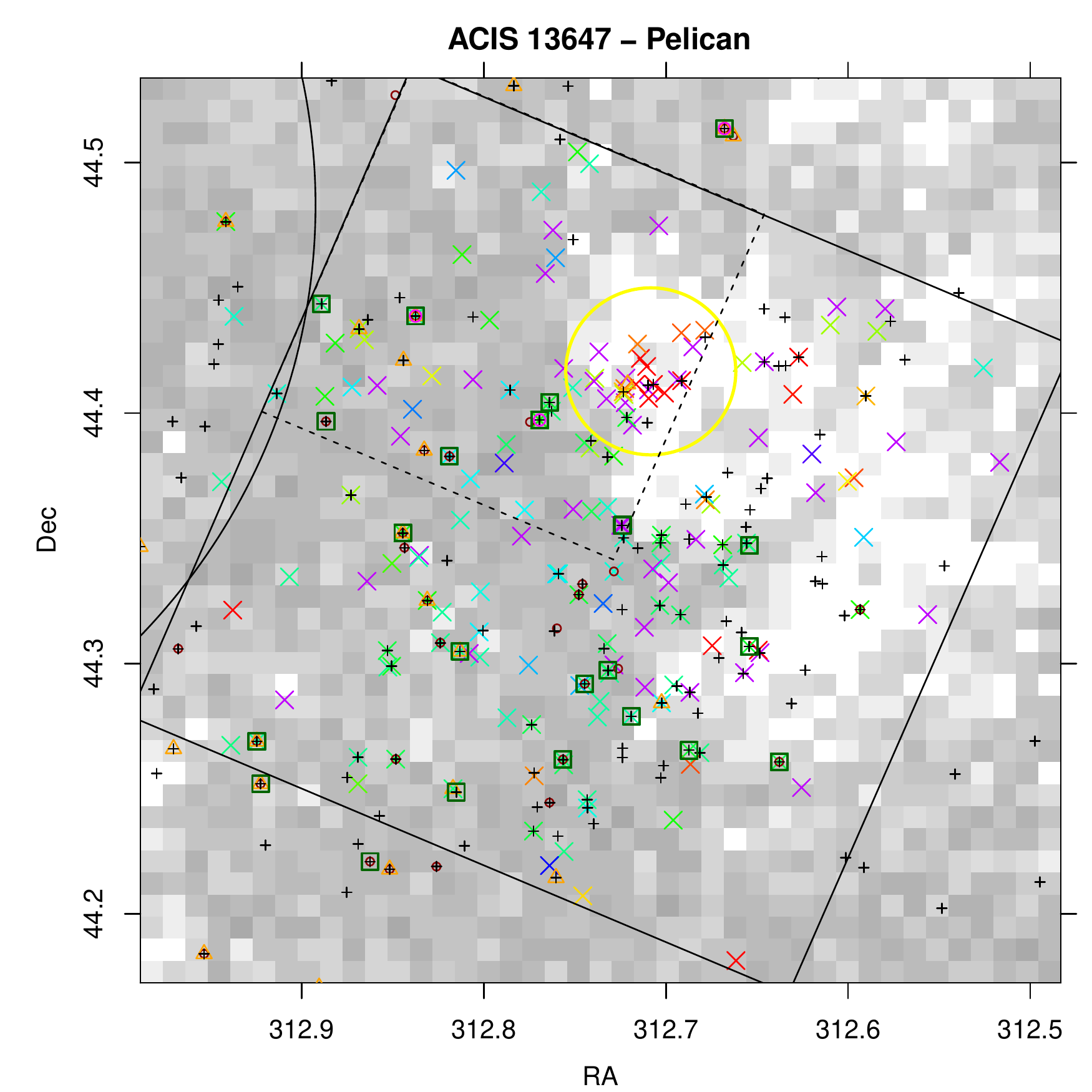} \\
\includegraphics[width=8.9cm]{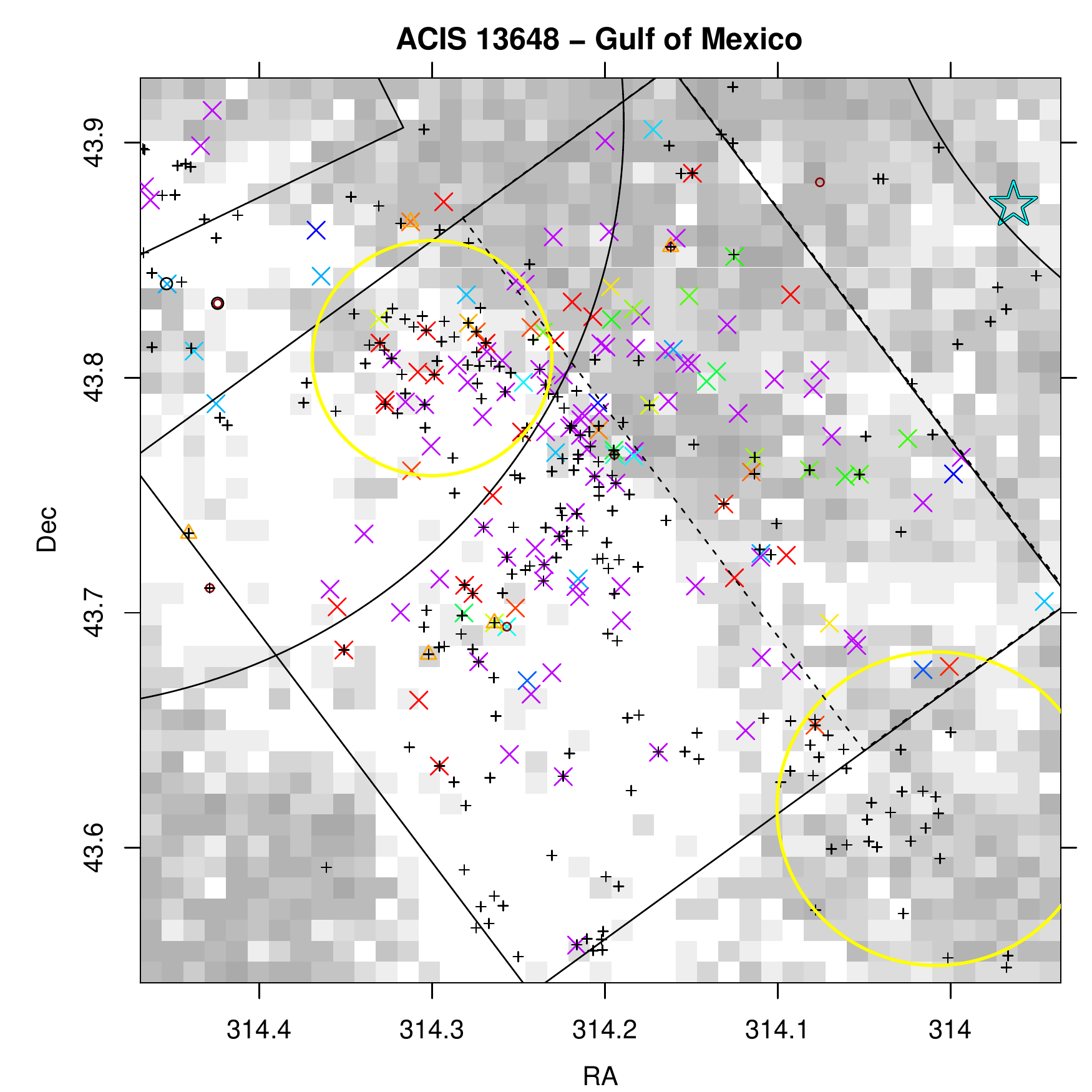}
\includegraphics[angle=90,width=8.9cm]{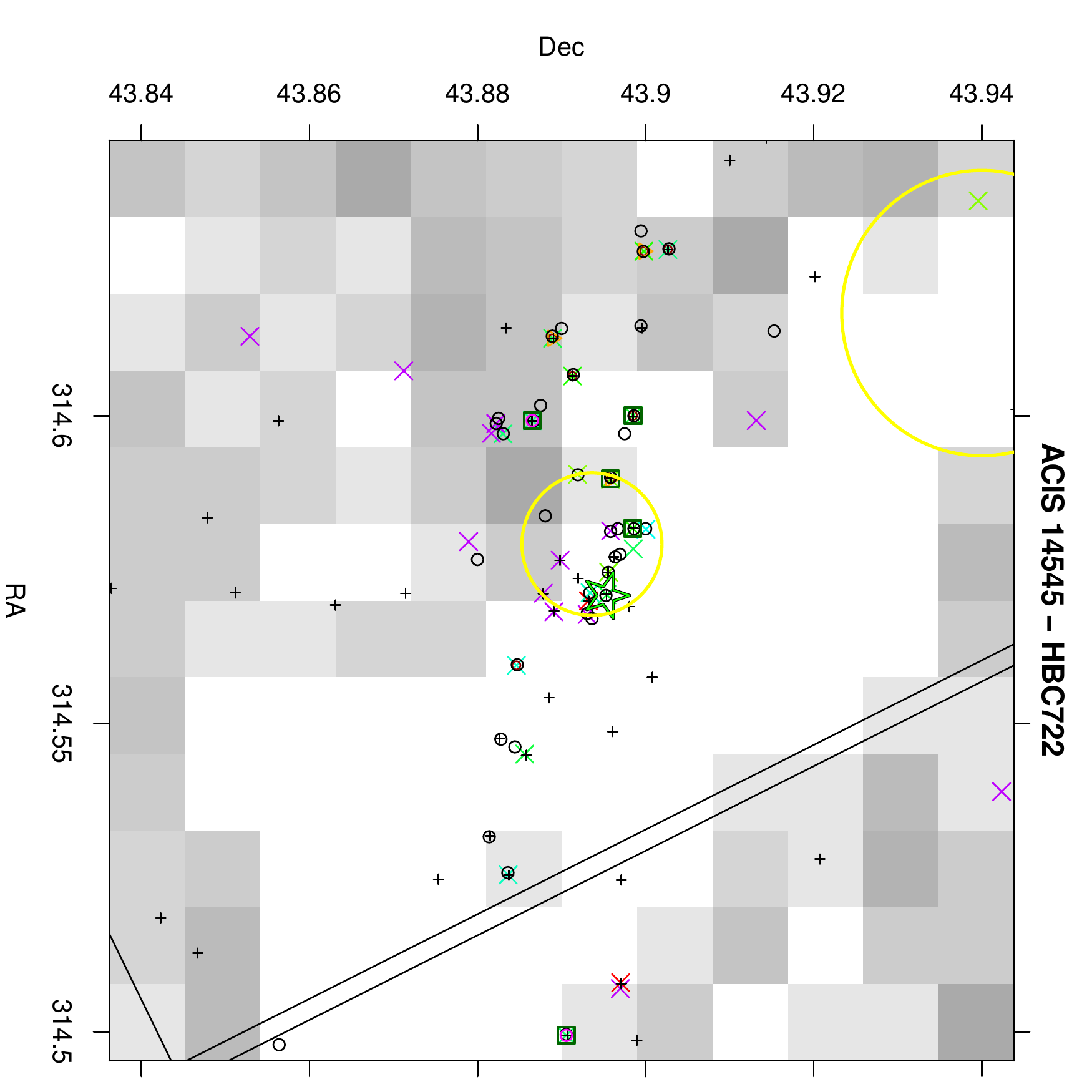}
\caption{
Continuation of Fig.~\ref{spatial-a}.
In the lower-right panel the zooming factor is much larger than in the
other panels, to help visually resolve the dense cluster in the center.
The 2-D bin size of the background image (and the density-grayscale
conversion) is the same in all panels here and in Fig.~\ref{spatial-a}.
\label{spatial-b}}
\end{figure*}

The ``Pelican hat'', as in RGS11, is the northernmost field in our survey
(Fig.~\ref{spatial-a}, top left), covered by XMM-Newton ObsId
0679580101. The region hosts a local concentration of YSOs (see RGS11),
nearly coincident with a region of enhanced obscuration, elongated along
N-S direction, through the center of the XMM-Newton FOV. Curiously, most 
X-ray detected sources in this field lie to the right of center (inside
the dashed rectangle in the Figure). Their
$J-H$ colors suggest they are an inhomogeneous group, either with very
different extinction values, and/or a wide range of masses. Very few of
these in-rectangle X-ray sources also possess a NIR excess. At the same
time, they are not spread uniformly across the X-ray FOV, as if they
were an unrelated field-star population\footnote{The XMM-Newton EPIC
camera sensitivity dependence on off-axis distance is flat enough that
such a central concentration of detections cannot be an instrumental
effect.
{ See also Sect.~\ref{sensitivity}.
}
}, but closer to the West side of the obscured region.
Therefore, these X-ray sources appear to correspond to a sub-population
of the NAP which is missed by YSO/CTTS searches, even using Spitzer or
mixed optical/IR criteria; they also lack strong \ha\ emission, as are
not coincident with IPHAS-selected CTTSs.
The properties of this group of X-ray sources are perhaps best
interpreted as a group of relatively older WTTS members of the NAP
population, and their spatial placement relative to the younger YSOs
(and the currently cloud overdensity seen from the star-count map) as an
indication of sequential star-formation episodes, proceeding from West
to East in this particular region of the NAP.

The upper-right panel of Fig.~\ref{spatial-a} (``Pelican'' field proper)
was covered by one new (0679580201) and one archival (0556050101)
XMM-Newton observations, and by Chandra ACIS-I ObsIds 13647 and~15592;
it hosts most of the H58 stars, and a large number of RGS11 YSOs (though
not the highest density). Most of the X-ray sources in the XMM-Newton
FOVs (those in the ACIS-I FOV will be discussed below)
have intermediate $J-H \sim 1-1.5$, except for 4-5 higher-reddening
X-ray sources falling in a more obscured region, and also coincident with
Cluster~\#6 from Cambr\'esy \e (2002; yellow circles).

Next, the lower-left panel of Fig.~\ref{spatial-a} (LDN~935) shows our
XMM-Newton field closest to the most oscured part of the NAP, and also
to its illuminating O5 star, which lies just outside of our (ObsId 0679580301)
FOV. In proceeding from the Pelican region towards the O5 star, there is
an obvious gradual decrease in the YSO spatial density. Also the S-E
part of the XMM-Newton FOV, closest to the O star, is almost devoid of
X-ray sources. The central and S-W part of this FOV (again marked with a
dashed rectangle) contains very few YSOs, but still a few tens X-ray
sources, again with a mixture of $J-H$ values.

The ``Gulf of Mexico'' region (lower-right panel of Fig.~\ref{spatial-a})
hosts a crowded YSO population, and also dense groups of X-ray sources,
which the spatial resolution of the XMM-Newton EPIC images was
insufficient to resolve. The region, also containing four of
Cambr\'esy's clusters, shows very definite spatial relationships between
the obscuration pattern, the YSO and X-ray populations. Its properties
are best examined by looking at the individual Chandra FOVs, as we do
below. To conclude with our XMM-Newton observed fields,
Fig.~\ref{spatial-b}, upper-left panel, shows the archival ObsId
0302640201, centered on the FU~Ori star V1057~Cyg (not detected in
X-rays), and containing inconspicuous, low-density populations of both
YSOs and X-ray sources. North of it, corresponding to ``continental North
America'', Figure~\ref{spatial-jh-mean} shows
that both the YSO density and the average reddening are very low:
although some hot, ionized gas must be there to produce the emission
nebula, the total dust column density must be low not to redden
background giants (main contributor to the star-count map); a low amount
of warm dust there is also evident from Fig.~\ref{wise}, so that we do
not expect active star formation in this region.

Our ACIS-I ``Pelican'' FOV (upper-right panel of Fig.~\ref{spatial-b})
comprises the highest density of H58 PMS
stars, and also probes with high spatial resolution the bright-rimmed
cloud giving the nebula its shape (see Fig.~\ref{wise}). This cloud is
so dense to obscure most background stars behind its illuminated rim (see
also Fig.~\ref{spatial-jh-mean}). This latter coincides with a cluster
from Cambr\'esy \e (2002), which is also detected as a small cluster of
highly reddened X-ray sources. In the N-E part of the ACIS FOV (dashed
square), the X-ray source density is larger than the YSO density, the
latter being much more numerous in the southern part of the same FOV.

Both the YSO spatial density and local obscuration reach their maximum
inside our ``Gulf of Mexico'' ACIS-I FOV (lower-left panel of
Fig.~\ref{spatial-b}), with YSO falling along arch-like patterns near
the border of the obscured region. Many tens X-ray sources are also
following the same pattern, indicating that they are also young stars
belonging to the NAP population. Some of them have large reddening,
especially those coincident with Cambr\'esy's Cluster~\#2 (northernmost
yellow circle). The obscuring material here, unlike that in the Pelican
nebula, does not coincide with the (southern) bright rim, as it should
be evident from comparing Figs.~\ref{wise} and~\ref{spatial-jh-mean}.
To the N-W of the most obscured region and the YSO cluster, the YSO
density drops abruptly but there are still several tens of X-ray
sources (dashed rectangle), which appear to be too little reddened to
belong to the
background, and too numerous to belong to the foreground of the NAP:
they again are good candidates as WTTS members of the NAP. It is worth
remarking that their spatial density increases towards the obscured
region, while it decreases slightly towards the N-W border of the FOV,
in the direction of the O5 star, only 5$^{\prime}$ away. As above, there
is not detectable increase of X-ray source density in the neighborhood
of the most massive star. 
It might be argued that the in-rectangle X-ray sources are too lightly
reddened to be NAP members, in a region dominated by strong obscuration,
and that they are unlikely to lie all on the near edge of the obscuring dust.
The reason for their apparent low reddening might then be their higher
mass ($M > 1 M_{\odot}$), causing reddening from $J-H$ to be
increasingly underestimated the higher the actual mass is.

Last, in the lower-right panel of Fig.~\ref{spatial-b} we show a zoom of
the central part of archive ACIS-S ObsId 14545, centered on the
recently-erupted FU~Ori object HBC722, surrounded by a dense group of
X-ray sources, known YSOs and PMS stars (H58, Armond \e 2011;
also coincident with Cambr\'esy's Cluster~\#3b): they lie all near the
border of a strong-obscuration region.

\subsection{X-ray spectra, lightcurves and luminosities}
\label{xlum}

\begin{figure}
\resizebox{\hsize}{!}{
\includegraphics[angle=90]{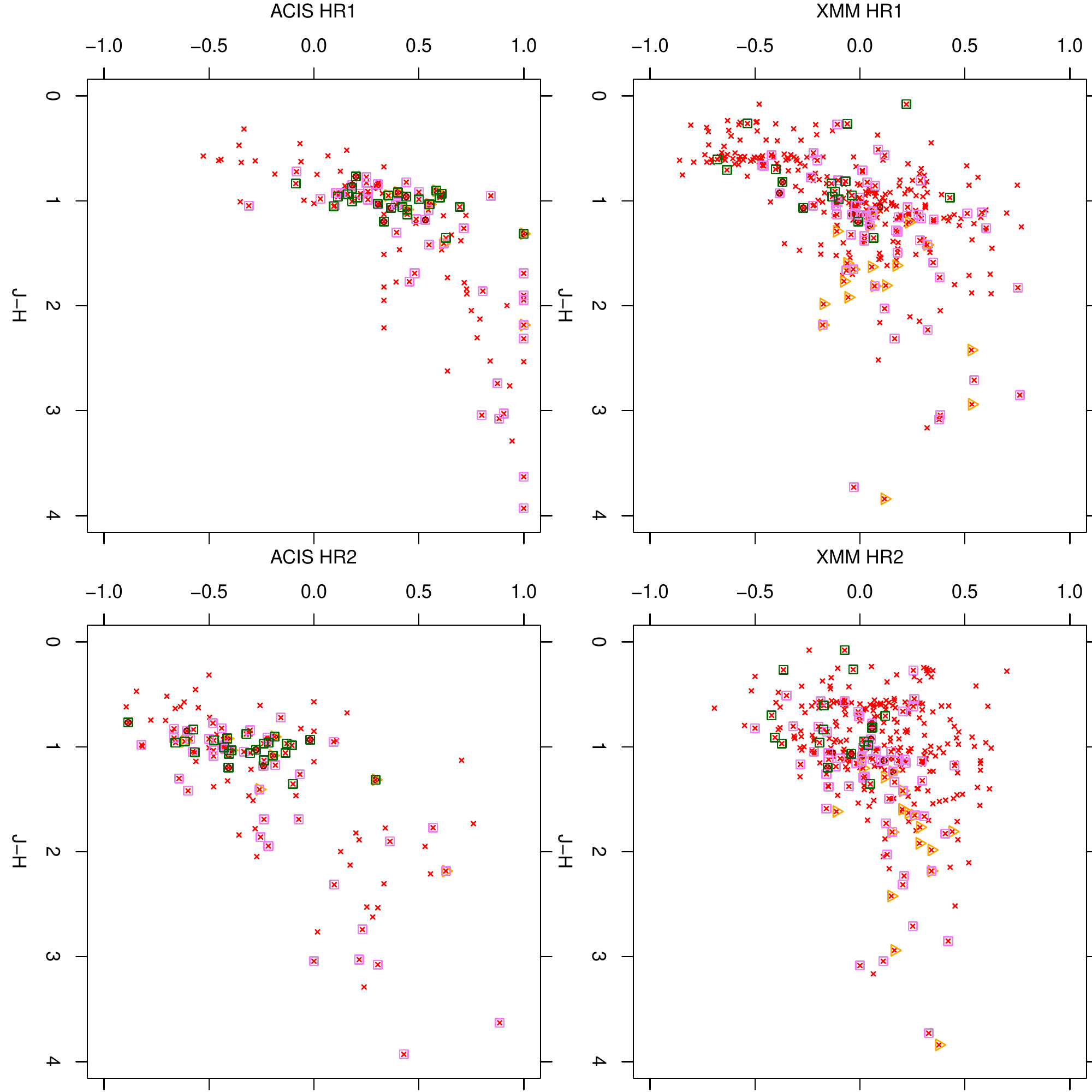}}
\caption{
Hardness ratios (left panels: $HR1$; right panels: $HR2$) vs.\ $J-H$,
for all ACIS (bottom panels) and XMM-Newton (top panels) X-ray sources.
Symbols as in Fig.~\ref{ri-rha}.
\label{hr}}
\end{figure}

\begin{figure*}
\includegraphics[width=18cm]{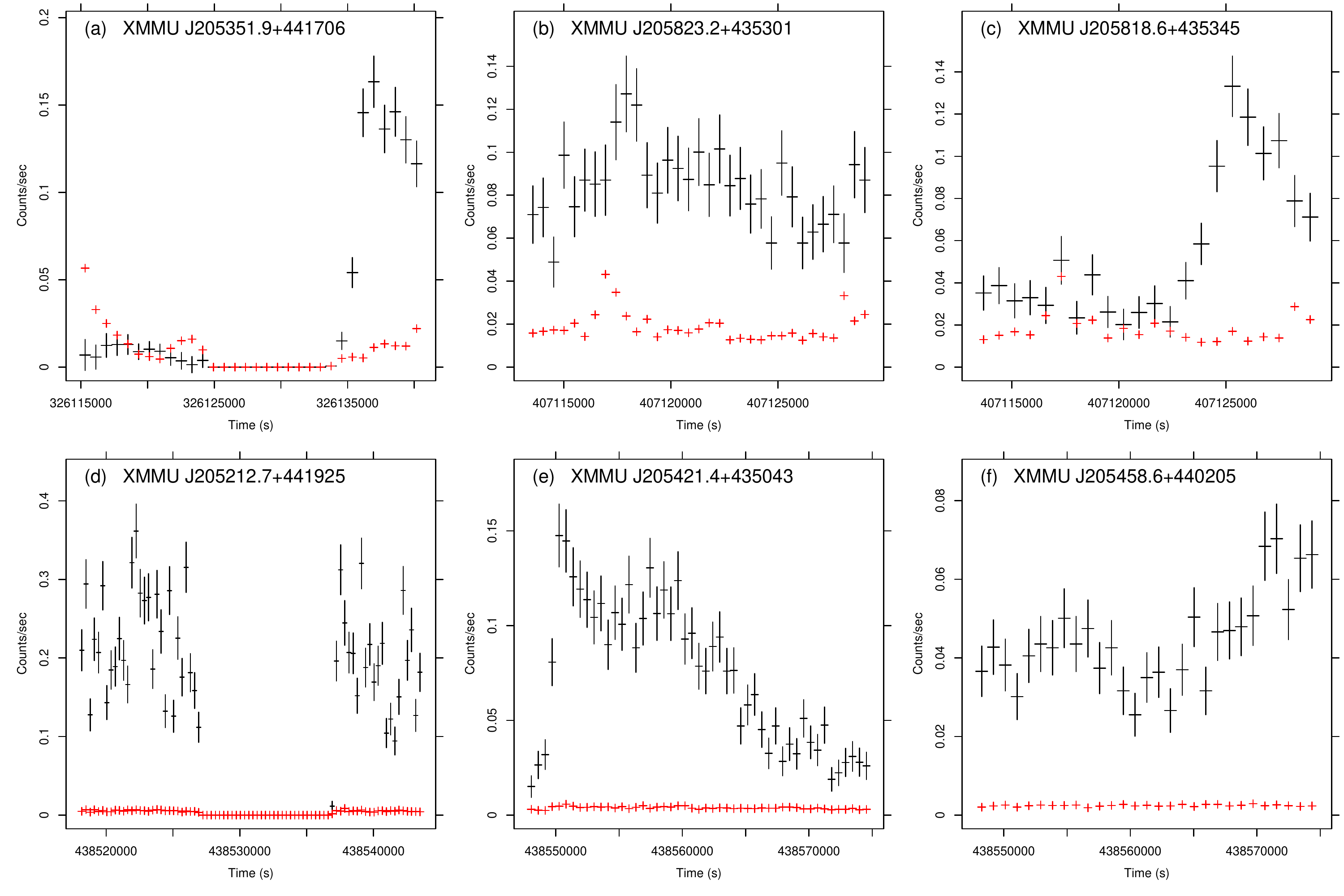}
\caption{
Examples of (background-subtracted) X-ray light curves of XMM-Newton
X-ray sources.  The observations of sources in panels $(a)$ and $(d)$
have interruptions in the middle, hence their zero count rates there. The
background count rate, scaled to the source area, is shown with red
symbols.
\label{x-ltc}}
\end{figure*}

Whenever X-ray sources are detected with a sufficient number of X-ray
counts, their X-ray spectra and lightcurves may help to understand their
nature.
Here we make use of hardness ratios and do not perform a detailed best
fit of the spectra of the brightest sources, deferring this to a future
paper.
Hardness ratios are here defined as $HR1 = (M-S)/(M+S)$, and
$HR2 = (H-M)/(H+M)$, where $S$, $M$, and $H$ are the source X-ray counts
in the soft (0.3-1.2 keV), medium (1.2-2.4 keV), and hard (2.4-8.0 keV)
bands, respectively.
$HR1$ is most affected by the line-of-sight absorption of soft X-ray
photons, while $HR2$ is affected by absorption only at very high column
densities, and more often is a measure of intrinsic spectrum hardness
(and high plasma temperatures for thermal sources).
Figure~\ref{hr} compares these hardness ratios with the $J-H$ color,
separately for XMM-Newton sources (with more than 50 X-ray counts) and
ACIS sources (with more than 20 counts). Since the instrumental
effective area has a different energy dependence for XMM-Newton and
Chandra ACIS, the respective $HR1,2$ must not be mixed together.
We have argued above that $J-H$ is dominated by reddening: this is
further supported by the fairly good correlation between $HR1$ and this
color (left two panels in Fig.~\ref{hr}); the X-ray sources associated
with a H58 star or with a RGS11 YSO follow the same relation as the other
sources. A much weaker correlation is seen between ACIS $HR2$ and $J-H$,
as expected if $HR2$ is only weakly related to absorption. No relation
is seen between XMM-Newton $HR2$ and $J-H$. Also when considering $HR2$,
the X-rays sources with a H58 star/YSO association are homogeneous with
the others. Nearly all of H58 stars have (XMM-Newton or ACIS) $HR2 \leq
0$, indicative of not-too-hot X-ray emitting plasma. Over the same $J-H$
range of the H58 stars, however, the upper-right panel of Fig.~\ref{hr}
shows that about one-half of the RGS11 YSOs have $HR2>0$, and thus even hotter
plasmas, in agreement with previous results on very young Class~I/II
protostars in Orion (e.g.\ Prisinzano \e 2008).
The group of unreddened X-ray sources at $J-H \sim 0.6$, already
mentioned with reference to Figs.~\ref{ri-jh} and~\ref{ri-ih}, is also
recognizable in the two upper panels of Fig.~\ref{hr}, their $HR1,2$
also being rather homogeneous. Their low $HR1$ values confirm their low
line-of-sight absorption, but not lower than several of the H58 CTTS;
their $HR2$ values average to $\sim 0$, like the bulk of YSOs, which
suggests that these X-ray sources have intrinsic spectra no different
from the bona-fide population of the NAP, despite that their low
reddening/absorption is compatible with being active foreground stars.

We have also examined the X-ray lightcurves of sources detected with
enough counts. Some examples are shown in Figure~\ref{x-ltc}, which
illustrates the type of variability found. When the count rate is large
enough, most sources are found to be variable; in many or most cases,
lightcurve shapes are complex, while ``simple'' flares with impulsive
rise and exponential decay are more rare. This is reminiscent of the
X-ray lightcurves of YSOs in the $\rho$~Oph star-forming region, from
the XMM-Newton DROXO survey (Pillitteri \e 2010).
Panel $e$ shows a probable (double-peaked) flare event, with a long
decay time of $\sim 10$~ksec, as found e.g.\ in the COUP Orion
lightcurves (Favata \e 2005).
Some of the XMM-Newton observations have long interruptions in the
middle (like in panels $a$ and $d$).
Large-amplitude variability on timescales of one day may also be seen
from the comparison of the co-pointed ACIS-I images ObsId 13647 and
15592, already shown in Fig.~\ref{xray-img-1}.

\begin{figure}
\resizebox{\hsize}{!}{
\includegraphics[]{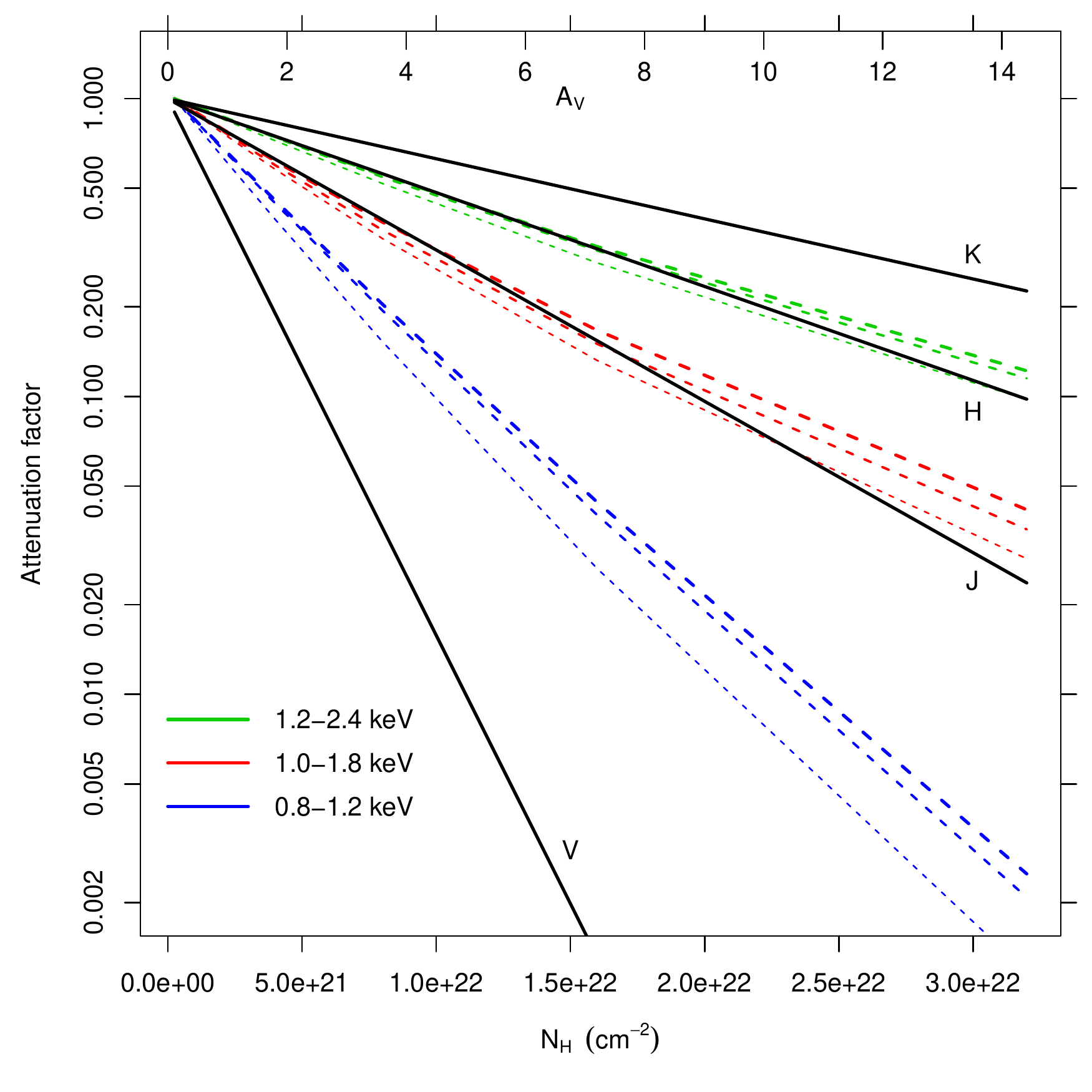}}
\caption{
Computed attenuation factors for X-ray and NIR emission and variable
line-of-sight absorption, as parameterized by the hydrogen column
density $N_H$ (bottom axis) and its equivalent extinction $A_V$ (top
axis). X-ray attenuation of the observed ACIS-I flux is computed for three
single-temperature emitting plasmas (short-dashed: 1~keV; mid-dashed:
1.5~keV; long-dashed: 2~keV), and three energy ranges, as labeled.
Optical and NIR attenuation is shown (using the $A_{\lambda}/A_V$
conversions from Rieke and Lebosfky 1985) with the black solid lines.
\label{x-ir-atten}}
\end{figure}

Since X-ray emission decays with stellar age, one possible way of
discriminating very young stars in the NAP from foreground/background
active non-members is through the use of X-ray-to-optical luminosity
ratios, which decrease with mass-dependent timescales (e.g.\ Preibisch
and Feigelson 2005). Ratios of X-ray to bolometric ($L_X/L_{bol}$) or
X-ray to visual ($L_X/L_V$) luminosities are commonly used. In our case,
the difficulty in obtaining accurate extinction values for individual
X-ray sources makes these ratios unsuitable, and
has prompted us to search a reddening-free equivalent
ratio. It is well known than while the dust absorption decreases
from optical to IR wavelengths, the X-ray absorption decreases from the
soft to the hard X-ray band, with photons of energy $E= 2$~keV being
absorbed like NIR photons with $\lambda = 2\mu$, for a typical
interstellar gas/dust ratio. Therefore, we searched
for an X-ray energy range where absorption is equivalent to that in the
$J$ NIR band, which as discussed above is convenient to study our NAP
stars. We made several experiments using the PIMMS
software\footnote{Portable, Interactive Multi-Mission Simulator, available at
http://heasarc.gsfc.nasa.gov/docs/software/tools/pimms.html}, for
different energy bands: 0.8-1.2~keV, 1.0-1.8~keV, and 1.2-2.4 keV,
sufficiently wide for being well characterized at the spectral resolution of
Chandra/ACIS detector.
Three single-temperature APEC thermal models were considered, in the
temperature range of Class~I/II/III YSOs. Hydrogen column
density ($N_H$) values were converted to optical extinction using the
relation $N_H = 2.22 \cdot 10^{21} A_V$ (Gorenstein 1975), while the
correspondence between $A_V$ and NIR extinction is explained in
Appendix~\ref{redden-law}. The results of the experiment is shown in
Figure~\ref{x-ir-atten}, where the ordinate shows the relative flux
attenuation. We see that the X-ray band between 1.0-1.8~keV is
absorbed by an amount similar to the $J$ band, within a factor of $\sim
50$\% up to $A_V \sim 10-12$. Absorption in the harder 1.2-2.4~keV band is also
very close to that in the $H$ band; however, X-ray counts are often less
in this higher-energy band than in the 1.0-1.8~keV band, and the NIR $H$
band is more easily affected by non-photosperic contributions than the
$J$ band, so that the pair [1.0-1.8~keV]/$J$ seems preferable to obtain an
absorption-independent flux ratio.

\begin{figure}
\resizebox{\hsize}{!}{
\includegraphics[angle=90]{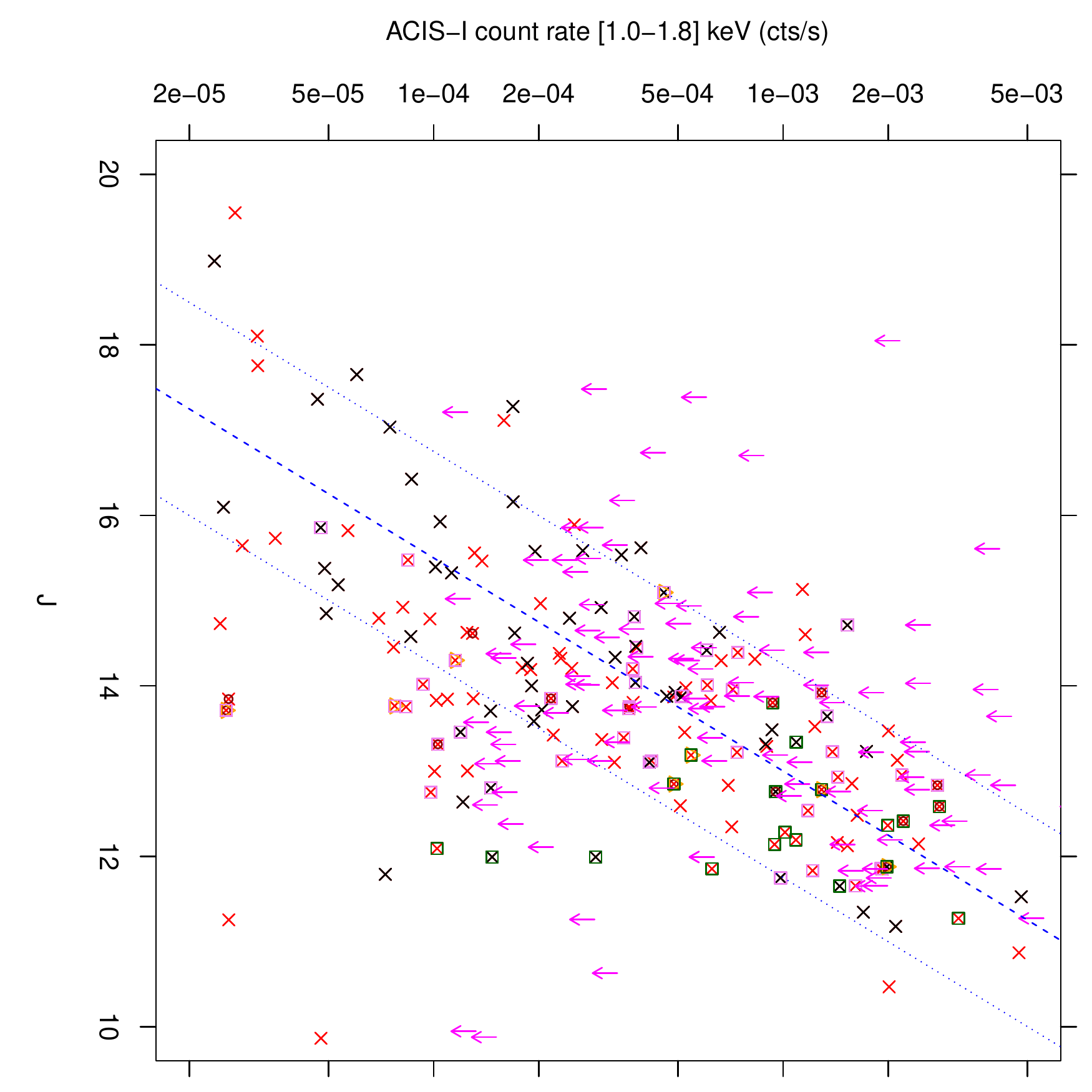}}
\caption{
ACIS-I count rate in the [1.0-1.8]~keV band vs.\ $J$ magnitude. The
dashed/dotted lines represent proportionality; the distance between the
dotted lines is an order magnitude difference in the X-ray to $J$ flux ratio.
Symbols as in Figs.~\ref{ri-rha} and~\ref{jh-hk}, except for black
crosses representing a spatially-selected subset of X-ray sources (see text).
Arrows indicate count-rate upper limits to undetected RGS11 YSOs.
\label{xrate-band-j}}
\end{figure}

A plot of X-ray count-rate in the 1.0-1.8~keV band vs.\ $J$ is shown in
Figure~\ref{xrate-band-j}, for ACIS-I sources with $J-H<2$,
corresponding roughly to $A_V=12$. In this diagram,
absorption causes datapoints to move parallel to the dashed line, and
might partially account for the correlation seen in the diagram, while
the spread above and below that line reflects the X-ray to photospheric
flux ratio. Most of the H58 stars have X-ray/NIR flux ratios within an
order of magnitude. Their average ratio is also close to that of the
bulk of NAP X-ray sources; only four of them show weaker X-ray/NIR flux
ratio. Also the RGS11 YSOs appear well mixed with the non-YSO X-ray
sources, and count-rate upper limits for X-ray undetected YSOs (arrows
in the Figure) are consistent with detections.
The black crosses in the diagram are X-ray sources in the
dashed rectangles in Figs.~\ref{spatial-a} and~\ref{spatial-b}: also these
sources do not appear to have particularly high or low X-ray/NIR flux
ratios, and are therefore homogeneous with the rest of NAP X-ray sources.

\begin{figure}
\resizebox{\hsize}{!}{
\includegraphics[angle=90]{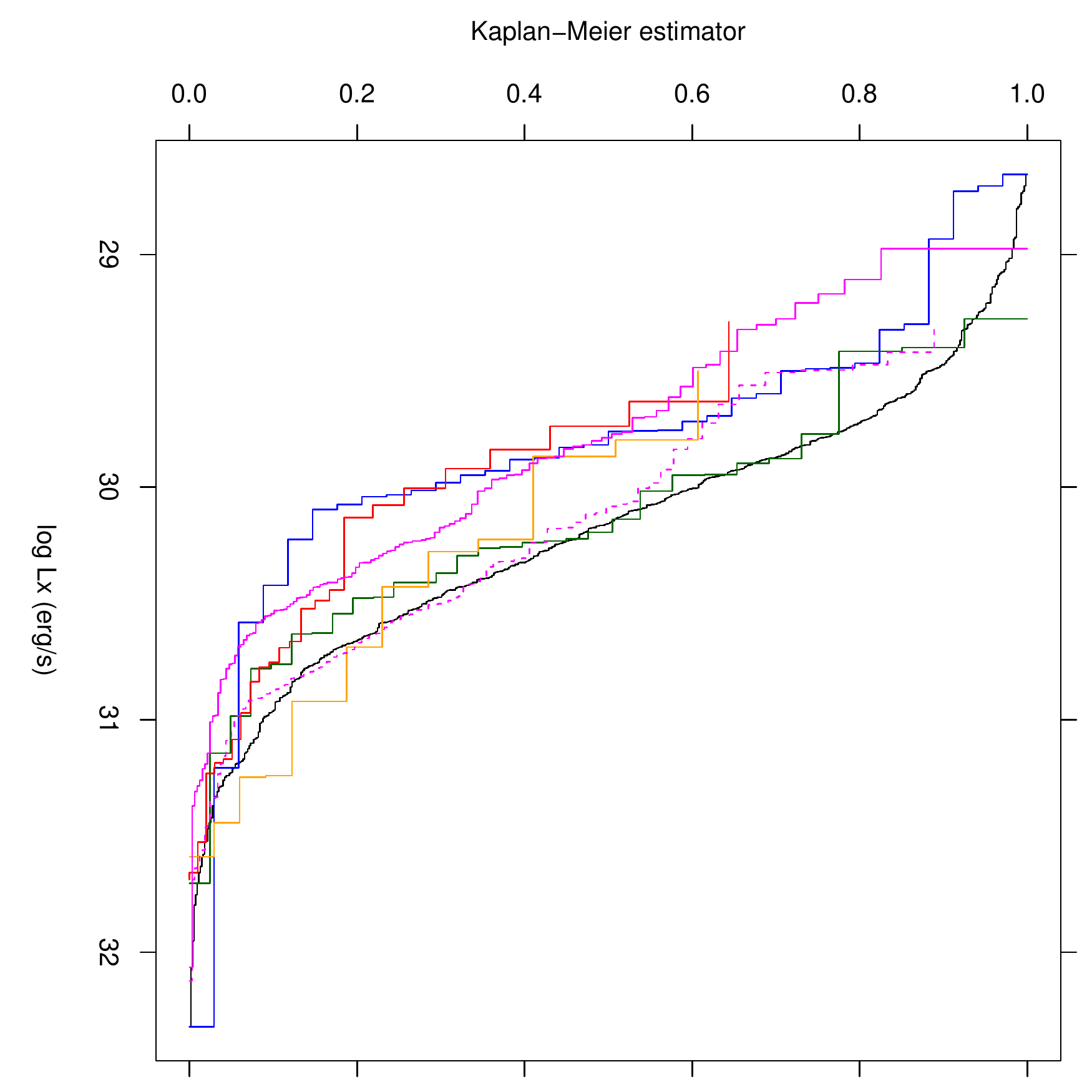}}
\caption{
Maximum-likelihood X-ray luminosity functions, for the entire NAP sample
(black), and different subsamples: unreddened M dwarfs (blue), H58 stars
(dark green), RGS11 YSOs of Class~II (solid magenta), Class~I (orange), and
flat-spectrum YSOs (red). The dashed magenta distribution refers to
Class~II YSOs, using different absorption $N_H$, see text.
\label{lx-lumfun}}
\end{figure}

\begin{figure*}
\sidecaption
\includegraphics[width=12cm]{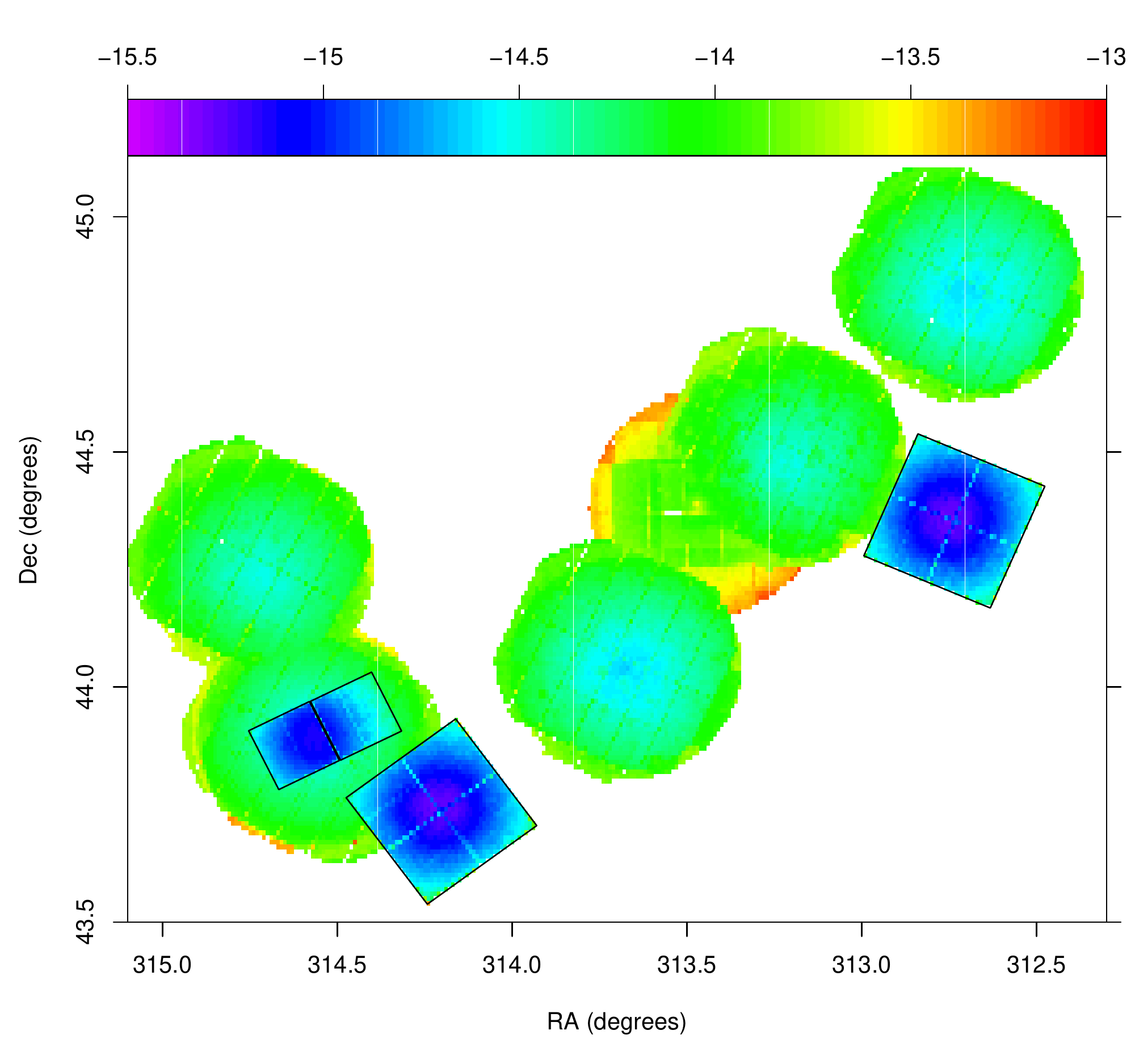}  
\caption{
Map of X-ray sensitivity across our survey. Color indicates $\log f_x$
(top-axis scale, in units of erg s$^{-1}$ cm$^{-2}$), where $f_x$ is the
minimum detectable X-ray flux at a reference energy of 1~keV.
At the NAP distance, $\log f_x = -15$ corresponds to an (unabsorbed)
X-ray luminosity $\log L_x \sim 28.6$ erg/s.
\label{sens-map}}
\end{figure*}

We have estimated X-ray luminosities $L_X$ for our X-ray sources, using PIMMS
to compute conversion factors between count-rate and flux. These are
weakly dependent on the emitting plasma temperature if this is above
1~keV, as often found in active PMS stars, but vary considerably as a
function of $N_H$, across the range of inferred absorption values for
our NAP sources. As above, we estimated $N_H$ for individual sources
from their $J-H$ color, and using the above conversion between this and
$A_V$, and between $A_V$ and $N_H$. Since we do not know individual
distances for possible foreground stars, we might be overestimating
$L_X$ for these latter. The derived $L_X$ distribution is shown in
Figure~\ref{lx-lumfun}, for both the total NAP X-ray sample and
selected subsamples.
The total sample and the X-ray emitting M dwarfs are X-ray selected
samples, and therefore do not include upper limits.
For independently-selected samples such as H58 stars and RGS11 YSOs we
have instead computed individual $L_X$ upper limits using PW(X)Detect;
these are used to compute maximum-likelihood X-ray luminosity functions,
using the {\em survival} R software
package\footnote{http://cran.r-project.org/web/packages/survival}, shown
in Fig.~\ref{lx-lumfun}.
With less than 20\% of objects detected in X-rays, our sampling of the
X-ray properties of flat-spectrum and Class~I YSOs is rather poor,
compared to the $>$60\% X-ray detections among Class~0-I YSOs in the ONC by
Prisinzano \e (2008); nevertheless, the $L_X$ medians for the
flat-spectrum and Class~I YSOs from Fig.~\ref{lx-lumfun} fall very
close to the $L_X$ median of Class~0-Ib objects in Prisinzano \e (2008;
note that YSO classification was made differently by these authors and
RGS11).
The X-ray luminosity function of RGS11 Class~II YSOs (33\% detection rate)
also has a median $L_X \sim 29.75$ (erg/s), which compares well to the
Preibisch and Feigelson (2005) results for the ONC M stars (median $L_X
\sim 29.55$ erg/s), but is significantly
lower than the median $L_X$ of more massive stars in the ONC (or NGC~2264, or
Chamaeleon: median $L_X \sim 29.9-30.5$ erg/s).
We have however found from our color-color diagrams that
only a minor fraction of our X-ray sources are consistent with M-star
colors. One possible solution to this contradiction is that we might be
strongly underestimating absorption $N_H$, and with it also $L_X$,
since we probably underestimated $E(J-H)$ as discussed above, if these
stars' masses are much above $\sim 0.5 M_{\odot}$.
To test this, we have computed the maximum $E(J-H)$ (and corresponding
$N_H$) for each star, assuming an intrinsic value $(J-H)_0 = 0.15$,
appropriate to $1.4 M_{\odot}$ stars at 100~Myr (Fig.~\ref{j-jh}), but
definitely not to $0.5 M_{\odot}$ stars: the
resulting X-ray luminosity function for Class~II YSOs is shown (dashed)
in Fig.~\ref{lx-lumfun}: the resulting median $\log L_X \sim 30.1$ is
closer to the ONC value, and even more to the values for NGC~2264 and
Chamaeleon stars in Preibisch and Feigelson (2005) for the mass range $0.5-0.9
M_{\odot}$, but remains a factor $\sim 2$ below the median $L_X$ for stars in
the mass range $0.9-1.2 M_{\odot}$ in the same clusters, for which $E(J-H)_0
= 0.15$ would be most appropriate. Therefore, reddening underestimates
may be only a partial explanation of the low X-ray luminosity function
of Class~II objects in the NAP.

The H58 stars, most of which are not M stars, and which are only lightly
absorbed, as seen above, have higher $L_X$ than the Class~II YSOs (solid
magenta line in Fig.~\ref{lx-lumfun}), well
consistent with PMS stars in the mass range $M = 0.5-0.9 M_{\odot}$ from
the different clusters shown by Preibisch and Feigelson (2005).

The unreddened M dwarfs have a $L_X$ distribution much lower than the
bulk of NAP sources; if they are NAP members, the lower $L_X$ would not
be surprising, since these stars are at the lower end of the mass
spectrum (Fig.~\ref{i-ri-0}) and their $L_X$ distribution of
Fig.~\ref{lx-lumfun} is close to that of similar-mass stars in the ONC
(e.g.\ Preibisch and Feigelson 2005). However, we cannot rule out that these
are M stars with ages similar to the Pleiades ($\sim$100 Myr),
3-10 times nearer to us, so that their actual
$L_X$ would be 10-100 times lower than shown in Fig.~\ref{lx-lumfun}:
also in this case their $L_X$ distribution would agree with the data
presented in Preibisch and Feigelson (2005; fig.4).

{

\subsection{X-ray sensitivity limits}
\label{sensitivity}

It is important to recognize to what extent the results presented above
may be affected by the uneven sensitivity of the available X-ray data.
As discussed in detail e.g.\ by Broos \e (2011), two distinct causes
contribute to the effect that an astrophysical object with a given
intrinsic X-ray flux (at the NAP distance) may be detected or not. The
first cause is instrumental, the detection sensitivity being not uniform
across the imaged FOV, because of both mirror vignetting and PSF
off-axis degradation. The second cause is of astrophysical origin and resides in
the uneven column density of absorbing matter along the line of sight
to the source, which was seen above to span a wide range inside the surveyed
region. Therefore, we try here to quantitatively examine both effects
individually.

The instrument-related effects are relatively easy to model, once the
background level of each individual X-ray observation is measured, and
properties like vignetting and PSF shape are known. If the spatial
distribution of X-ray sources is wide enough, this gives rise to what
was termed the ``egg-crate effect'' by Broos \e (2011), since the
smaller PSF near the (ACIS-I) FOV center results in a larger spatial
density of detections there, with respect to FOV borders.
We have computed nominal minimum detectable fluxes $f_x$ (for point sources)
for the entire NAP
survey area using the capabilities of PW(X)Detect: the resulting map is
shown in Figure~\ref{sens-map}, including both ACIS and EPIC pointings.
In order to compare different detectors in the same map, we cannot use 
purely instrumental quantities (counts/sec), but physical ones, scaling
counts to erg/cm$^{2}$ with the adoption of the nominal instrument effective
area at a representative energy of 1~keV.
Since the two ACIS-I ObsIds 13647 and~15592 were analyzed together and the
ACIS-S ObsId 14545 has higher sensitivity than ACIS-I datasets, all ACIS
fields span approximately the same range of $f_x$.
Instead, the various XMM-Newton EPIC exposures are of unequal duration (see
Table~\ref{table.xmm}), and their $f_x$ is more sensitive to
background than ACIS datasets: therefore, despite the more uniform PSF
they collectively span a wider range of $f_x$.
This is more cleary seen in the $f_x$ histograms of
Figure~\ref{sens-hist}, separately for ACIS and EPIC data: most of the
width of the EPIC histogram is related to ObsId-to-ObsId variations,
while most of the width of the ACIS histogram is related to
center-to-border effects, the various ACIS FOV having similar $f_x$
ranges.

The strong center-to-border $f_x$ variation in ACIS datasets is
responsible for the egg-crate effect, when this is found. However, the
distribution of X-ray detections in NAP ACIS-I FOVs (Figure~\ref{spatial-b})
does not show such a centrally-symmetric shape, so we may conclude
that any egg-crate effect is not primarily responsible for the observed
spatial distribution of NAP X-ray sources. As far as other effects are
concerned, it is interesting to observe from Figure~\ref{xrate-band-j}
(including only ACIS-I data) that upper limits span a wide range of
almost two orders of magnitude, because they are computed on the basis
of the individual source absorption values. The fact that they are even
more frequent in the upper part of the diagram emphasizes that
absorption, more than uneven detector properties, is the most important
factor against their detection. This emerges also clearly from a
comparison between the range in the ordinates of
Figure~\ref{xrate-band-j} of 2.5 orders of magnitude, and that spanned
by the instrumental ACIS sensitivity (one order of magnitude) in the red
histogram of Figure~\ref{sens-hist}.

\begin{figure}
\resizebox{\hsize}{!}{
\includegraphics[]{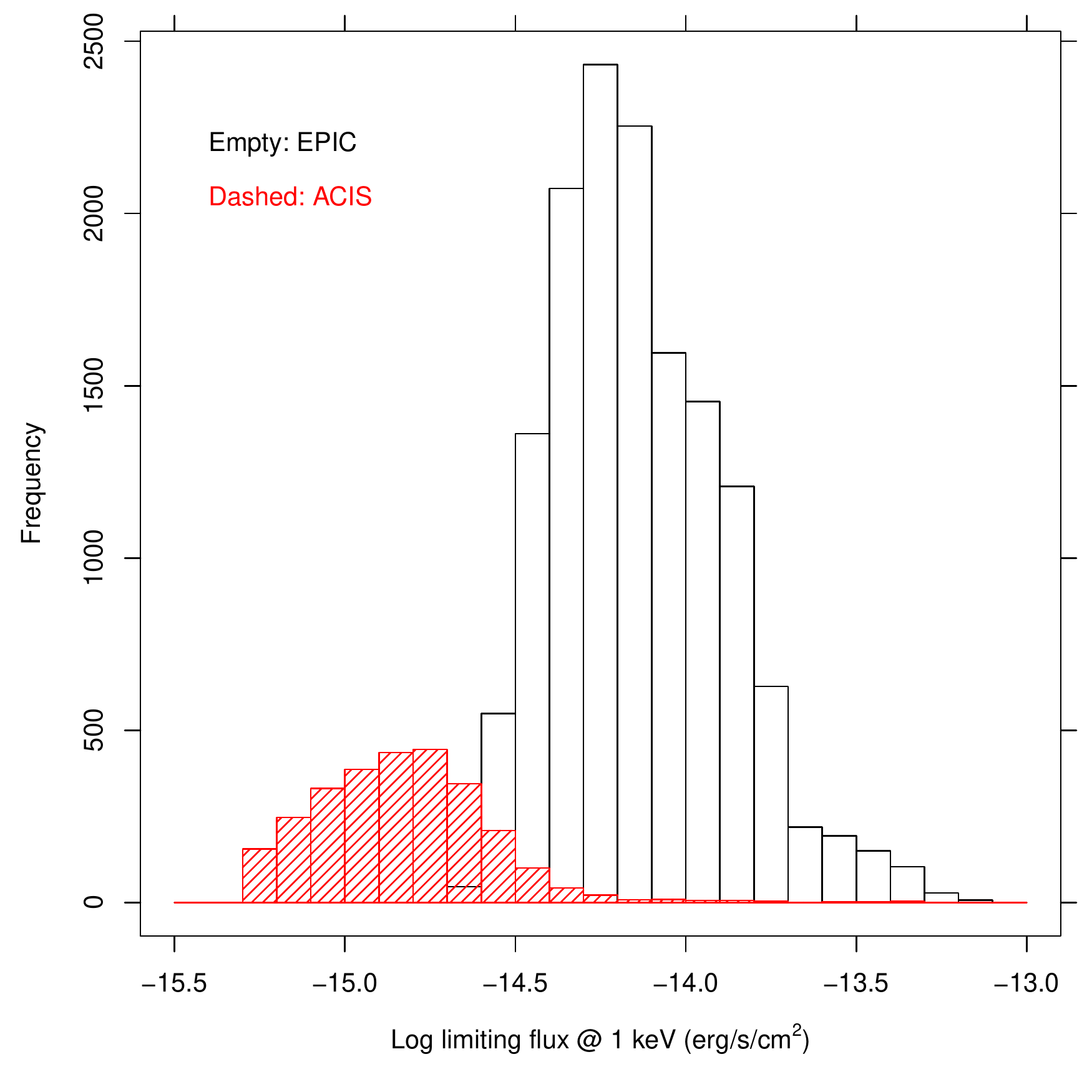}}
\caption{
Histograms of minimum detectable flux $f_x$ from the same spatial grid
as in Fig.~\ref{sens-map}, shown separately for ACIS (red) and EPIC data
(black).
\label{sens-hist}}
\end{figure}

\begin{figure}
\resizebox{\hsize}{!}{
\includegraphics[]{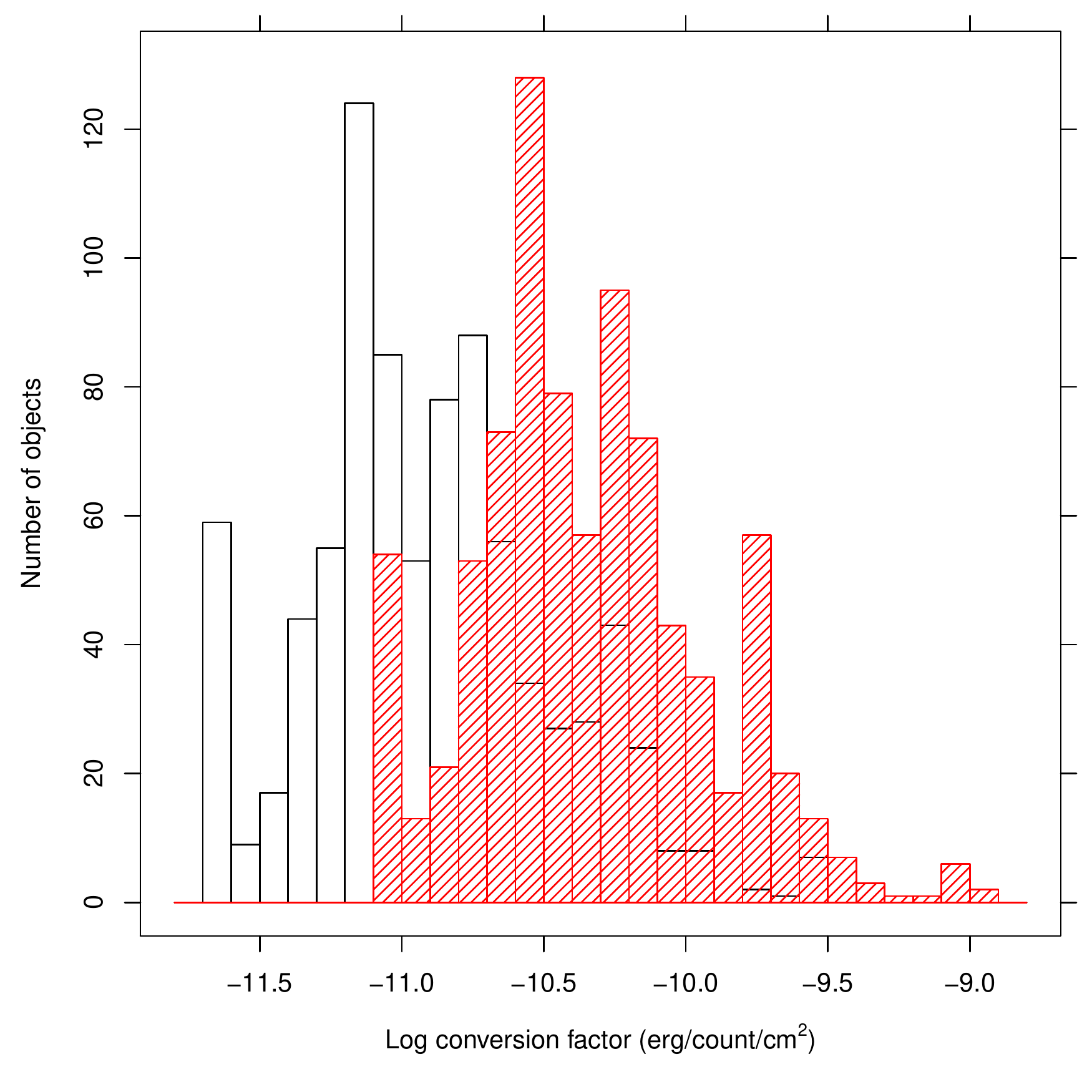}}
\caption{
Distribution of computed count-rate to flux conversion factors for all
observed NIR objects (either X-ray detected of not), separately for ACIS
(red) and EPIC (black) observations.
\label{convfact-hist}}
\end{figure}

The distribution of absorption-based (from $J-H$ colors) X-ray
count-rate-to-flux conversion factors is shown in
Figure~\ref{convfact-hist}, separately for ACIS and EPIC detectors (and
regardless of whether each detector actually imaged a given source), under
the same assumptions about the intrinsic source spectrum as in the
previous subsection.
For both instruments, the bulk of stars span approximately 1.5 orders of
magnitude,
in agreement with
conclusions given just above. While the total NAP cloud absorption shows
definite spatial properties (Figure~\ref{spatial-jh-mean}), the X-ray
colors (Figures~\ref{xray-img-1} to~\ref{xray-img-3})
and NIR colors (Figures~\ref{spatial-a} and~\ref{spatial-b}) of X-ray
sources are not generally correlated with the local total depth.
This suggests that X-ray sources are located at various depths in
the cloud, and the extrapolation of the undetected source density on the
basis of the local density of detections is here at risk of being misleading.
Such an approach was instead used by e.g.\ Kuhn \e (2015) in their X-ray study
of several massive SFRs, in which absorption is more uniform and source
statistics (per FOV) is much higher than in the NAP, permitting a
believable reconstruction of the undetected population.
Contrarily to the data studied by Kuhn \e (2015), in the NAP dataset
the flux completeness limit (a flux above which we are
confident to have detected near 100\% of sources) is near the upper end
of the detected source range, as can be understood from
Figure~\ref{xrate-band-j} where upper limits are found up to the top of
the count-rate scale; this also renders the approach adopted by those
authors unapplicable here.
The conclusion drawn from Figure~\ref{convfact-hist} is that the
intrinsic X-ray luminosities of undetected sources may range up to 1.5
orders of magnitude above their unabsorbed value, whose spatial
distribution was shown in Figure~\ref{sens-map}.
The peaks of the instrumental-sensitivity and conversion-factor
histograms, taken together, correspond to an intrinsic X-ray luminosity
$\log L_x \sim
29.9$ erg/s, detectable over most of the NAP fields. This $L_x$ value
being typical of low-mass PMS stars, we conclude that our NAP detection
list can be considered representative of the young star population in
this SFR.

\begin{figure}
\resizebox{\hsize}{!}{
\includegraphics[]{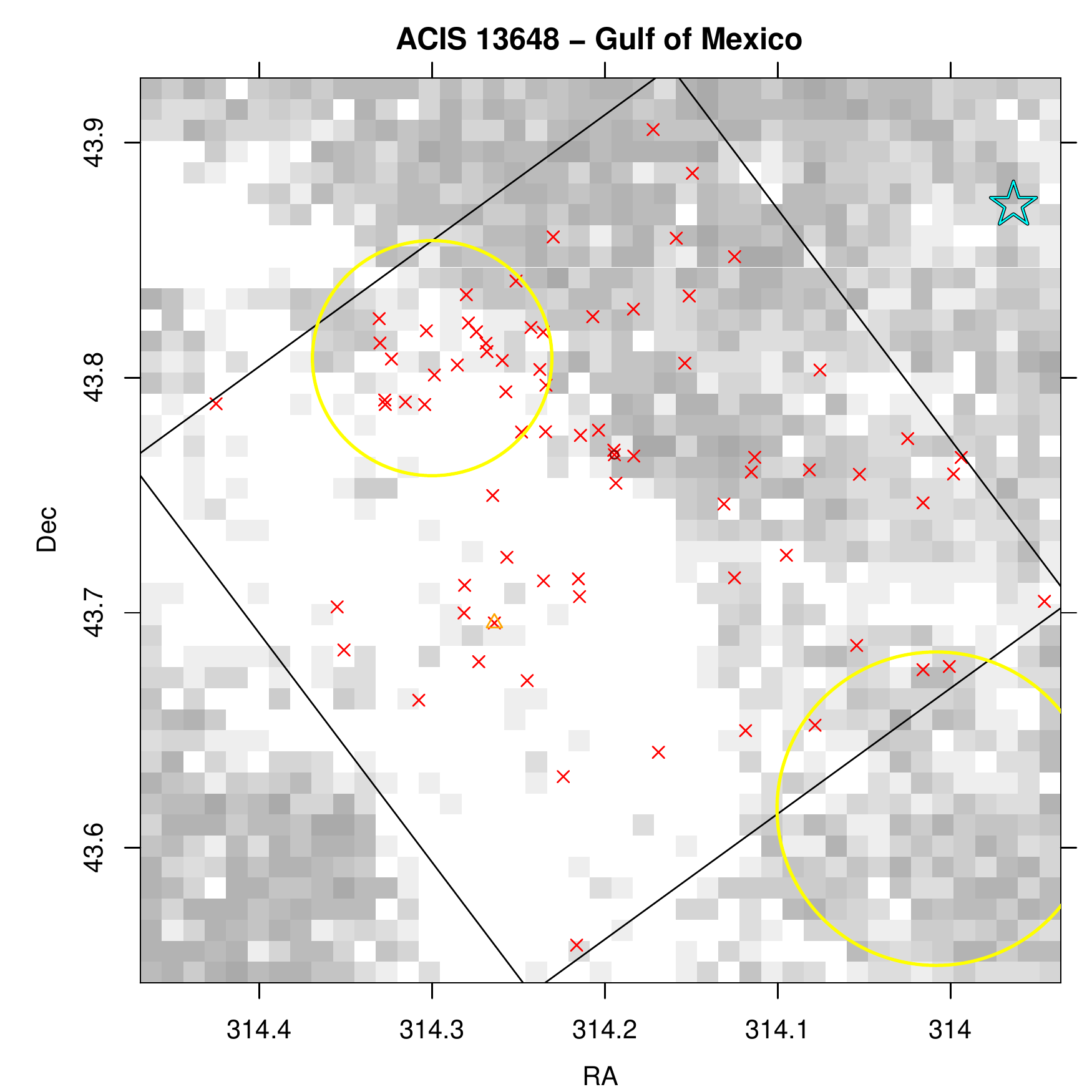}}
\caption{
Analogous of Fig.~\ref{spatial-b} (lower-left panel), but leaving out
all X-ray detections with less than 15 counts. No optical/NIR objects
are plotted here in order to show better the X-ray source distribution.
The big yellow circles are Cambresy's IR clusters, and the cyan star
symbol is the O5 star, as in Fig.~\ref{spatial-b}.
\label{spatial-7-threshold}}
\end{figure}

While a proper reconstruction of the space and flux distributions
of undetected NAP sources is unfeasible, we may nevertheless mitigate the
instrumental effects described above, by considering only detections
above a uniform threshold for a given X-ray FOV.
Figure~\ref{spatial-7-threshold} shows the detected X-ray sources in the
``Gulf of Mexico'' ACIS FOV, already shown in Fig.~\ref{spatial-b}
(lower-left panel), but limited to sources with more than 15 counts, and
therefore detectable across the whole FOV (thus free from
egg-crate effects). Although the source statistics becomes smaller, the
non-uniform spatial distribution of detections is still clear,
especially near one of the Cambresy's NIR clusters. Therefore, the main
conclusions of the previous subsections are not dramatically affected by
instrumental factors.
}

\section{Discussion}
\label{discuss}

The results reported in the above sections emphasize once again the
known complexity of the NAP star formation region, made of several
subgroups of young stars (Cambr\'esy \e 2002, RGS11), with no obvious
connection between them. Although it was very difficult to assess
individual membership for our X-ray detected sources, above results show
that many (or most) of them exhibit properties (X-ray/NIR flux ratios,
X-ray variability and spectrum hardness) consistent with young PMS stars.
The non-detection in the X-rays of most YSOs, even those of Class~II
whose X-ray properties are best known, remains puzzling, since optical/NIR
color-color diagrams seem to rule out for most of them the low mass
values ($M<0.5
M_{\odot}$) that their X-ray luminosity function would imply. A better
characterization of these objects would be desirable.

Perhaps the spatial distribution of our X-ray sources is our most
interesting result. Although the illuminating O5 star 2MASS J20555125+4352246
does not fall in any of our X-ray FOVs, we observed its immediate
neighborhood, a few arcmin away from it: there is no sign of any
concentration of X-ray sources around it, in agreement with the finding
by its discoverers that it is remarkably isolated (Comer\'on and
Pasquali 2005). This isolation refers therefore not only to comparably
massive stars, but to stars down to solar-like masses, to which our
survey is sensitive even at the extinction of the O5 star.
This is very different from most star-forming regions, where the
highest-mass stars are immediately surrounded by low-mass PMS stars.
It would be interesting to investigate if this is a runaway star, with
implications on the timescales on which it might have interacted with
the molecular cloud; although no such information is available now, the
Gaia data should provide it in a short time frame.

The two sub-regions with the highest YSO spatial density, the Gulf of
Mexico and Pelican, are also those
with the largest number of X-ray detections (although this might be
somewhat biased by having used there the most sensitive ACIS detectors).
In the Gulf of Mexico, closest to the O5 star, there is a clear spatial
trend: YSOs are clustered near the border of the dark cloud which
faces the O star (and not inside the cloud), and tens of X-ray sources
follow the same pattern; moreover, several other tens X-ray sources
are found between the YSO cluster(s) and the O star, forming an
apparently distinct layer. It is interesting to remark that in Figure~20
of RGS11 the Class~III stars fall in this same intermediate region. This
is highly suggestive of a time sequence, with formation of stars
proceeding southwards, under the strong influence of the massive O star
nearby. The X-ray sources without a YSO counterpart, as well as the
Class~III stars, would thus be the oldest, with younger stars being
currently formed in the outer layers of the dense dust cloud.
Toujima \e (2011) also suggested triggered star formation in NGC~7000,
on the basis of molecular-line observations.
The disk-free X-ray detected stars are not necessarily much older than the
others (no such separation is found in our de-reddened CMD), since when
exposed to strong UV irradiation disk lifetimes are likely to be
significantly shortened (e.g.\ Guarcello \e 2007, 2009).
In other places in the NAP, a similar segregation between the YSO and
X-ray source distributions is observed, as in the ``Pelican hat'', but
its origin is much less clear, since the triggerer of the possible time
sequence is not obvious.
Moreover, we recover at least three of the Cambr\'esy \e (2002) NIR
clusters in X-rays, confirming their existence.

The assessment of membership for the X-ray detected sources was difficult,
as discussed above for each of the possible indicators. Only $\sim 29$\% of
all X-ray sources have \ha\ or NIR excesses (either from this work or from
RGS11 and H58), and are therefore high-probability members individually.
Most other X-ray sources have however X-ray properties, colors and
extinction very similar to the former, and are probable members as well.
The foreground extinction towards the NAP is very small, as we find from
color-color diagrams and in agreement with Laugalys \e (2006, 2007), and
therefore foreground X-ray sources may only be found at near-zero
reddening: we found only a few tens such sources above.
Extinction rises rather abruptly in the NAP, although this may be
position-dependent as we have examined above (in agreement with Cambr\'esy
\e 2002). Background X-ray sources, behind such large absorbing column, are
unlikely to be coronal sources, but may be distant compact objects or even
AGNs, both characterized by large X-ray-to-optical flux ratios. We
tentatively identify these classes of sources with our unidentified X-ray
sources (115 out of 721 total detections), which must have
large X-ray-to-optical flux ratios if they are
missed in both of the very deep IPHAS/UKIDSS catalogs.
To summarize, most of the NAP X-ray sources with an optical/NIR counterpart,
except perhaps a significant fraction of the unreddened ones, are good
candidates as young NAP members.

\section{Summary}
\label{concl}

We have performed the first extensive, wide-area X-ray survey of the
North-America and Pelican star-forming region, using both Chandra/ACIS and
XMM-Newton/EPIC detectors. More than 700 unique X-ray sources have been
detected, of which $\sim 85$\% have a counterpart in at least one of the
IPHAS/UKIDSS/2MASS catalogs. Only $\sim 29$\% of X-ray sources are
identified with previously known YSOs or CTTS, or with newly found
\ha-emission or NIR-excess stars. We argued that most of the optically/NIR
identified X-ray sources are probable members of the star-forming region.
A few tens unreddened X-ray sources are probable active foreground low-mass
stars.

In the Gulf-of-Mexico region, the respective spatial properties of
dark obscuration, YSOs, and X-ray sources without YSO association are
suggestive of a sequential star-formation scenario, with the O5 star
illuminating the entire nebula as the probable triggerer.  Unlike most
star-forming regions, this most massive star appears isolated even in
X-ray images. The large-scale spatial distribution of the X-ray detected member
candidates follows qualitatively that of the known YSOs, with several
sub-clusters, apparently physically unconnected.

Detailed follow-up (spectroscopic) observations will be needed to confirm
individual membership of the X-ray sources found here.

\begin{appendix}

\section{2MASS - UKIDSS calibration}
\label{ukidss-2mass}

\begin{figure}
\resizebox{\hsize}{!}{
\includegraphics[angle=90]{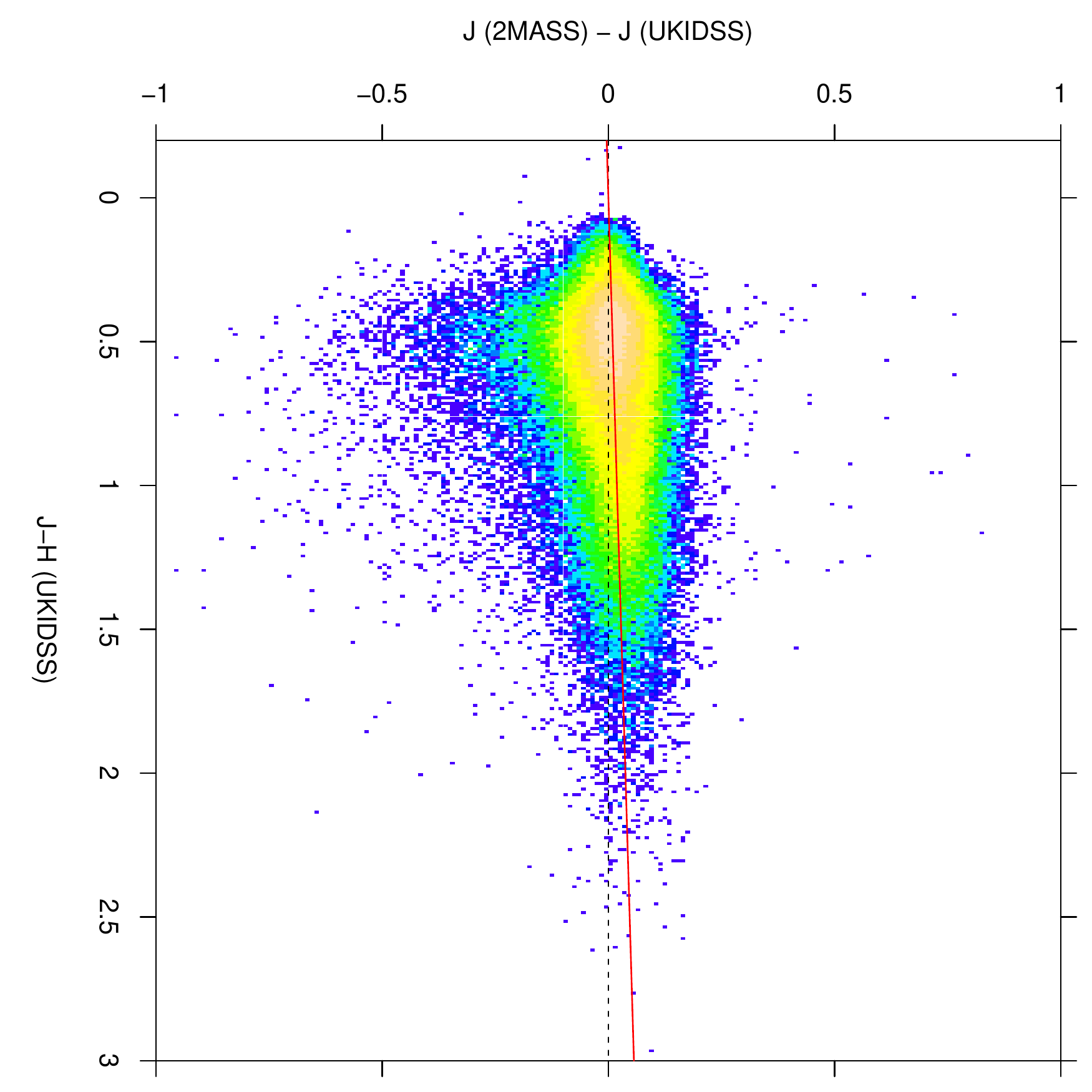}}
\caption{
2MASS-UKIDSS comparison: difference $J_{2MASS} - J_{UKIDSS}$ vs.\
UKIDSS $J-H$ color (2-D histogram).
The black dashed line represents $J_{2MASS} - J_{UKIDSS}=0$, while the red
solid line is a best-fit to the data.
\label{j-jh-2mass-ukidss}}
\end{figure}

\begin{figure}
\resizebox{\hsize}{!}{
\includegraphics[angle=90]{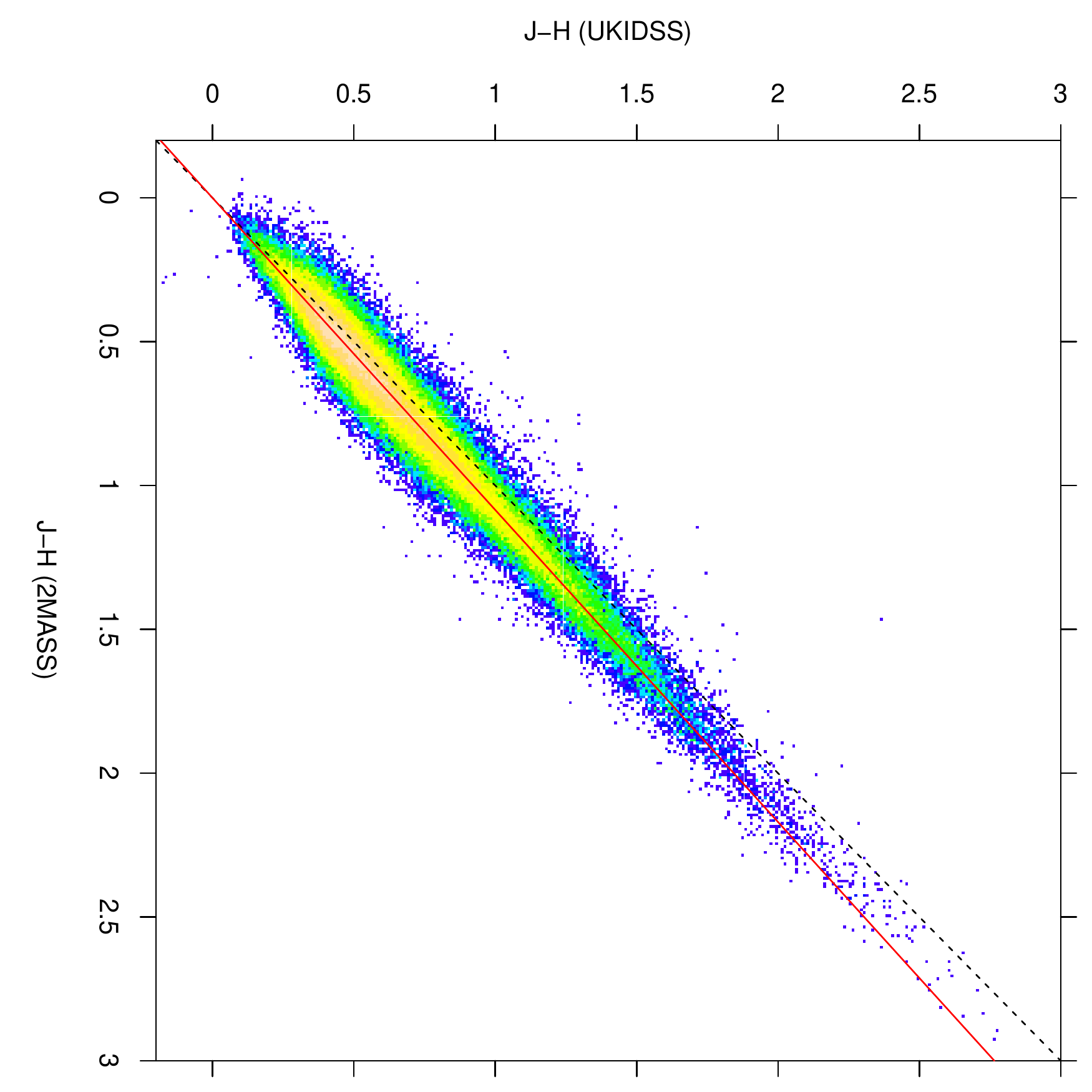}}
\caption{
Comparison between $J-H$ colors from 2MASS and UKIDSS.
The black dashed line represents $(J-H)_{2MASS}=(J-H)_{UKIDSS}$, while the red
solid line is a best-fit to the data.
\label{jh-2mass-ukidss}}
\end{figure}

The UKIDSS $J,H,K$ photometric system is calibrated to the 2MASS system
(Hodgkin \e 2009); however, the calibration was made at low reddening, and
we find that small adjustments are required to match the two catalogs at
large reddening values, typical of the NGC~7000/IC~5070 region.
The difference between 2MASS and UKIDSS $J$ magnitudes is shown
in Figure~\ref{j-jh-2mass-ukidss}: although small, this difference is often
significant compared to the very small errors in the UKIDSS magnitudes.
Only objects with errors $<0.07$~mag in both $J$ bands, and errors
$<0.1$~mag in $(J-H)_{UKIDSS}$, are shown.
A least-square fit to the data (with fixed zero intercept) has the form:
\begin{equation}
J_{2MASS} - J_{UKIDSS} = 0.0188 \; (J-H)_{UKIDSS}.
\end{equation}

\begin{figure}
\resizebox{\hsize}{!}{
\includegraphics[angle=90]{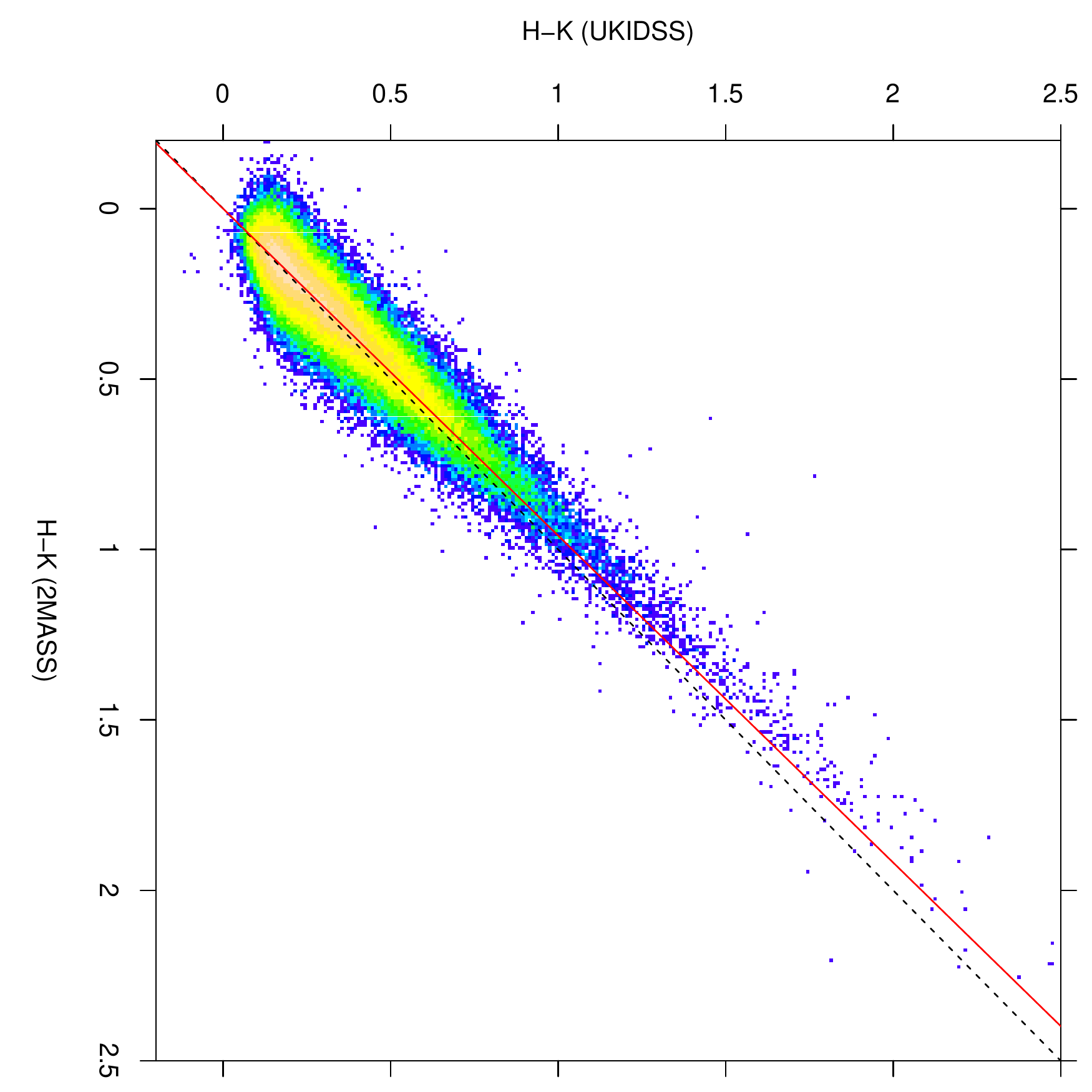}}
\caption{
Same as Fig.~\ref{jh-2mass-ukidss}, for $H-K$.
\label{hk-2mass-ukidss}}
\end{figure}

The $J-H$ colors differ between 2MASS and UKIDSS more appreciably, however,
by almost 8\% (Figure~\ref{jh-2mass-ukidss}, where
only objects with errors $<0.1$~mag in both $(J-H)_{UKIDSS}$ and
$(J-H)_{2MASS}$ are shown).
The best fit to the data (with zero intercept) has the form:
\begin{equation}
(J-H)_{UKIDSS} = 0.9217 \; (J-H)_{2MASS}.
\end{equation}
Similarly, Figure~\ref{hk-2mass-ukidss} shows the comparison between the
respective $H-K$ colors, with the best fit here given by:
\begin{equation}
(H-K)_{UKIDSS} = 1.04236 \; (H-K)_{2MASS}.
\end{equation}

All UKIDSS magnitudes were therefore rescaled according to the
$(J-H)_{UKIDSS}$ color, to match the 2MASS photometric system even at large
reddening. At low reddening (low $J-H$), this recalibration is irrelevant,
and does not interfere with the one presented by Hodgkin \e (2009).
After recalibration, we merged the 2MASS and UKIDSS catalogs into a single
NIR source catalog, choosing the more precise UKIDSS magnitudes when
available.

\section{IPHAS - VPHAS$+$ calibration}
\label{iphas-vphas}

\begin{figure}
\resizebox{\hsize}{!}{
\includegraphics[angle=90]{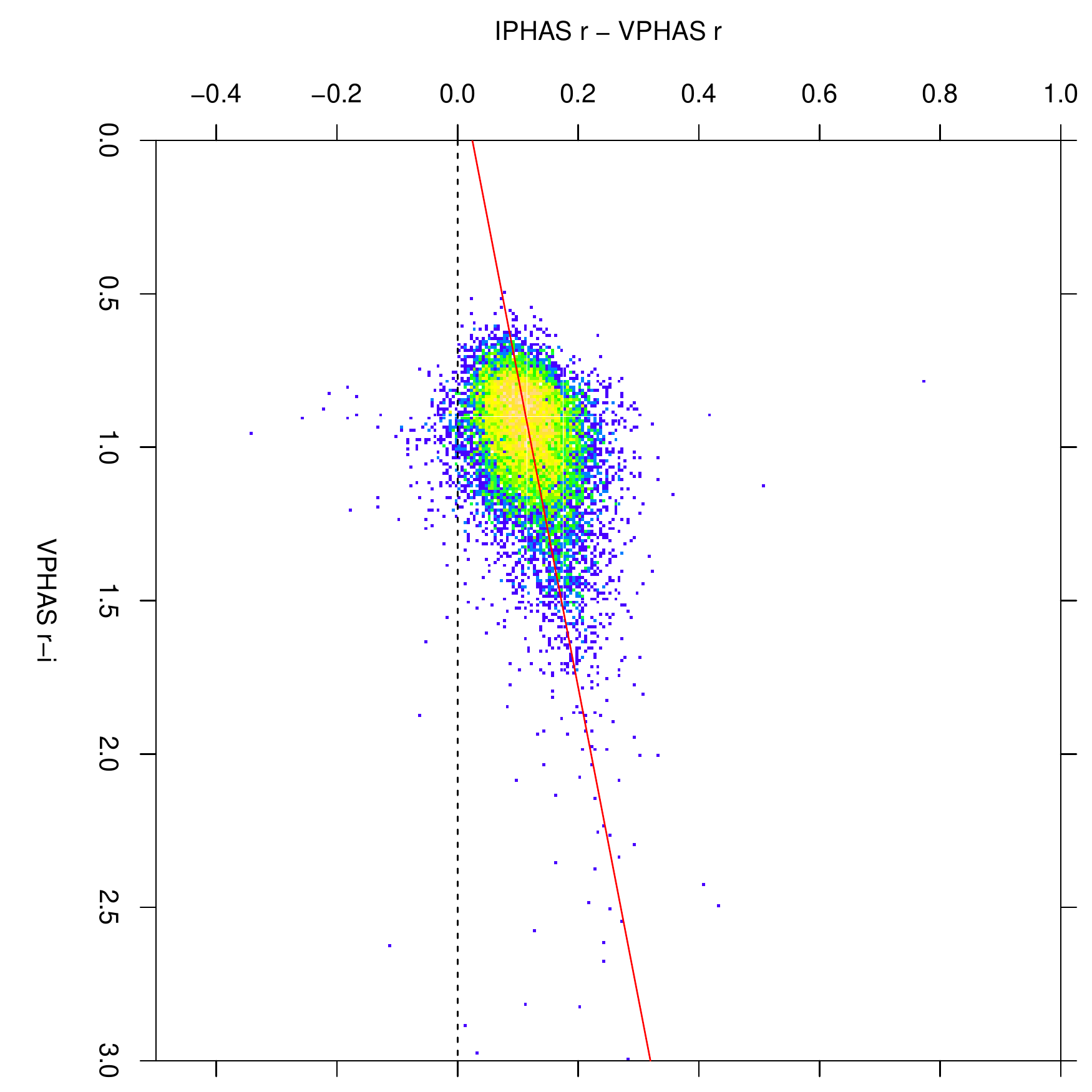}}
\caption{
IPHAS-VPHAS$+$ comparison: difference $r_{IPHAS} - r_{VPHAS}$ vs.\
VPHAS$+$ $r-i$ color (2-D histogram).
The black dashed line represents $r_{IPHAS} - r_{VPHAS}=0$, while
the red solid line is a best-fit to the data.
\label{r-iphas-vphas}}
\end{figure}

\begin{figure}
\resizebox{\hsize}{!}{
\includegraphics[angle=90]{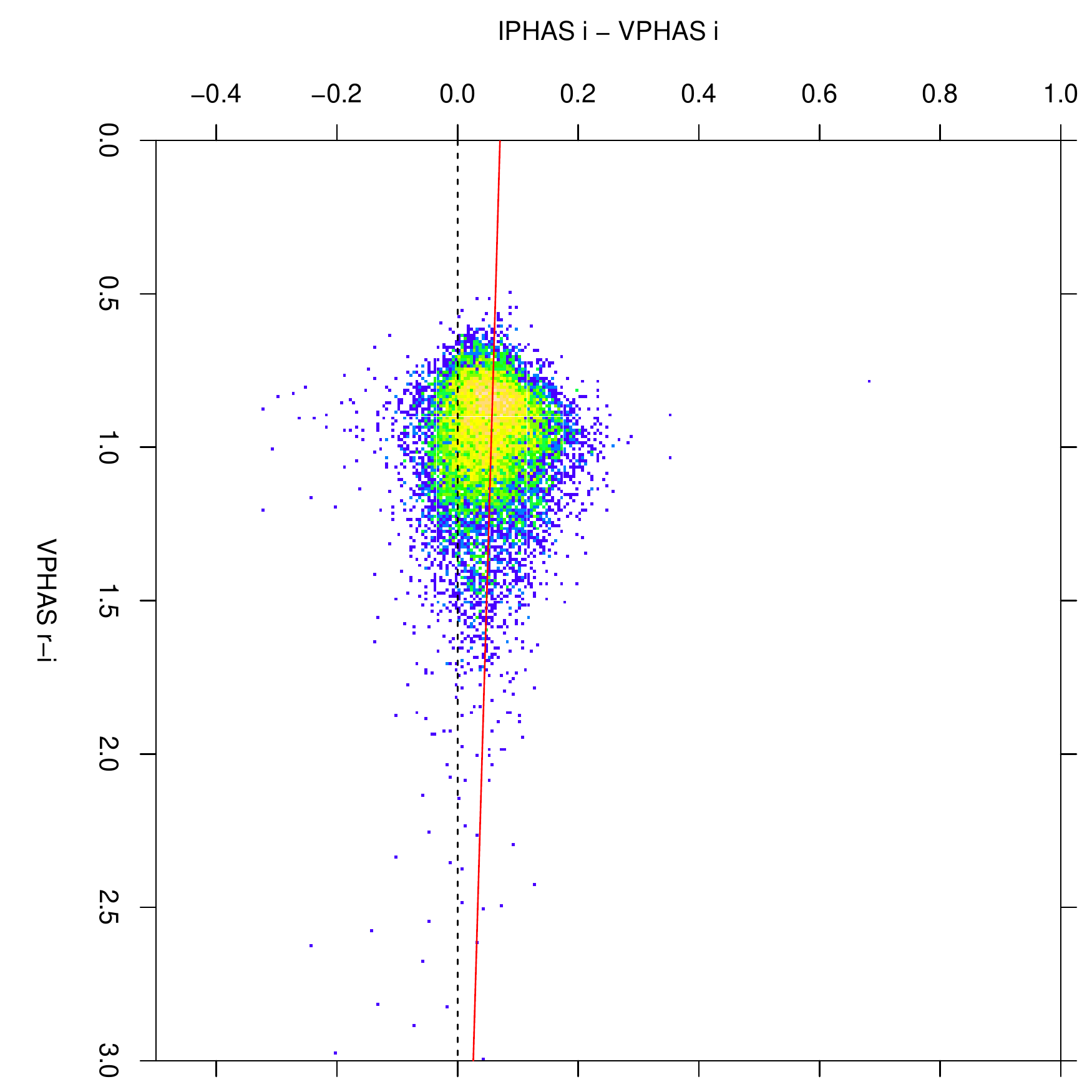}}
\caption{
Same as Fig.~\ref{r-iphas-vphas}, for the $i$ band.
\label{i-iphas-vphas}}
\end{figure}

Although the IPHAS $r,i$ photometric system is similar to the Sloan Survey
system, the formulae in Barentsen \e (2014) indicate that significat
differences exist between the two systems. Moreover, no recipe is given to
convert the IPHAS $r,i$ magnitudes into Johnson/Cousin $VRI$ magnitudes. Such a
recipe is instead given by Drew \e (2014) for the VPHAS$+$ $ugriz$
magnitudes, which are closely related to the IPHAS filter system. The BHAC
evolutionary tracks and isochrones, moreover, are currently available for
the $UBVRI$ filters, but not for either the Sloan or the IPHAS/VPHAS$+$
filters, so we attempt here to recalibrate the IPHAS $r,i$ magnitudes in
terms of the VPHAS$+$ $r,i$, whose relation with $VRI$ is known from Drew
\e (2014). We consider an equatorial field where both IPHAS and VPHAS$+$
data are available, to study their photometric differences.
The results of this experiment are shown in Figures~\ref{r-iphas-vphas}
and~\ref{i-iphas-vphas}, where
only objects with errors $<0.05$~mag in all bands are shown.
The best-fit relations shown in these Figures are:
\begin{equation}
r_{IPHAS} - r_{VPHAS} = 0.02448 + 0.0984 \; (r-i)_{VPHAS}
\end{equation}
\begin{equation}
i_{IPHAS} - i_{VPHAS} = 0.070283 -0.01477 \; (r-i)_{VPHAS}.
\end{equation}

These relations (together with those given in Drew \e 2014)
were used to convert the BHAC tracks/isochrones in the
Johnson/Cousin filters into the appropriate IPHAS $r,i$ magnitudes and colors.

\section{Reddening law}
\label{redden-law}

\begin{figure}
\resizebox{\hsize}{!}{
\includegraphics[]{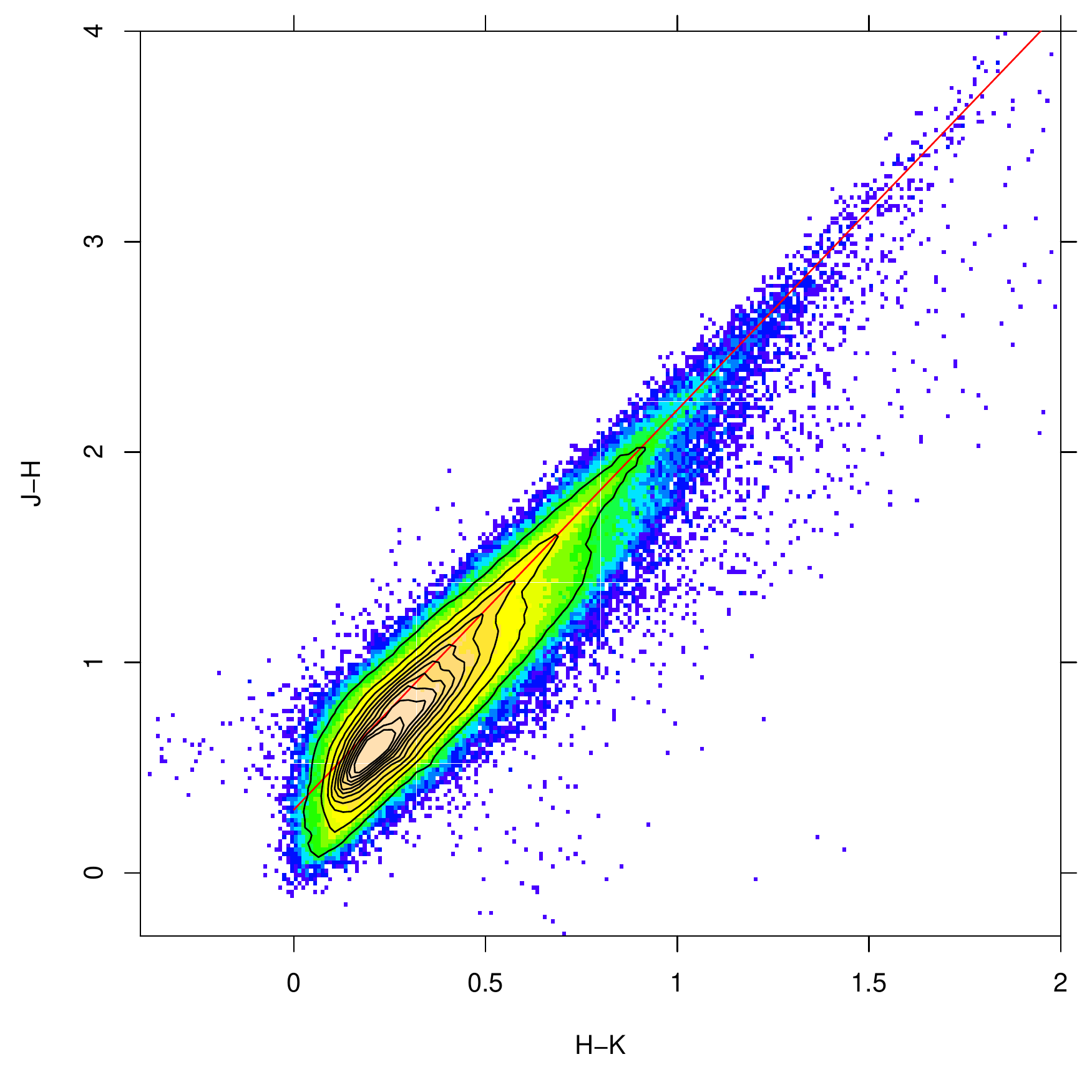}}
\caption{
Slope of reddening vector ($E(J-H)/E(H-K)= 1.9$) in the $(J-H,H-K)$ plane
(red line), obtained from the giant-star sequence,
which is best seen from the iso-density contours (black).
\label{jh-hk-law}}
\end{figure}

\begin{figure}
\resizebox{\hsize}{!}{
\includegraphics[]{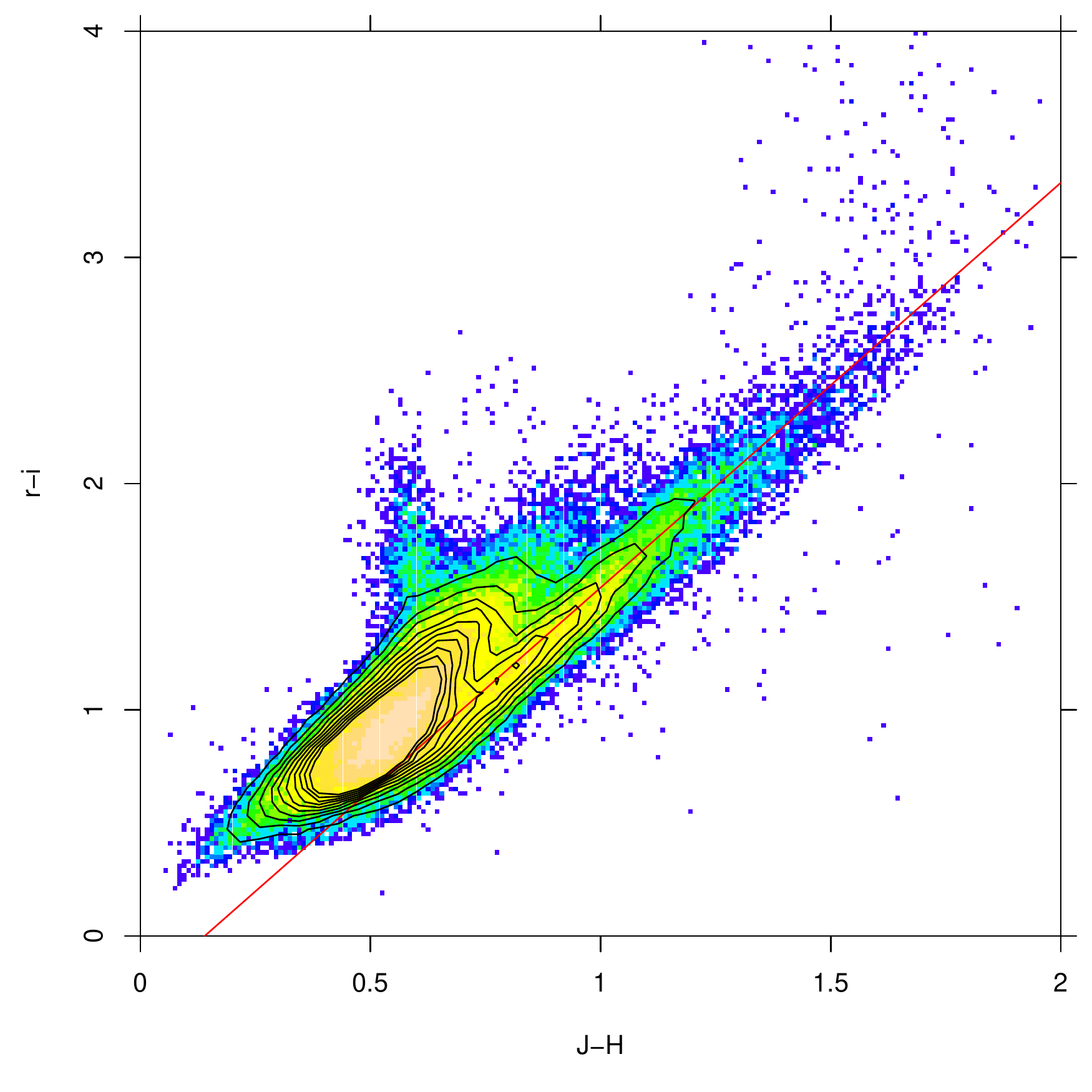}}
\caption{
Same as Fig.~\ref{jh-hk-law}, for the $(r-i,J-H)$ plane:
here the reddened giant-star sequence is particularly well distinct from the
reddened main sequence, while the unreddened M-dwarf locus protrudes
upwards near $J-H \sim 0.6$.
The reddening-vector slope is here $E(r-i)/E(J-H)= 1.79$.
\label{ri-jh-law}}
\end{figure}

We study here the empirical reddening law for the NGC~7000/IC~5070 region,
in the $r,i,J,H,K$ bands. With respect to previous studies, we benefit here
from an extensive optical/NIR catalog, reaching down to faint magnitudes
with small photometric errors. Thanks to these properties, we are able to
discern in several color-color diagrams the giant-star locus from the
dwarfs locus.
Figure~\ref{jh-hk-law} shows the $(J-H,H-K)$ diagram of all sources in the
studied region: the reddened red-giant sequence starts at $J-H \sim 0.8$,
whereas the dwarf sequence starts at $J-H \sim 0$, but the giants being
more luminous remain observable up to larger reddening values. We have
therefore visually fitted the giant sequence with the red line in the
Figure (with slope $E(J-H)/E(H-K)= 1.9$). It should be remarked that at low
reddening dwarfs dominate the distribution because of their higher spatial
density, but at large reddening giants dominate being observable up to
larger distances: it is therefore incorrect to derive the reddening vector
slope by simply joining the low and high ends of the color distribution
with a straight line, since this would mix together stars with different
intrinsic NIR colors, and the resulting slope would be higher than the true
one. The upper envelope of datapoints in the Figure also agrees with the
slope we estimate from the giant sequence.

Figure~\ref{ri-jh-law} is a mixed optical/NIR $(r-i,J-H)$ diagram, where we
can discern well the reddened giant sequence (the lowest one, including the
highest-reddening stars) from the dwarf sequence. Protruding upwards, the
M-dwarf sequence is also clearly seen (cf.\ Fig.~\ref{ri-jh}).
The reddening vector has a slope of $E(r-i)/E(J-H)= 1.79$.

\begin{figure}
\resizebox{\hsize}{!}{
\includegraphics[]{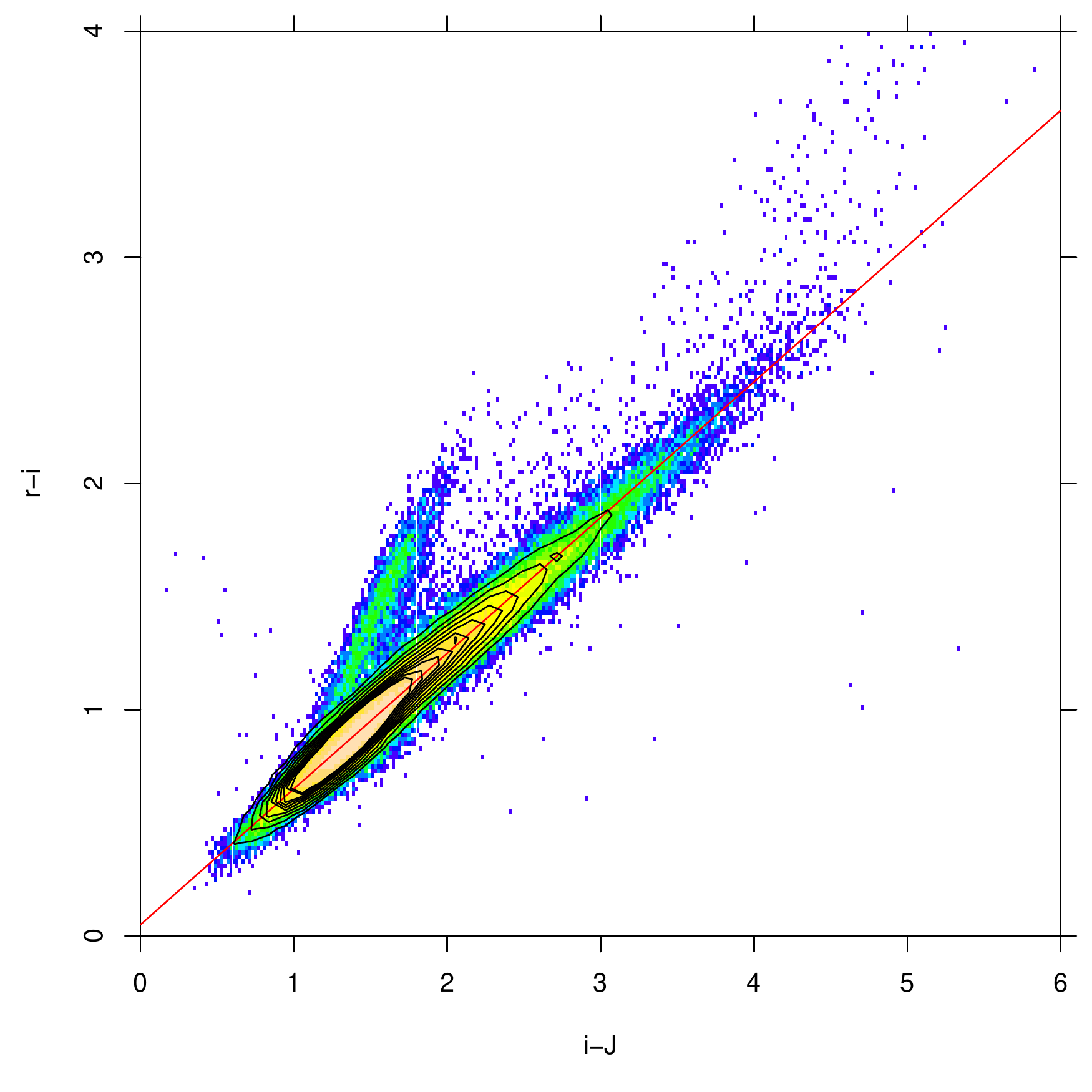}}
\caption{
Same as Fig.~\ref{jh-hk-law}, for the $(r-i,i-J)$ plane:
the reddening-vector slope is here $E(r-i)/E(i-J)= 0.6$.
\label{ri-ij-law}}
\end{figure}

In the $(r-i,i-J)$ diagram of Figure~\ref{ri-ij-law}, instead, the dwarf
and giant sequence are not distinct, but the M-dwarfs are. The
reddening-vector slope is here $E(r-i)/E(i-J)= 0.6$.

In order to connect these reddening laws to the $V$-band extinction, and to
obtain the total-to-selective extinction ratio $A_i/E(r-i)$ for the IPHAS
filters, we need two more equations. Assuming that the $J$ band is least
affected by peculiar reddening laws, we therefore take from Rieke and
Lebofsky (1985) the ratios $A_J/A_V=0.282$ and $E(J-H)/A_V=0.107$. The set
of relative extinctions $A_\lambda/A_V$ for the NAP thus becomes:
\begin{equation}
A_r/A_V = 0.793
\end{equation}
\begin{equation}
A_i/A_V = 0.601
\end{equation}
\begin{equation}
A_J/A_V = 0.282
\label{ajav}
\end{equation}
\begin{equation}
A_H/A_V = 0.175
\label{ahav}
\end{equation}
\begin{equation}
A_K/A_V = 0.1187
\end{equation}

where equations~\ref{ajav} and~\ref{ahav} are merely the Rieke and
Lebofsky (1985) relations, included for convenience.

\end{appendix}

\begin{acknowledgements}
We wish to thank an anonymous referee for his/her helpful suggestions.
I.P.\ acknowledges support by Chandra fund 16617814 contract GO2-13021X.
The scientific results reported in this article are based
on observations made by the Chandra and XMM-Newton X-ray Observatories.
This paper also makes use of data obtained as part of the INT Photometric \ha\
Survey of the Northern Galactic Plane (IPHAS, www.iphas.org) carried out at
the Isaac Newton Telescope (INT). The INT is operated on the island of La
Palma by the Isaac Newton Group in the Spanish Observatorio del Roque de
los Muchachos of the Instituto de Astrofisica de Canarias. All IPHAS data
are processed by the Cambridge Astronomical Survey Unit, at the Institute
of Astronomy in Cambridge. The bandmerged DR2 catalogue was assembled at
the Centre for Astrophysics Research, University of Hertfordshire,
supported by STFC grant ST/J001333/1.
This publication makes use of data products from the Wide-field Infrared
Survey Explorer, which is a joint project of the University of
California, Los Angeles, and the Jet Propulsion Laboratory/California
Institute of Technology, funded by the National Aeronautics and Space
Administration.
This research makes use of the SIMBAD database,
operated at CDS, Strasbourg, France.
We also make heavy use of R: A language and environment for statistical
computing. R Foundation for Statistical Computing, Vienna, Austria.
(http://www.R-project.org/).
\end{acknowledgements}

\bibliographystyle{aa}

\begin{thebibliography}{}

\bibitem[Armond et al.(2011)]{2011A&A...528A.125A} Armond, T., Reipurth, B., Bally, J., \& Aspin, C.\ 2011, \aap, 528, A125 

\bibitem[Aspin(2011)]{2011AJ....141..196A} Aspin, C.\ 2011, \aj, 141, 196 

\bibitem[Bally \& Reipurth(2003)]{2003AJ....126..893B} Bally, J., \& Reipurth, B.\ 2003, \aj, 126, 893 

\bibitem[Bally et al.(2014)]{2014AJ....148..120B} Bally, J., Ginsburg, A., Probst, R., et al.\ 2014, \aj, 148, 120 

\bibitem[Baraffe et al.(2015)]{2015A&A...577A..42B} Baraffe, I., Homeier,
D., Allard, F., \& Chabrier, G.\ 2015, \aap, 577, A42

\bibitem[Barentsen et al.(2014)]{2014MNRAS.444.3230B} Barentsen, G.,
Farnhill, H.~J., Drew, J.~E., et al.\ 2014, \mnras, 444, 3230

\bibitem[Broos et al.(2011)]{2011ApJS..194....2B} Broos, P.~S.,
Townsley, L.~K., Feigelson, E.~D., et al.\ 2011, \apjs, 194, 2

\bibitem[Cambr{\'e}sy et al.(2002)]{2002AJ....123.2559C} Cambr{\'e}sy, L., Beichman, C.~A., Jarrett, T.~H., \& Cutri, R.~M.\ 2002, \aj, 123, 2559 

\bibitem[Comer{\'o}n \& Pasquali(2005)]{2005A&A...430..541C} Comer{\'o}n, F., \& Pasquali, A.\ 2005, \aap, 430, 541 

\bibitem[Corbally et al.(2009)]{2009BaltA..18..111C} Corbally, C.~J., Strai{\v z}ys, V., \& Laugalys, V.\ 2009, Baltic Astronomy, 18, 111 

\bibitem[Covey et al.(2011)]{2011AJ....141...40C} Covey, K.~R., Hillenbrand, L.~A., Miller, A.~A., et al.\ 2011, \aj, 141, 40 

\bibitem[Damiani et al.(1997)]{1997ApJ...483..350D} Damiani, F., Maggio,
A., Micela, G., \& Sciortino, S.\ 1997, \apj, 483, 350

\bibitem[Damiani et al.(1997)]{1997ApJ...483..370D} Damiani, F., Maggio,
A., Micela, G., \& Sciortino, S.\ 1997, \apj, 483, 370

\bibitem[Damiani et al.(2006)]{2006A&A...459..477D} Damiani, F.,
Prisinzano, L., Micela, G., \& Sciortino, S.\ 2006, \aap, 459, 477

\bibitem[Dunham et al.(2012)]{2012ApJ...755..157D} Dunham, M.~M., Arce, H.~G., Bourke, T.~L., et al.\ 2012, \apj, 755, 157 

\bibitem[Drew et al.(2005)]{2005MNRAS.362..753D} Drew, J.~E., Greimel, R.,
Irwin, M.~J., et al.\ 2005, \mnras, 362, 753

\bibitem[Drew et al.(2014)]{2014MNRAS.440.2036D} Drew, J.~E.,
Gonzalez-Solares, E., Greimel, R., et al.\ 2014, \mnras, 440, 2036

\bibitem[Favata et al.(2005)]{2005ApJS..160..469F} Favata, F.,
Flaccomio, E., Reale, F., et al.\ 2005, \apjs, 160, 469

\bibitem[Favata \& Micela(2003)]{2003SSRv..108..577F} Favata, F., \&
Micela, G.\ 2003, \ssr, 108, 577

\bibitem[Gorenstein(1975)]{1975ApJ...198...95G} Gorenstein, P.\ 1975,
\apj, 198, 95

\bibitem[Green et al.(2011)]{2011ApJ...731L..25G} Green, J.~D., Evans, N.~J., II, K{\'o}sp{\'a}l, {\'A}., et al.\ 2011, \apjl, 731, L25 

\bibitem[Guarcello et al.(2007)]{2007A&A...462..245G} Guarcello, M.~G.,
Prisinzano, L., Micela, G., et al.\ 2007, \aap, 462, 245

\bibitem[Guarcello et al.(2009)]{2009A&A...496..453G} Guarcello, M.~G.,
Micela, G., Damiani, F., et al.\ 2009, \aap, 496, 453

\bibitem[Guarcello et al.(2010)]{2010A&A...521A..18G} Guarcello, M.~G.,
Damiani, F., Micela, G., et al.\ 2010, \aap, 521, A18

\bibitem[G{\"u}del et al.(1997)]{1997ApJ...483..947G} G{\"u}del, M.,
Guinan, E.~F., \& Skinner, S.~L.\ 1997, \apj, 483, 947

\bibitem[Guieu et al.(2009)]{2009ApJ...697..787G} Guieu, S., Rebull, L.~M., Stauffer, J.~R., et al.\ 2009, \apj, 697, 787 

\bibitem[Herbig(1958)]{1958ApJ...128..259H} Herbig, G.~H.\ 1958, \apj, 128, 259 

\bibitem[Hodgkin et al.(2009)]{2009MNRAS.394..675H} Hodgkin, S.~T., Irwin,
M.~J., Hewett, P.~C., \& Warren, S.~J.\ 2009, \mnras, 394, 675

\bibitem[Kalari et al.(2015)]{2015MNRAS.453.1026K} Kalari, V.~M., Vink,
J.~S., Drew, J.~E., et al.\ 2015, \mnras, 453, 1026

\bibitem[K{\'o}sp{\'a}l et al.(2011)]{2011A&A...527A.133K} K{\'o}sp{\'a}l, {\'A}., {\'A}brah{\'a}m, P., Acosta-Pulido, J.~A., et al.\ 2011, \aap, 527, A133 

\bibitem[Kuhn et al.(2015)]{2015ApJ...802...60K} Kuhn, M.~A., Getman,
K.~V., \& Feigelson, E.~D.\ 2015, \apj, 802, 60

\bibitem[Laugalys et al.(2006)]{2006BaltA..15..327L} Laugalys, V., Strai{\v z}ys, V., Vrba, F.~J., et al.\ 2006, Baltic Astronomy, 15, 327 

\bibitem[Laugalys et al.(2006)]{2006BaltA..15..483L} Laugalys, V., Strai{\v z}ys, V., Vrba, F.~J., et al.\ 2006, Baltic Astronomy, 15, 483 

\bibitem[Laugalys et al.(2007)]{2007BaltA..16..349L} Laugalys, V., Strai{\v z}ys, V., Vrba, F.~J., et al.\ 2007, Baltic Astronomy, 16, 349 

\bibitem[Lawrence et al.(2007)]{2007MNRAS.379.1599L} Lawrence, A.,
Warren, S.~J., Almaini, O., et al.\ 2007, \mnras, 379, 1599

\bibitem[Lee et al.(2015)]{2015ApJ...807...84L} Lee, J.-E., Park, S., Green, J.~D., et al.\ 2015, \apj, 807, 84 

\bibitem[Liebhart et al.(2014)]{2014A&A...570L..11L} Liebhart, A., G{\"u}del, M., Skinner, S.~L., \& Green, J.\ 2014, \aap, 570, L11 

\bibitem[Lin et al.(2012)]{2012ApJ...756...27L} Lin, D., Webb, N.~A., \&
Barret, D.\ 2012, \apj, 756, 27

\bibitem[Miller et al.(2011)]{2011ApJ...730...80M} Miller, A.~A., Hillenbrand, L.~A., Covey, K.~R., et al.\ 2011, \apj, 730, 80 

\bibitem[Pillitteri et al.(2010)]{2010A&A...519A..34P} Pillitteri, I.,
Sciortino, S., Flaccomio, E., et al.\ 2010, \aap, 519, A34

\bibitem[Pillitteri et al.(2013)]{2013ApJ...768...99P} Pillitteri, I.,
Wolk, S.~J., Megeath, S.~T., et al.\ 2013, \apj, 768, 99

\bibitem[Preibisch \& Feigelson(2005)]{2005ApJS..160..390P} Preibisch,
T., \& Feigelson, E.~D.\ 2005, \apjs, 160, 390

\bibitem[Prisinzano et al.(2005)]{2005A&A...430..941P} Prisinzano, L.,
Damiani, F., Micela, G., \& Sciortino, S.\ 2005, \aap, 430, 941

\bibitem[Prisinzano et al.(2008)]{2008ApJ...677..401P} Prisinzano, L.,
Micela, G., Flaccomio, E., et al.\ 2008, \apj, 677, 401

\bibitem[Rebull et al.(2011)]{2011ApJS..193...25R} Rebull, L.~M., Guieu, S., Stauffer, J.~R., et al.\ 2011, \apjs, 193, 25 

\bibitem[Reipurth \& Schneider(2008)]{2008hsf1.book...36R} Reipurth, B.,
\& Schneider, N.\ 2008, Handbook of Star Forming Regions, Volume I, 4, 36

\bibitem[Rieke \& Lebofsky(1985)]{1985ApJ...288..618R} Rieke, G.~H., \&
Lebofsky, M.~J.\ 1985, \apj, 288, 618

\bibitem[Semkov et al.(2010)]{2010A&A...523L...3S} Semkov, E.~H., Peneva, S.~P., Munari, U., Milani, A., \& Valisa, P.\ 2010, \aap, 523, L3 

\bibitem[Semkov et al.(2012)]{2012A&A...542A..43S} Semkov, E.~H., Peneva, S.~P., Munari, U., et al.\ 2012, \aap, 542, A43 

\bibitem[Skinner et al.(2007)]{2007AAS...211.6227S} Skinner, S.~L.,
Briggs, K.~R., Guedel, M., \& Sokal, K.~R.\ 2007, Bulletin of the
American Astronomical Society, 39, 62.27 

\bibitem[Strai{\v z}ys et al.(2008)]{2008BaltA..17..125S} Strai{\v z}ys, V., Corbally, C.~J., \& Laugalys, V.\ 2008, Baltic Astronomy, 17, 125 

\bibitem[Toujima et al.(2011)]{2011PASJ...63.1259T} Toujima, H., Nagayama, T., Omodaka, T., et al.\ 2011, \pasj, 63, 1259 

\bibitem[Welin(1973)]{1973A&AS....9..183W} Welin, G.\ 1973, \aaps, 9, 183 

\bibitem[Witham et al.(2008)]{2008MNRAS.384.1277W} Witham, A.~R.,
Knigge, C., Drew, J.~E., et al.\ 2008, \mnras, 384, 1277

\bibitem[Wright et al.(2010)]{2010AJ....140.1868W} Wright, E.~L.,
Eisenhardt, P.~R.~M., Mainzer, A.~K., et al.\ 2010, \aj, 140, 1868-1881

\end{thebibliography}

\end{document}